\documentclass[useAMS,usedcolumn,eps,psfig,graphics]{mn2e}
\usepackage{epsfig, graphicx,natbib2}

\newcommand{\ltsimeq}{\raisebox{-0.6ex}{$\,\stackrel
        {\raisebox{-.2ex}{$\textstyle <$}}{\sim}\,$}}
\newcommand{\gtsimeq}{\raisebox{-0.6ex}{$\,\stackrel
        {\raisebox{-.2ex}{$\textstyle >$}}{\sim}\,$}}

\def\grtsim{\mathrel{\hbox{\rlap{\hbox{\lower2pt\hbox{$\sim$}}}\raise2pt\hbox{$>$}}}}
\def\lesssim{\mathrel{\hbox{\rlap{\hbox{\lower2pt\hbox{$\sim$}}}\raise2pt\hbox{$<$}}}}

\def\degree{\nobreak\ifmmode{^\circ}\else{$^\circ$}\fi}

\newcommand{\mc}{\multicolumn}

\newcommand{\aap}{A\&A}

\newcommand{\aj}{AJ}
\newcommand{\apj}{ApJ}
\newcommand{\apjl}{ApJL}
\newcommand{\apjs}{ApJS}

\newcommand{\araa}{ARA\&A}
\newcommand{\mnras}{MNRAS}
\newcommand{\nat}{Nat}

\newcommand{\pasp}{PASP}

\begin{document}

\topmargin -0.5in 

\title[A population of high-redshift type-2 quasars-I]{A population of high-redshift type-2 quasars-I. Selection Criteria and Optical Spectra}
\author[A. Mart\'\i nez-Sansigre et al.]{Alejo Mart\'\i nez-Sansigre$^{1}$\thanks{E-mail: a.martinez-sansigre1@physics.oxford.ac.uk (AMS)}, 
Steve Rawlings$^{1}$, Mark Lacy$^{2}$, Dario Fadda$^{2}$,
\and  
Matt J. Jarvis$^{1}$, Francine R. Marleau$^{2}$, Chris Simpson$^{3,4}$, Chris J. Willott$^{5}$\\
\footnotesize\\
$^{1}$Astrophysics, Department of Physics, University of
Oxford, Keble Road, Oxford OX1 3RH, UK\\
$^{2}$Spitzer Science Center, California Institute of Technology, MS220-6, 1200 E. California Boulevard, Pasadena, CA 91125, USA\\
$^{3}$Department of Physics, University of Durham, South Road, Durham DH1 3LE, UK\\
$^{4}$Astrophysics Research Institute, Liverpool John Moores University, Twelve Quays House, Egerton Wharf, Birkenhead CH41 1LD\\
$^{5}$Herzberg Institute of Astrophysics, National Research Council, 5071 West Saanich Rd, Victoria, B.C. V9E 2E7, Canada\\
}

\pagerange{\pageref{firstpage}--\pageref{lastpage}} \pubyear{}

\maketitle

\label{firstpage}

\begin{abstract}
We discuss the relative merits of mid-infrared and X-ray selection of
type-2 quasars. We describe the mid-infrared, near-infrared and radio
selection criteria used to find a population of redshift $z\sim2$
type-2 quasars which we previously argued suggests that most
supermassive black hole growth in the Universe is obscured \citep
{2005Natur.436..666M}. We present the optical spectra obtained from
the William Herschel Telescope, and we compare the narrow emission
line luminosity, radio luminosity and maximum size of jets to those of
objects from radio-selected samples. This analysis suggests that these
are genuine radio-quiet type-2 quasars, albeit the radio-bright end of
this population. We also discuss the possibility of two different
types of quasar obscuration, which could explain how the $\sim$2-3:1
ratio of type-2 to type-1 quasars preferred by modelling our
population can be reconciled with the $\sim$1:1 ratio predicted by
unified schemes.
\end{abstract}


\section{Introduction}

For a long time, a population of obscured (type-2) radio-quiet quasars
has been postulated from modelling the hard (energies $\geq$ 1 keV)
X-ray background as the sum of the emission from extragalactic
sources. \citet{1995A&A...296....1C} assumed a population of absorbed
active galactic nuclei (AGN: in this paper we use the term to
encompass both Seyferts and quasars), with the same intrinsic spectrum
and evolution as unobscured AGN, but with a range of obscuring
Hydrogen column densities, $N_{\rm H}$, between $10^{25}$ and
$10^{29}$ m$^{-2}$. The resulting best-fit ratio for absorbed ($N_{\rm
H} > 10^{26.5}$ m$^{-2}$) to unabsorbed ($N_{\rm H} < 10^{26.5}$ m$^{-2}$)
AGN was found to be $\sim$3:1.  The model assumed
the same distribution of obscuring column densities and redshifts for the
low-X-ray luminosity objects ($L_{\rm X} < 10^{37}$ W; the Seyferts)
and the high-X-ray luminosity objects ($L_{\rm X} > 10^{37}$ W; the
quasars), including a population of highly obscured AGN (the
Seyfert-2s and type-2 quasars).

According to unified schemes \citep{1993ARA&A..31..473A}, type-2 quasars would
consist of quasars with the symmetry axes perpendicular to the
observer's line of sight, so that the dusty torus around the accretion
disk would be viewed edge on, obscuring the optically-bright accretion
disk. The gas present in this torus will obscure the X rays via
photoelectric absorption and Compton scattering. This unfavourable
orientation means type-2 quasars do not outshine their host galaxy at
(rest-frame) ultra-violet or optical wavelengths, as is the case with
the unobscured (type-1) quasars. Type-2 quasars are therefore
indistinguishable from normal galaxies in optical imaging surveys. If the
torus has a half-opening angle of $\sim$ 40\degree, as appears to be
the case in radio-loud samples \citep[e.g.][]{2000MNRAS.316..449W}, the
population of type-2s is expected to be comparable in size to the
type-1 population and not to outnumber it by a factor of $\sim$3.

A modification to the standard unified scheme is the ``receding
torus'' model \citep {1991MNRAS.252..586L} in which the more luminous
AGN sublimate the dust in the inner edge of the torus out to larger
distances than the less luminous AGN. This leads to a larger opening
angle for more luminous AGN, so the fraction of type-1 quasars
increases as a function of bolometric luminosity. Such luminosity
dependence of the type-2 to type-1 ratio is indeed found, for example,
in radio-selected samples by \citet {2000MNRAS.316..449W}, in
X-ray-selected samples by \citet {2003ApJ...598..886U}, and in
spectroscopically selected samples by \citet {2005MNRAS.360..565S}.

More recent modelling of the hard X-ray background has consistently
shown signs of a large population of obscured AGN \citep
{1999MNRAS.309..862W,2005MNRAS.357.1281W} and synthesis models have
generally required a type-2 to type-1 ratio $\sim 3-4$:1 \citep
{2001A&A...366..407G,2003ApJ...598..886U,2005ApJ...630..115T}. However,
the bulk of the sources required to fit the unresolved X-ray
background have moderate redshifts ($z = 0.5-1$) and X-ray
luminosities ($L_{\rm X} \sim 10^{36}$\,W) characteristic of Seyferts
rather than quasars. The latest studies of the hard X-ray background
indeed suggest a luminosity-dependent Compton-thin-type-2 to type-1
ratio, which decreases to $\ltsimeq$1:1 at the higher luminosities
corresponding to quasars \citep{2003ApJ...598..886U,
2005ApJ...630..115T}. The true type-2 to type-1 ratio will remain
unknown until the Compton-thick population is unveiled.

Deep hard X-ray surveys have now revealed a large population of
obscured AGN \citep [e.g.][]{2003AJ....126..539A}, but again most of these
sources have moderate X-ray luminosities ($L_{\rm X} < 10^{37}$\,W)
and are better described as Seyfert-2s rather than type-2
quasars. Amongst objects with X-ray luminosities large enough to
qualify as quasars, the ratio of type-2 to type-1 objects appears to
be only $\sim$1:1 \citep[e.g.][] {2004ApJS..155...73Z}, although redshift
completeness at faint optical magnitudes (m$_{\rm R}> 23$, Vega) may
still be an issue. Despite the great progress with the deep hard X-ray
surveys, at high energies (above 6-8 keV) only $\sim 50\%$ of the
X-ray background is accounted for by individual sources, suggesting a
substantial population of Compton-thick AGN which contribute to the
hard X-ray background but are missing from current flux-limited X-ray
surveys \citep {2005MNRAS.357.1281W}. Indeed, a good example is the
type-2 quasar IRAS FSC 10214+4724: this object, found in an IRAS 60 $\mu$m
survey \citep {1991Natur.351..719R}, has recently been found to be
Compton-thick \citep {2005MNRAS.357L..16A}.

An alternative strategy for looking for type-2 objects is  to
look for the mid-infrared emission. IRAS FSC 10214 +4724, at $z \sim
2.3$, is a classic type-2 object \citep[e.g][]
{1998MNRAS.298..321S}. However, due to the limited sensitivity of
IRAS, this object was only detected due to a huge gravitational lens
magnification ($\sim$ 50) of the flux \citep
{1995ApJ...450L..41B}. The dramatic increase of mid-infrared
sensitivity allowed by the Spitzer Space Telescope \citep
{2004ApJS..154....1W} means that similar objects can now be found
without the `benefit' of gravitational lensing, and mid-infrared 
selection should be sensitive to the Compton-thick
quasars missed by X-rays.

Following the success of mid-infrared selection in finding type-2
radio-quiet quasars at $z \sim 0.5$ \citep {2004ApJS..154..166L,
2005MmSAI..76..154L}, we combined data from the Spitzer MIPS
instrument at 24 $\mu$m \citep{2004ApJS..154...66M,Fadda:2006mx}, IRAC at 3.6 $\mu$m \citep {2005ApJS..161...41L}
and from the Very Large Array (VLA) at 1.4 GHz
\citep{2003AJ....125.2411C} and devised strict selection criteria to
hunt for higher redshift type-2s \citep{2005Natur.436..666M}. The
trade-off between our method and X-ray selection, is that our quasars
are intrinsically more luminous than AGN found in the deep hard
X-ray surveys.

The format of this paper is as follows. In Section \ref{sec:Xray} we
compare X-ray and mid-infrared selection of obscured quasars. In
Section \ref{sec:selcrit} we explain in detail the selection criteria
which we used to find our sample of type-2 quasars. Section
\ref{sec:obs} deals with the observations while Section
\ref{sec:spectra} describes the optical spectra and Section
\ref{sec:photoz} compares the spectroscopic and photometric
redshifts. In Section \ref{sec:composite} we make the first composite
radio-quiet type-2 spectrum. Section \ref{sec:radio} is dedicated to
comparing our sample to the well studied radio-loud samples. Finally,
Section \ref{sec:discussion} summarises and discusses the implications
of this population of type-2 quasars. Throughout this paper we adopt a
$\Lambda$CDM cosmology with the following parameters: $h = H_{0} /
(100 ~ \rm km ~ s^{-1} ~ Mpc^{-1}) = 0.7$; $\Omega_{\rm m} = 0.3$;
$\Omega_{\Lambda} = 0.7$.

\section{Comparison between X-ray and Mid-infrared Selection}\label{sec:Xray}

In the unified scheme for AGN \citep {1993ARA&A..31..473A}, the
central engine is surrounded by a torus of dust which will obscure the
broad-line region from certain lines-of-sight. The optical
classification for quasars is that type-1s should show broad lines
(and strong optical-UV continuum) while the type-2s are those objects
where the dusty torus is obscuring the broad line region, so only
narrow lines are seen. This classification is concerned with the
optical/UV properties, and \citet {1999MNRAS.306..828S} found that the
dividing extinction for radio-selected (3CR and 3CRR) objects
corresponded to $A_{\rm V} \gtsimeq 5$, with a low fraction of objects
with $A_{\rm V}$ in the range 5-15 and a large fraction with $A_{\rm
V} \gtsimeq 15$. Note also that the \citet {1999MNRAS.306..828S} study
found that most type-1s ($A_{\rm V} \ltsimeq 5$) had $A_{\rm V} \sim
0$, with only a small fraction lightly reddened ($A_{\rm V} \sim 1-5$).
Gas and dust are responsible for obscuring the X-rays emitted from the
central engine, and the geometrical distribution of the gas is almost
certainly different from that of the dust. It is possible to imagine
lines-of-sight which ``graze'' the dusty torus, leading to small
amounts of extinction in the optical, but which go through significant
additional amounts of gas. Such a line-of-sight would lead to a quasar
classified as ``reddened'' (not type-2) in the optical ($A_{\rm V}
\simeq 1-5$) but as type-2 ($N_{\rm H} \gtsimeq 10^{26}$\,m$^{-2}$) in
X-rays. The converse situation, where an optically obscured quasar is
barely absorbed in X-rays, is harder to envisage: any line of sight
that passes through significant amounts of dust will lead to
significant absorption in the X-rays. It is therefore unlikely that
optical type-2 quasars will have negligible X-ray absorption. However,
in edge-on sources, the nuclear soft X-ray emission can plausibly be
scattered from an optically thin material located above or below the
plane of the torus, leading to an optically-obscured AGN with a soft
X-ray spectrum. Finally, obscuring material in the host galaxy of the
AGN can further complicate interpretation. Independently of the
orientation of the torus, dust and gas in the host galaxy can cause
extinction and absorption with a huge range in $A_{\rm V}$ and $N_{\rm
H}$.

X-ray and optical definitions of obscured AGN are therefore slightly
different and the range of gas-to-dust ratios found by comparing
dust-reddening in the optical or near-infrared and X-ray absorption
suggests they are not always matched \citep[e.g.]
[]{2003MNRAS.339..397W,2005ApJ...627...75U,2005ApJ...634..183W}.  We
note that we have used a mid-infrared plus radio selection, but since
the role of the radio criterion was to avoid non-AGN contaminants (see
Section~\ref{sec:selcrit}), in this Section we can proceed to consider
our technique as basically mid-infrared selection. The effect of
adding the radio flux density cuts is to constrain the type-2 quasars
selected to those at the high end of the radio-to-optical correlation
for radio-quiet quasars of \citet{2003MNRAS.346..447C}, as described
in in Section \ref {sec:selcrit} and in \citet
{2005Natur.436..666M}. It is of course plausible that radio luminosity
correlates in some complicated way with joint mid-infrared, X-ray
detectability but we ignore this possibility here.

This section aims to compare the ``merits'' of mid-infrared and X-ray
selection by considering a model quasar (described in section \ref
{sub:modelquasar}) with varying amounts of dust extinction (section
\ref{sub:dust}) and different gas-to-dust ratios (section
\ref{sub:gastodust} ). We also consider the effects of different types
of dust (section \ref{sub:dusttype}) and the orientation dependence of
24-$\mu$m emission (section \ref{sub:orientation}).

\subsection{Model quasar}\label{sub:modelquasar}

We choose this quasar to have
$M_{\rm B} = -25.7$ since this corresponds to the break in the optical
quasar luminosity function at $z = 2$ \citep {2004MNRAS.349.1397C},
the redshift of interest in the analysis in \citet
{2005Natur.436..666M}.  The model quasar is assumed to have an
intrinsic unreddened type-1 spectral energy distribution (SED), which we take
from \citet {1995MNRAS.272..737R}. This model covers the range between
the (rest-frame) far-infrared and 4 keV. We need to complement this
with a spectral energy distribution or SED that goes into the hard X-rays, so we assume
the \citet {1994MNRAS.270L..17M} spectrum with the form $L_{\nu}
\propto \nu^{-0.9}\rm exp(-\nu/\nu_{c})$ with $\rm h\nu_{c} =
360~$keV. This intrinsic unabsorbed type-1 spectrum is chosen as it was used by
\citet {1999MNRAS.309..862W} for the models we use for absorbed X-ray
spectra.  This is practically a flat X-ray
SED, and in the range probed by observations at 24 $\mu$m (for $z \leq
5$), the SED is also flat ($L_{\nu} \propto \nu^{-1}$ so $\nu L_{\nu}
\propto \nu^{0}$). We choose the normalisation to match the \citet
{1995MNRAS.272..737R} SED at 2 keV.

\subsection{Obscuration by dust}\label{sub:dust}

Starting with this type-1 SED, we proceed to model the obscuration of
it with dust and gas. In recent times, the type of dust present in
type-1 quasars \citep [from the SDSS sample of][]
{2002AJ....123.2945R} has been described as ``grey'' and alternatively
as similar to that of the Small Magellanic Clouds (SMC). Grey dust
shows a relatively flat extinction law in the optical and UV, while
SMC-type dust is characterised by a steep increase in the extinction
at UV wavelengths. Milky Way (MW)-type dust has a UV extinction law
which sits in between that of grey and SMC. \citet
{2003AJ....126.1131R} and \citet {2004AJ....128.1112H} have found the
dust in the SDSS type-1 sample to be closest to that found in the SMC, while
\citet {2004MNRAS.348L..54C} and \citet {2004ApJ...616..147G} argued
that extinction curves flatten in the UV (grey dust). However,
\citet{2005ApJ...627L.101W} has shown this implication of grey dust to
be a selection effect.  The main
selection effect arises from measuring dust-reddening from a composite
constructed from a flux limited sample: the UV part of the composite
is constructed from the higher redshift quasars, which will only make
the sample if they are intrinsically brighter and have very little
extinction due to dust. This leads to a negative correlation between
$E(B-V)$ and redshift, so that different parts of the composite
spectrum are dominated by quasars with different amounts of dust. The
result is a derived dust extinction curve which appears to flatten in
the UV (see Willott, 2005, for more details).

\begin{figure}
\begin{center}
\psfig{file=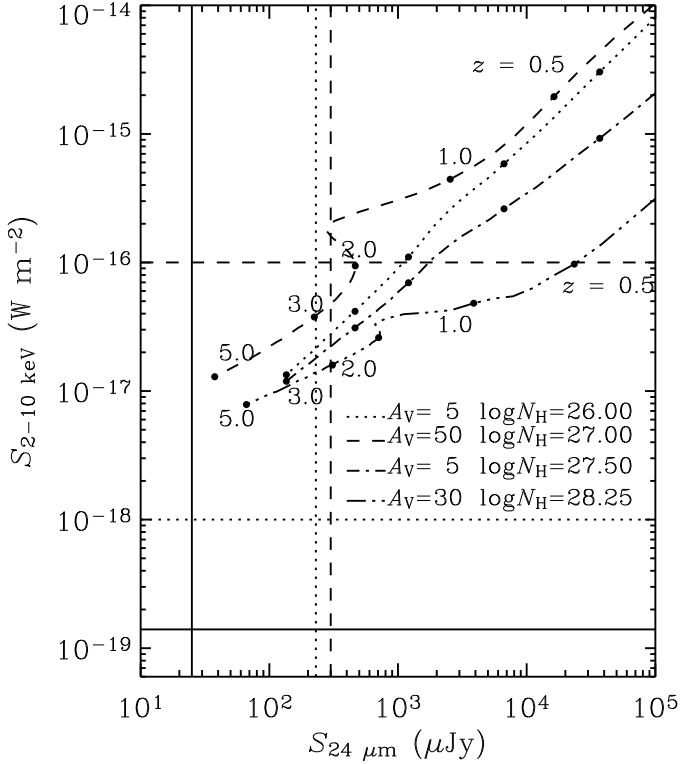,width=15cm,angle=0} 
\caption{\noindent Hard X-ray flux ($S_{2-10 \rm keV}$) versus
24-$\mu$m flux density ($S_{24~\mu \rm m}$) as a function of redshift
for the model quasar described in Section~\ref{sec:Xray}.  The tracks
have the redshifts labelled at the positions marked with black dots at
$z=$ 0.5, 1, 2, 3 and 5, and they show the effect on the quasar of
MW-type dust with two different gas-to-dust ratios: $N_{\rm H} = 1.9
\times 10^{25}$ m$^{-2} \times A_{\rm V}$ (Milky-Way, top two lines)
and $N_{\rm H} = 5 \times 10^{26}$ m$^{-2} \times A_{\rm V}$ (bottom
two lines). Two tracks are drawn for each gas-to-dust ratio, $A_{\rm
V}= 5$ and 50 for the MW gas-to-dust ratio and $A_{\rm V}= 5$ and 30
for the higher gas-to-dust ratio. The legend shows the combination of
$A_{\rm V}$ and $N_{\rm H}$ [in log${10}(N_{\rm H}/\rm m^{-2})$] for
each track. For the larger $A_{\rm V}$, the 9.7 $\mu$m silicate
absorption feature can be seen as a ``meander''.  The dashed straight
lines mark the limits for deep surveys covering large areas ($\sim
4-5$ deg$^{2}$), while the solid straight lines show the limits for
the GOODS fields, ultra-deep surveys which cover $\sim$ 0.1
deg$^{2}$. As an example of the large area Spitzer surveys, we have
taken the flux density limit for 24-$\mu$m from the FLS (flux limit
from Mart\'\i nez-Sansigre et al., 2005, catalogue from Fadda et al.,
2006), while the X-ray flux is taken from the ASCA survey of \citet
{1999ApJ...518..656U}. For the deeper GOODS surveys, the 24-$\mu$m
flux density limit is as quoted by \citet{2006ApJ...640..603T} and the
X-ray flux limit is from \citet{2003AJ....126..539A}. We note that the
latter is actually in the 2-8 keV band (not 2-10 keV quoted here) but
this relatively small difference does not alter our argument. Finally,
the dotted lines are for the SXDF limit, a deep survey which has
$\sim$1 deg$^{2}$ of deep X-ray data with 24-$\mu$m data from SWIRE
\citep [][Ueda et al. in prep]{2003PASP..115..897L}.}
\label{fig:Xraycomp}
\end{center}
\end{figure}

We use the extinction curves from \citet {1992ApJ...395..130P}, with
the extinction at (observed) 24 $\mu$m given in terms of the
extinction at rest-frame visual band, $A_{\rm V}$, and a particular
dust type. The quasars in the SDSS sample are all type-1s, and the
dust discussed in the above paragraph is not necessarily associated
with the torus. The SMC has sub-solar metallicity, while the nuclear
regions around quasars have solar or super-solar metallicities, with
dust properties probably closer to the MW.  The effects of different
types of dust are secondary compared to the gas-to-dust ratio, yet
important, so we assume MW-dust in Figure~\ref{fig:Xraycomp} and
discuss the effect of large magellanic cloud (LMC) and SMC-type dust
in Section \ref{sub:dusttype} and in Figure~\ref{fig:dustcomp}.  Dust
extinction is small at rest-frame 24~$\mu$m, yet the extreme obscuring
columns present in the torus and the large redshifts discussed here
mean that dust obscuration is not necessarily negligible at observed
24 $\mu$m. In particular, the 9.7 $\mu$m silicate absorption feature
can have a very important effect on observed 24-$\mu$m flux density at
$z \sim 1.5$ (and this feature is significantly deeper for SMC-type
dust than for MW-type dust, Figure~\ref{fig:dustcomp}).

\subsection{X-ray absorption}\label{sub:gastodust}

In contrast to dust extinction, which increases with redshift, X-ray
absorption decreases at higher frequencies, meaning X-ray samples
benefit from a ``negative K-correction''. The X-ray spectra of heavily
absorbed quasars can only be modelled correctly via Monte-Carlo
simulations, so we use the models of \citet
{1999MNRAS.309..862W}. These assume the intrinsic unabsorbed type-1
spectrum metioned earlier, and assume an Fe abundance 5-times solar,
which provides a better fit to the hard X-ray backround. The
particular models which we use here ignore the reflection component
of X-ray spectra, as it has little effect on our discussion, and have
a 2\%  scattered component. The obscuring column of Hydrogen, responsible
for both types of absorption, is given here as a function of $A_{\rm
V}$ by assuming a gas-to-dust ratio.

The Milky Way is known to have a gast-to-dust ratio such that $N_{\rm
H} = 1.9 \times 10^{25}$ m$^{-2} \times A_{\rm V}$ \citep[][ note this is per
magnitude of extinction in V band]{1978ApJ...224..132B}, while the SMC
has $N_{\rm H} = 1.8 \times 10^{26}$ m$^{-2}  \times A_{\rm V}$
\citep{1985A&A...149..330B}.  Many AGN have been found to have higher
gas-to-dust ratios than the MW \citep
{2001A&A...365...28M,2004ApJ...610..140W}. We therefore choose two
representative values for the gas-to-dust ratio. The MW value is
sometimes used as a ``standard'' gas-to-dust ratio. To show the
effects of higher gas-to-dust ratios, we note that, for example, NGC
3281 has a value of $N_{\rm H}= 3.2 \times 10^{26}$ m$^{-2} \times A_{\rm V}$
\citep {1998ApJ...509..653S}, while \citet {2005ApJ...627...75U} study 
one object with $N_{\rm H} \sim 7.7 \times 10^{26}$ m$^{-2} \times A_{\rm V}$
(FTM1004+1229, assuming a factor $R = 3.0$, the average between the SMC
and MW factors, to convert between $E(B-V)$ and $A_{\rm V}$). We
therefore choose $N_{\rm H} = 5 \times 10^{26}$ m$^{-2} \times A_{\rm V}$ as a
realistic higher value to illustrate the effect of different
gas-to-dust ratios. This is not claiming we encompass the whole range
in gas-to-dust ratios, but is chosen to illustrate how mid-infrared and
X-ray selection are affected by variations in gas-to-dust ratio. The higher value of gas-to-dust ratio
$N_{\rm H} = 5 \times 10^{26}$ m$^{-2} \times A_{\rm V}$ is high enough for an
$A_{\rm V} = 20$ type-2 quasar to be Compton-thick.

\subsection{Comparison}\label{sub:comparison}

Figure \ref {fig:Xraycomp} shows the X-ray flux in the 2-10 keV band
($S_{2-10 \rm keV}$) versus the 24-$\mu$m flux density ($S_{24 \mu \rm
m}$) for type-2 quasars with a range in $A_{\rm V}$. MW-type dust is
used to obscure the 24-$\mu$m emission, and two different gas-to-dust ratios
are used. The top set of curves are for a MW gas-to-dust ratio
($N_{\rm H} = 1.9 \times 10^{25}$ m$^{-2}  \times A_{\rm V}$), while the bottom
curves are for $N_{\rm H} = 5 \times 10^{26}$ m$^{-2} \times A_{\rm V}$.  For
the higher value of $A_{\rm V}$, the 9.7 $\mu$m silicate absorption
feature can be seen as a ``meander''. The caption in Figure \ref
{fig:Xraycomp} details the survey limits used for comparative purposes.

The effect of dust at observed 24 $\mu$m is small in the range $5
\ltsimeq A_{\rm V} \ltsimeq 50$.  However, it is important to note
that at $A_{\rm V} \sim 50$, type-2s will start to drop out of
24-$\mu$m selection (for the larger area surveys) at $z \sim 1.5$ (due
to the silicate absorption) and at $z \gtsimeq 2.5$ due to significant
dust extinction at the shorter rest-frame wavelengths. For larger
values of $A_{\rm V}$, the width of the silicate absorption line means
type-2s in the range $1.3 \leq z \leq 1.7$ drop out of the sample. As
long as the $A_{\rm V}$ is $< 50$, the model type-2 quasar would be
detectable by the large area 24-$\mu$m surveys at $z \ltsimeq 3$
\citep[FLS and SWIRE surveys: ][ respectively]
{Fadda:2006mx,2003PASP..115..897L}.

In the $\sim 5$ deg$^{2}$ X-ray ASCA survey \citep
{1999ApJ...518..656U} with flux limit 10$^{-16}$ W m$^{-2}$, the
model type-2 quasars with $N_{\rm H} \ltsimeq 2\times10^{27}$ m$^{-2}$ are
hardly affected by the absorption, and would be detected at $z
\ltsimeq 2$. For $N_{\rm H} \sim 3\times10^{27}$ m$^{-2}$ only $z
\ltsimeq 1.5$ type-2s are detected, and the ASCA survey would not have
the sensitivity to detect Compton-thick quasars with $N_{\rm H}
\gtsimeq 1\times10^{28}$ m$^{-2}$ above $z=0.5$.

Since 24-$\mu$m selection becomes poor for objects with $A_{\rm V} \gtsimeq
 50$ and X-ray selection fails for objects with $N_{\rm H} \gtsimeq
 10^{24}$ m$^{-2}$, we can see clearly how the gas-to-dust ratio
 determines which of the two selections is better. For surveys
 covering a large area ($\sim 4-5$ deg$^{2}$), 24-$\mu$m selection
 will be able to pick out type-2s with $N_{\rm H} \geq 10^{28}$
 m$^{-2}$ as long as they had $A_{\rm V} \ltsimeq 50$.  Therefore, for
 AGN with gas-to-dust ratios like that of the MW or higher, 24-$\mu$m selection
 would fare better than X-ray selection in these surveys. However, we do
 not know exactly the selection biases in measuring these gas-to-dust
 ratios, so the possibility of large numbers of AGN with gas-to-dust
 ratios lower than the MW is not excluded.

For  such a gas-to-dust ratio, significantly lower than the MW
(e.g. a tenth of the MW value), only quasars with $A_{\rm V} \gtsimeq
50$ would be classified as type-2s by X-ray observations. An extreme 
$A_{\rm V} \gtsimeq 5000$ would be required for the quasars to be
Compton-thick. Such objects would not be detected by the larger area  mid-infrared
samples, and therefore X-ray selection would be more sensitive.

This model type-2 would be detected in the  GOODS fields at the whole
range of redshifts concerned here, both at 24 $\mu$m and at 2-10 keV
\citep[survey fluxes from ][ for X-ray and 24-$\mu$m respectively]{2003AJ....126..539A,2006ApJ...640..603T}. So although the
Compton-thick quasars have not been detected in the ASCA survey, they
are, in theory X-ray detectable in the GOODS survey.  The reason they
have not been found in large numbers is  probably an issue of space density, as our
model type-2 is rare and large areas are required to find significant
numbers of these objects. GOODS therefore has the required
sensitivity to find hard-X-ray-selected Compton-thick type-2 quasars
at all redshifts $z \ltsimeq 5$, but they do not cover the area
required for this.

\begin{figure}
\begin{center}
\psfig{file=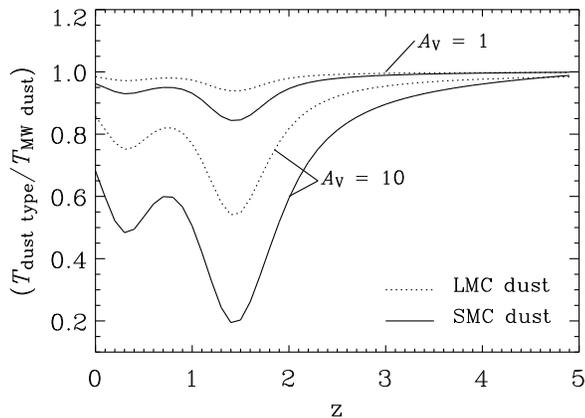,width=8.5cm,angle=0} 
\caption{\noindent Curves showing the ratio of transmission for two
dust types (LMC and SMC) over the
transmission for MW-type dust, using the dust models of
\citet{1992ApJ...395..130P}, and given two values of $A_{\rm V}$. The
transmission is here defined as $T = 10^{-A_{\lambda}/2.5}$. We use
$\lambda = 23.7~\mu \rm m/(1+z)$, the central wavelength of the MIPS
instrument at 24 $\mu$m, accounting for the redshift. $A_{\lambda}$
is parametrised as a function of $A_{\rm V}$ and is dependent on the
type of dust. We plot and label curves for $A_{\rm V} = $1 and 10, and
include LMC as well as SMC-type dust for comparative purposes.  The
different extinction curves show that, given an $A_{\rm V}$, the
different types of dust would have significantly different
mid-infrared properties, and in particular, type-2 quasars with
SMC-type dust will be more difficult to detect in a 24-$\mu$m survey
than those with MW-dust. At $z \grtsim 3$ the difference becomes
small, but around $z \sim 1.5$ SMC-type dust has very deep silicate
absorption feature which would make type-2s drop out of flux-limited
surveys. }
\label{fig:dustcomp}
\end{center}
\end{figure}

\begin{figure}
\begin{center}
\psfig{file=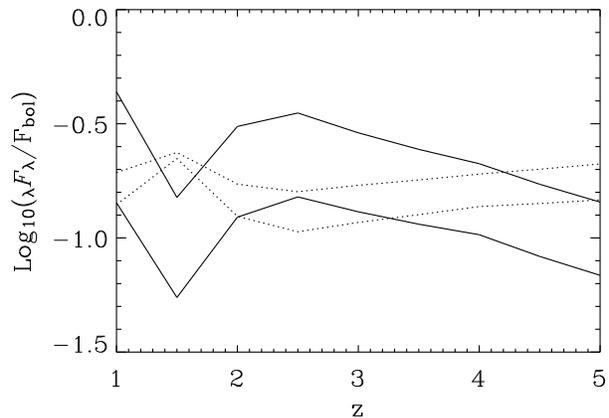,width=8.5cm,angle=0} 
\caption{\noindent Estimate of $\lambda F_{\lambda}$ at $\lambda = 24
\mu$m (observed) as a fraction of $F_{\rm bol}$ for two \citet
{1994MNRAS.268..235G} models with the following parameters: the
density has no radial dependence ($\beta=0$), a covering fraction of
0.8, ratio of outer radius $r_{\rm m}$ to inner radius of torus
$r_{\rm o}$ (where $r_{\rm o} \sim 1$ pc), $r_{\rm m}/r_{\rm o} = 100$
(top two curves) and $r_{\rm m}/r_{\rm o} = 1000$ (bottom two curves). The equatorial optical
depth at 0.3 $\mu$m, $\tau_{\rm e}$ is 80 (corresponding to $A_{\rm V}
\approx 50$). The dotted line is for an AGN viewed pole-on (a type-1)
while the solid line is for an edge-on view (type-2).  The reason why
the curves only have a few points is that the \citet
{1994MNRAS.268..235G} models have  few data points at the
relevant range of wavelengths, and therefore further smoothing of the
curves would  be an overinterpretation. This shows that the
24-$\mu$m flux is roughly a constant fraction of $F_{\rm bol}$ between
$2 \lesssim z \lesssim 4$ for both type-1 and type-2s. In this range
of redshifts, 24-$\mu$m is therefore a good wavelength to select AGN
independent of orientation. Once again we see the problems caused by
the silicate absorption at $z \sim 1.5$. At very high redshifts ($z
\grtsim 4$), observed 24-$\mu$m is now a relatively short wavelength
in the rest-frame, meaning obscuration by dust starts to have an
increasingly large effect.  The curves for both $r_{\rm m}/r_{\rm o}$
ratios are very similar, except for a normalisation, showing this
argument holds for a range of torus sizes.}
\label{fig:l24_lbol}
\end{center}
\end{figure}

Since the hard X-ray background is dominated by $z = 0.5-1$ Seyfert-2s,
and since Seyferts have a higher space density than quasars, a large
area is not as crucial. However, Seyfert-2s will have typical fluxes
$\ltsimeq 10-100$ times fainter than our model type-2 quasar, and
therefore, at all redshifts Seyfert-2s with $N_{\rm H} \geq 10^{28}$
m$^{-2}$ will have $S_{\rm X} \ltsimeq 10^{-19}-10^{-18}$ W m$^{-2}$. This was,
of course, already known, as the deep X-ray surveys in GOODS have not
been able to find a population of Seyfert-2s large enough to account
for the hard X-ray backround. This has been explained by a bias
against the most heavily obscured AGN both by X-ray observations and
optical spectroscopy \citep {2004ApJ...616..123T}, and our argument is
consistent with this.

\begin{figure}
\begin{center}
\psfig{file=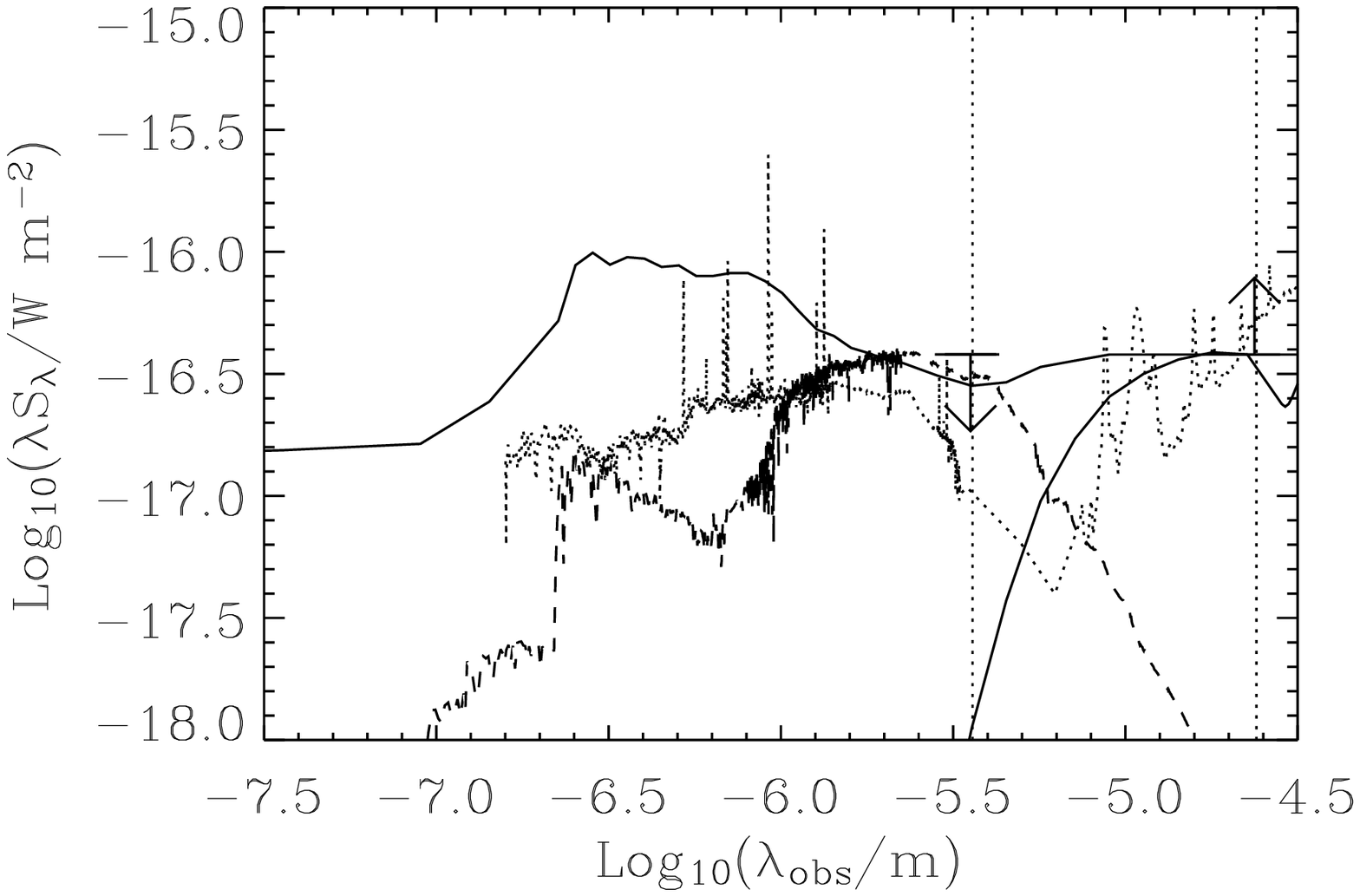,width=8.5cm,angle=0} 
\caption{\noindent Mid-infrared selection criteria plotted on 
observed spectral energy distributions (SED). The 24-$\mu$m criterion
is shown as a lower limit, and the 3.6-$\mu$m criterion as an upper
limit. The solid lines represent quasars at $z = 2$ bright enough to
make the 24-$\mu$m flux cut. The upper solid line is a type-1 quasar
\citep[SED from ][] {1995MNRAS.272..737R}, while the lower solid line
is a type-2 with $A_{\rm V}= 5$. A type-1 quasar just bright enough to
make the 24-$\mu$m cut will be too bright at 3.6 $\mu$m (once the host
galaxy contribution is added).  The dashed SED represents a $2.6L^{*}$
elliptical galaxy at $z=1.4$  \citep[from][] {2003MNRAS.344.1000B},
where it becomes just faint enough to make it fall below the 3.6-$\mu$m cut.
The dotted line is the ULIRG NGC 7714 redshifted to $z=0.4$ \citep[SED from][]{2004ApJS..154..188B}, which would meet our
mid-infrared criteria if it lay between $z=0.25$ and $z=0.4$.   }
\label{fig:sed}
\end{center}
\end{figure}

\begin{figure*}
\begin{center}
\hbox{
\hspace{-. cm}
\psfig{file= 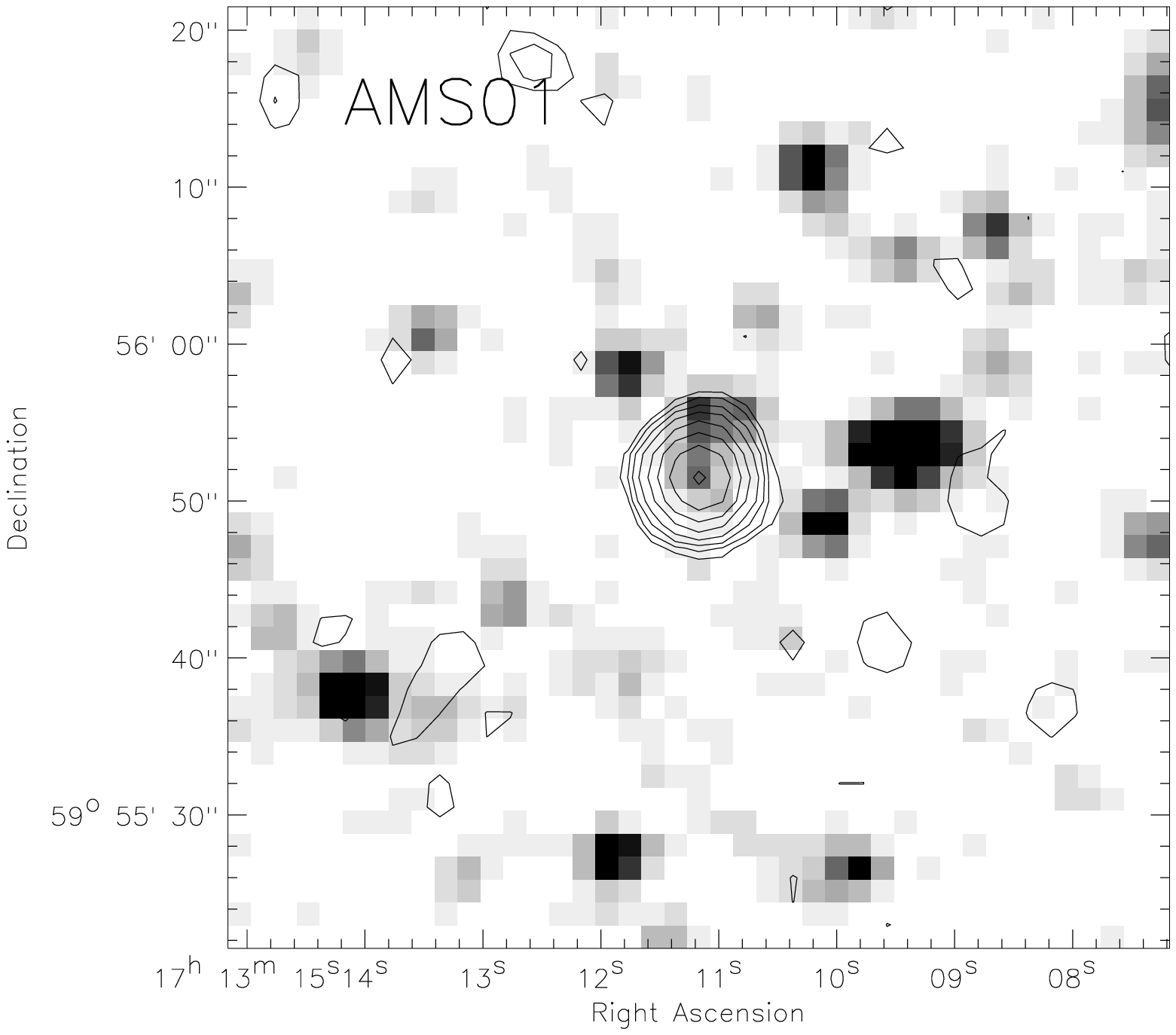,width=4.5cm,angle=0} 
\hspace{-.5cm}
\psfig{file= 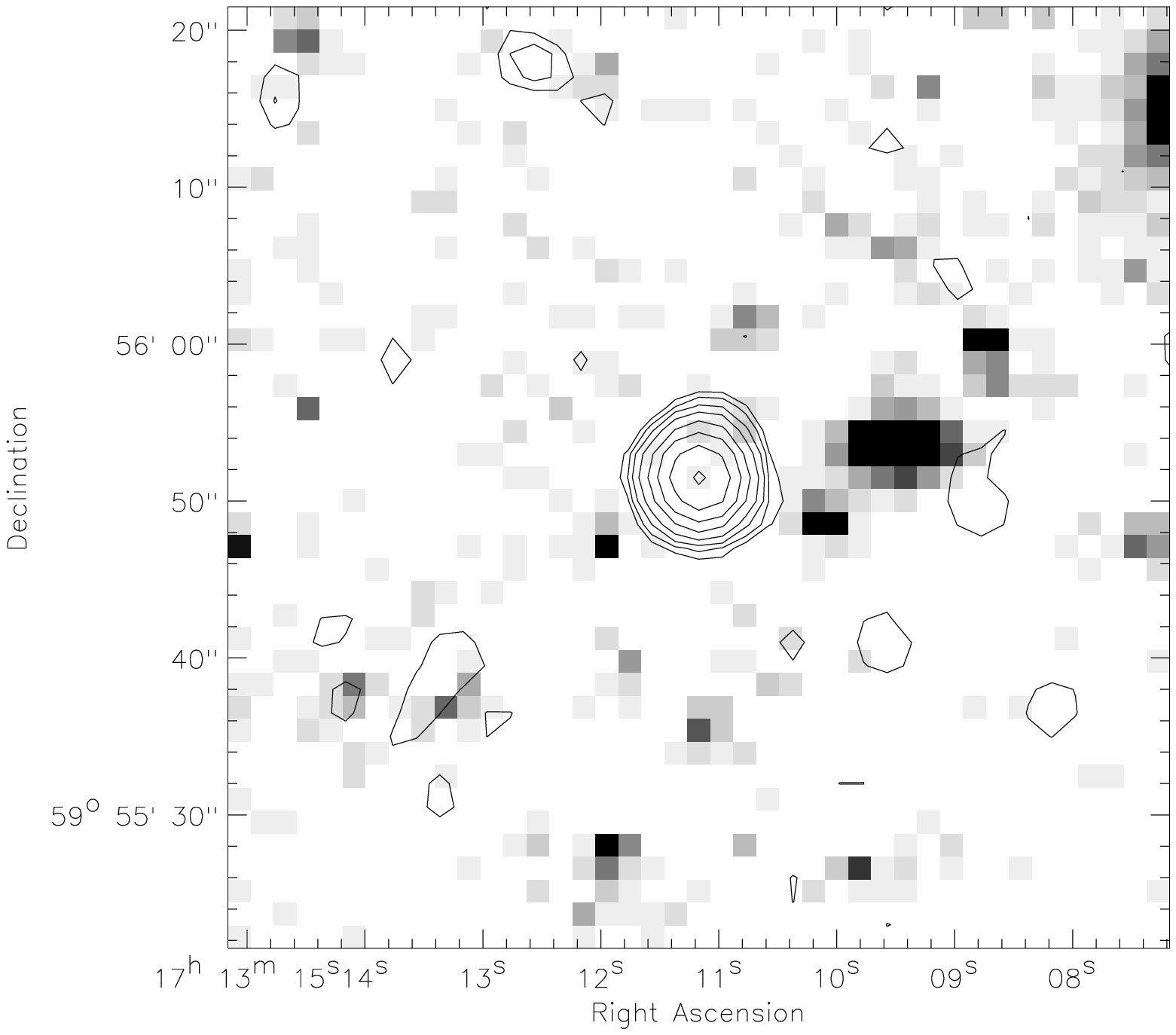,width=4.5cm,angle=0} 
\hspace{0. cm}
\psfig{file= 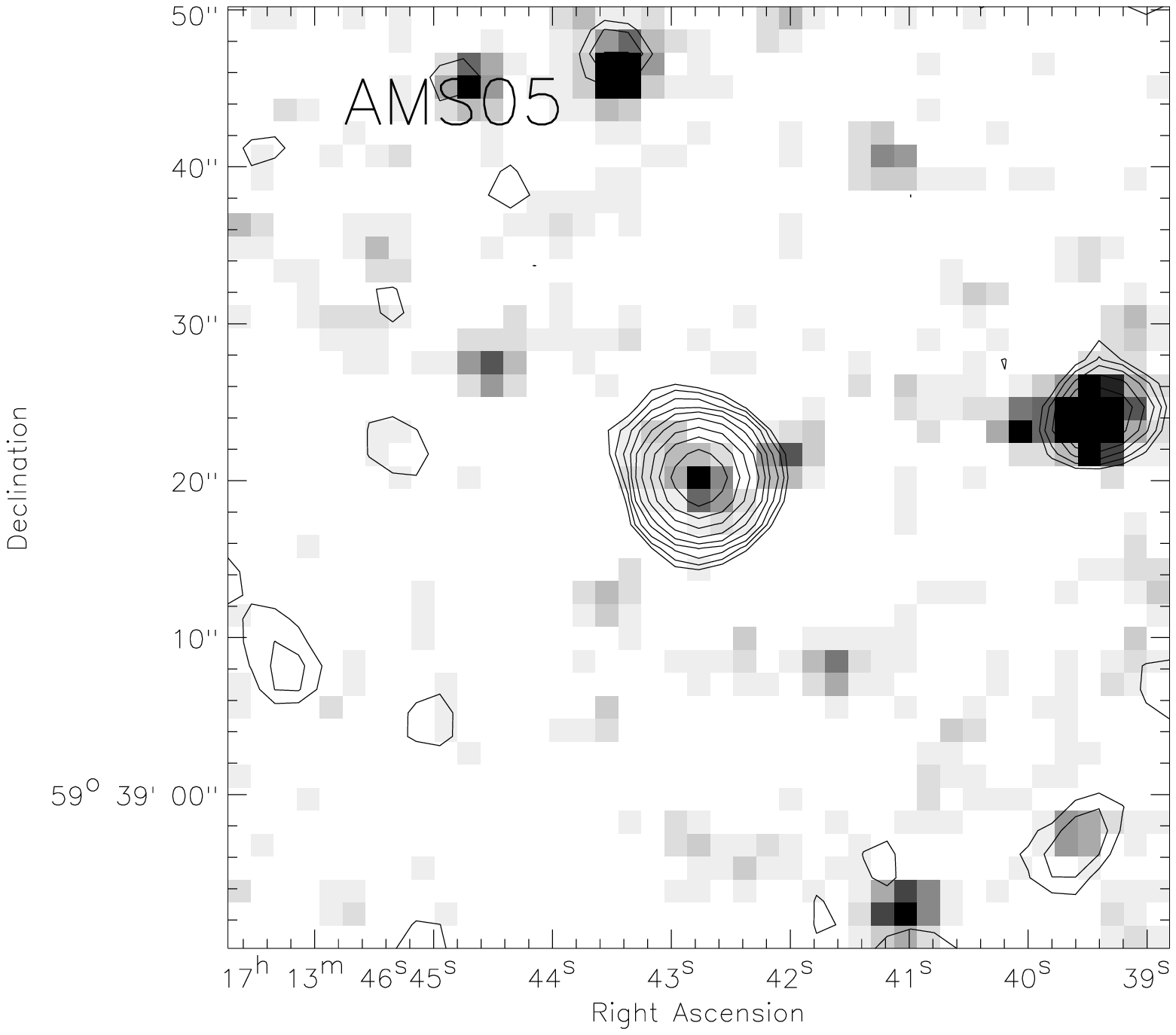,width=4.5cm,angle=0} 
\hspace{-.5cm}
\psfig{file= 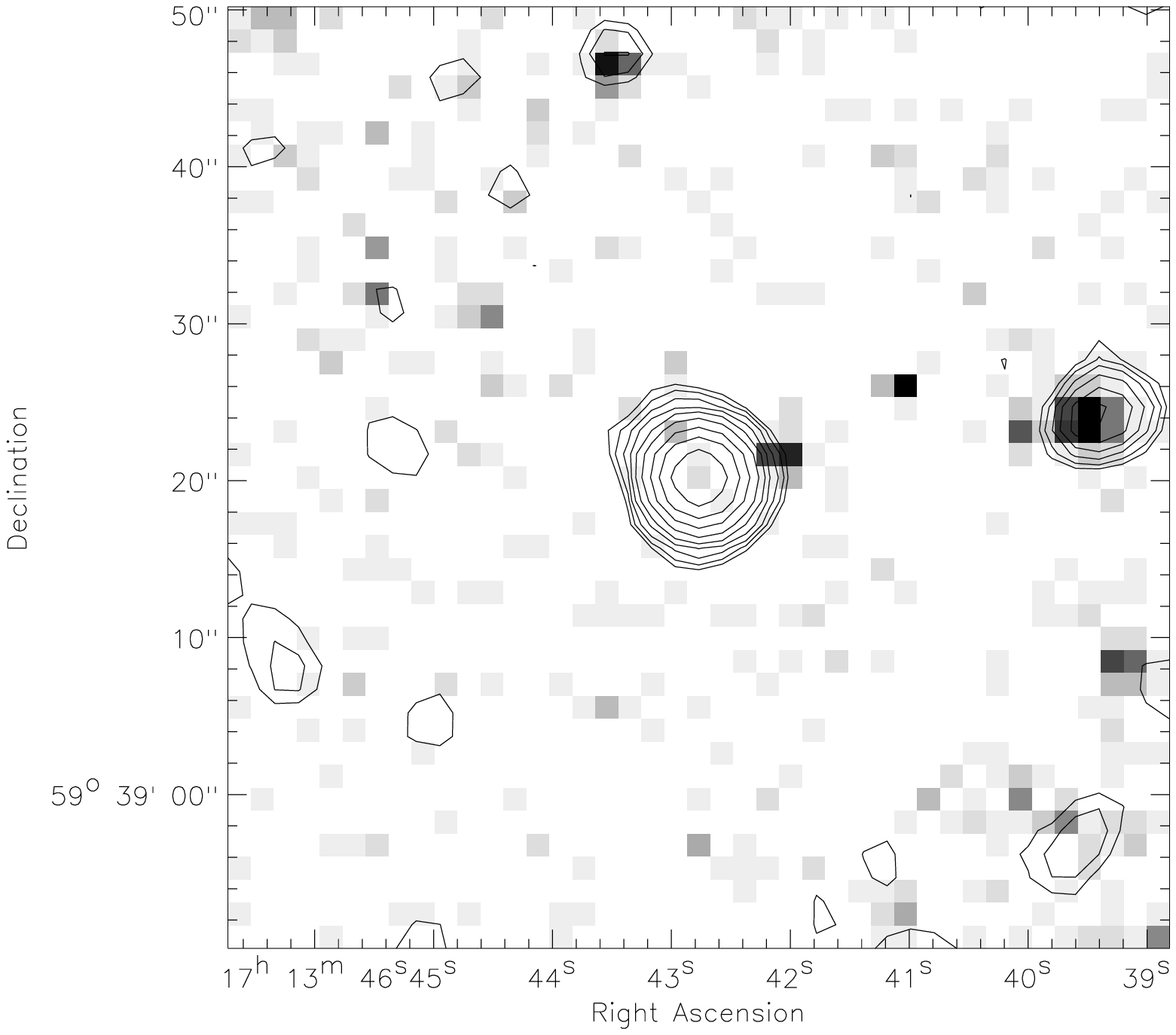,width=4.5cm,angle=0} 
}
\vspace{0.5cm} 
\hbox{
\hspace{-. cm}
\psfig{file= 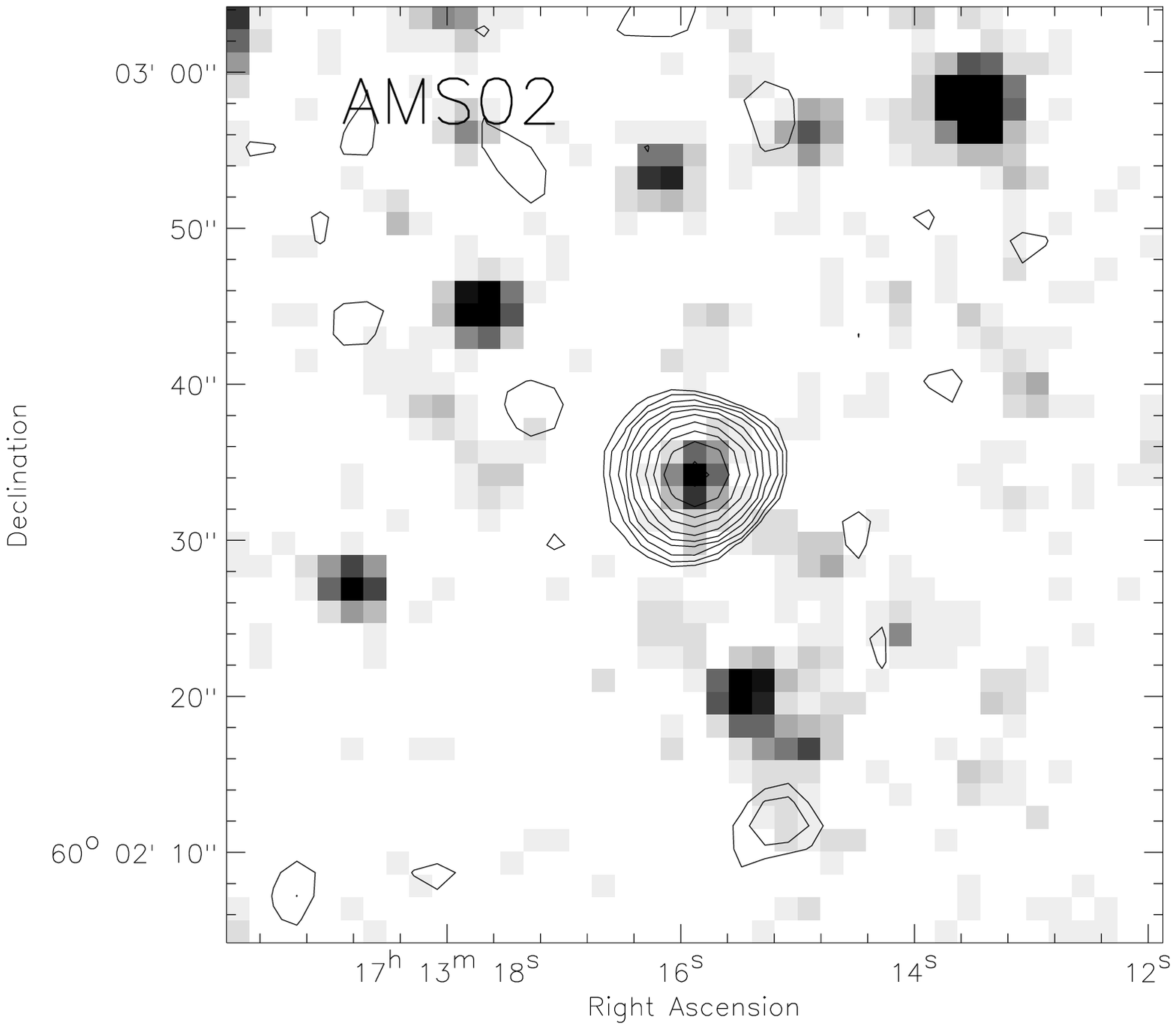,width=4.5cm,angle=0} 
\hspace{-.5cm}
\psfig{file= 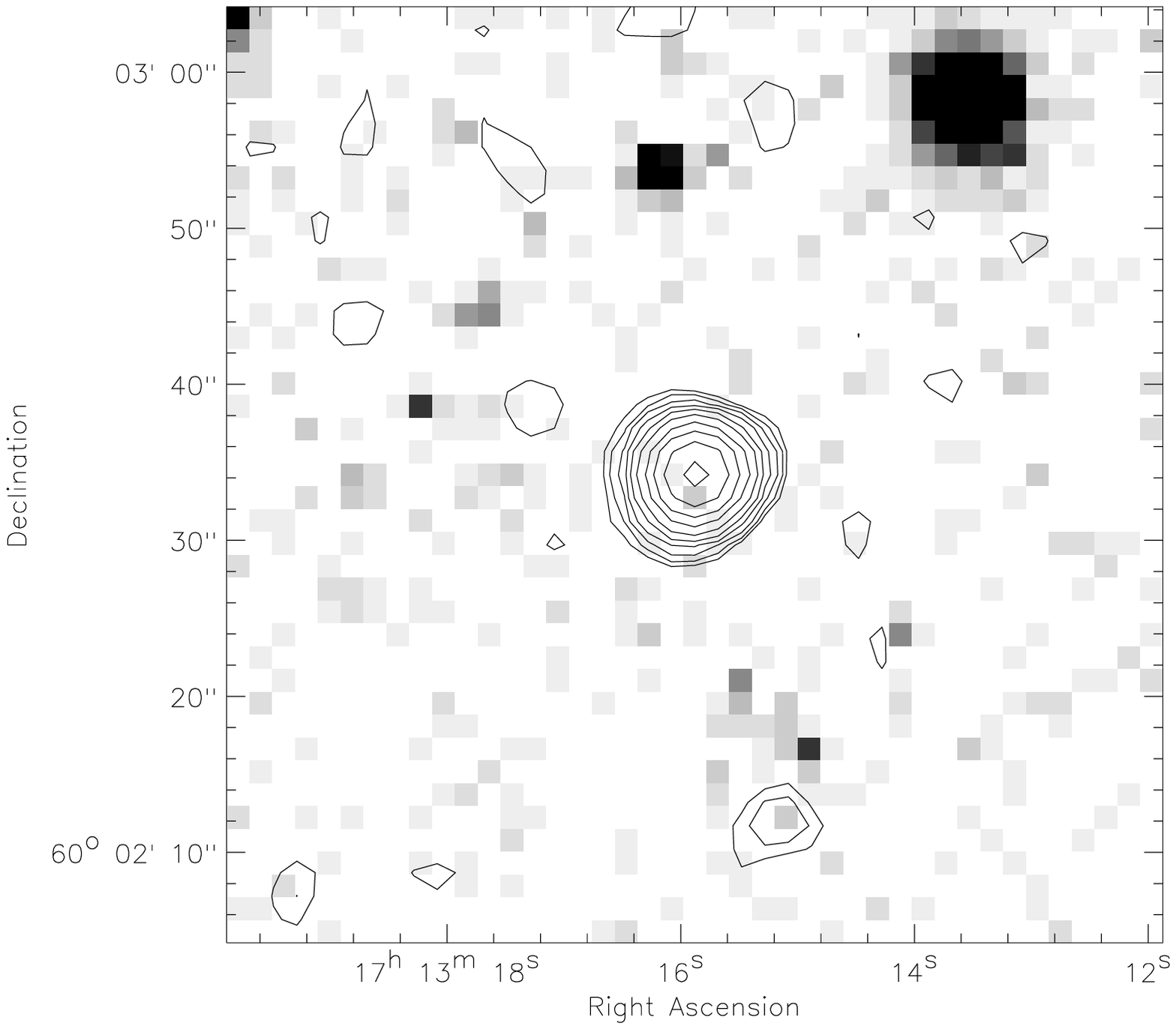,width=4.5cm,angle=0} 
\hspace{0.cm}
\psfig{file= 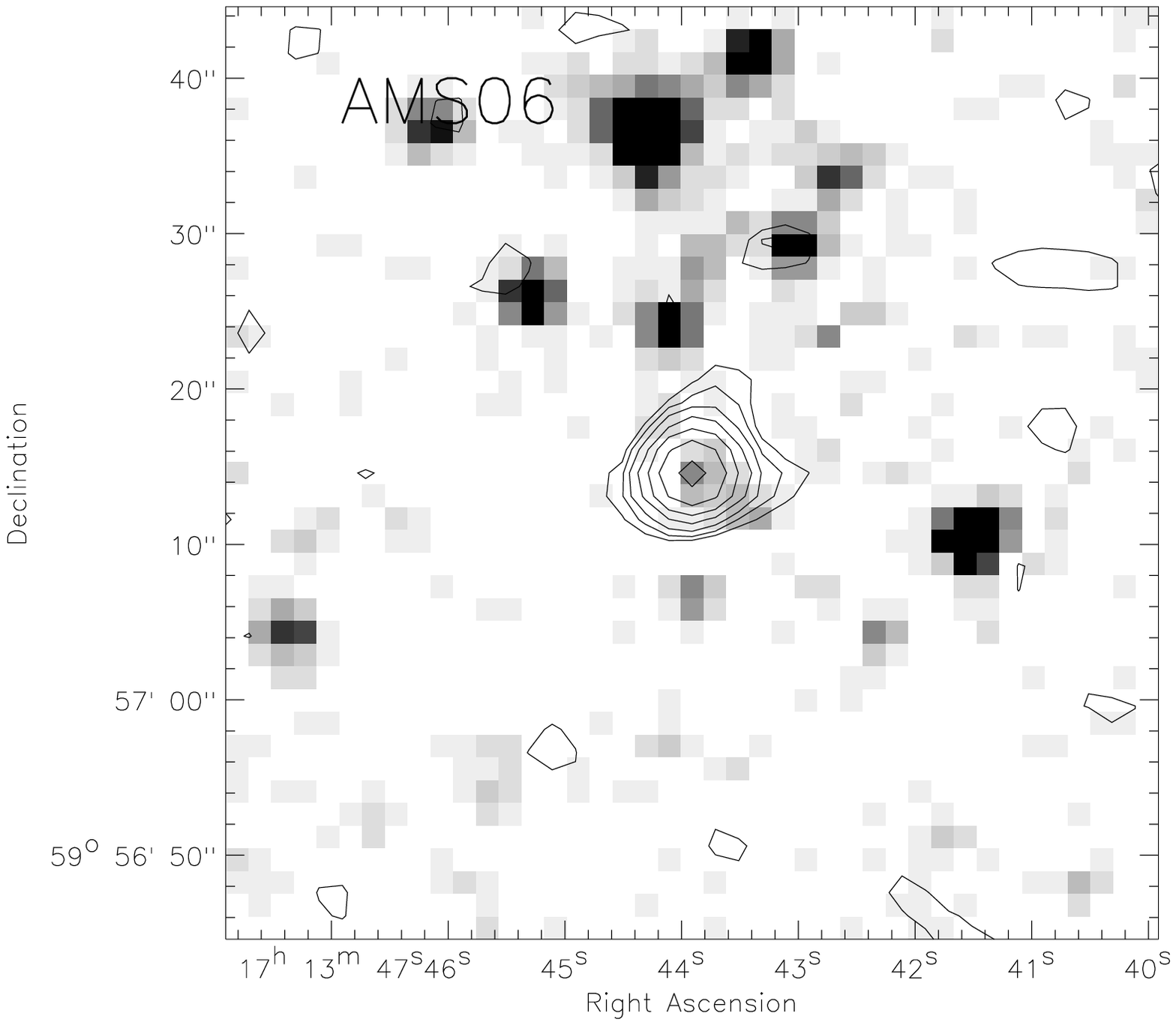,width=4.5cm,angle=0} 
\hspace{-.5cm}
\psfig{file= 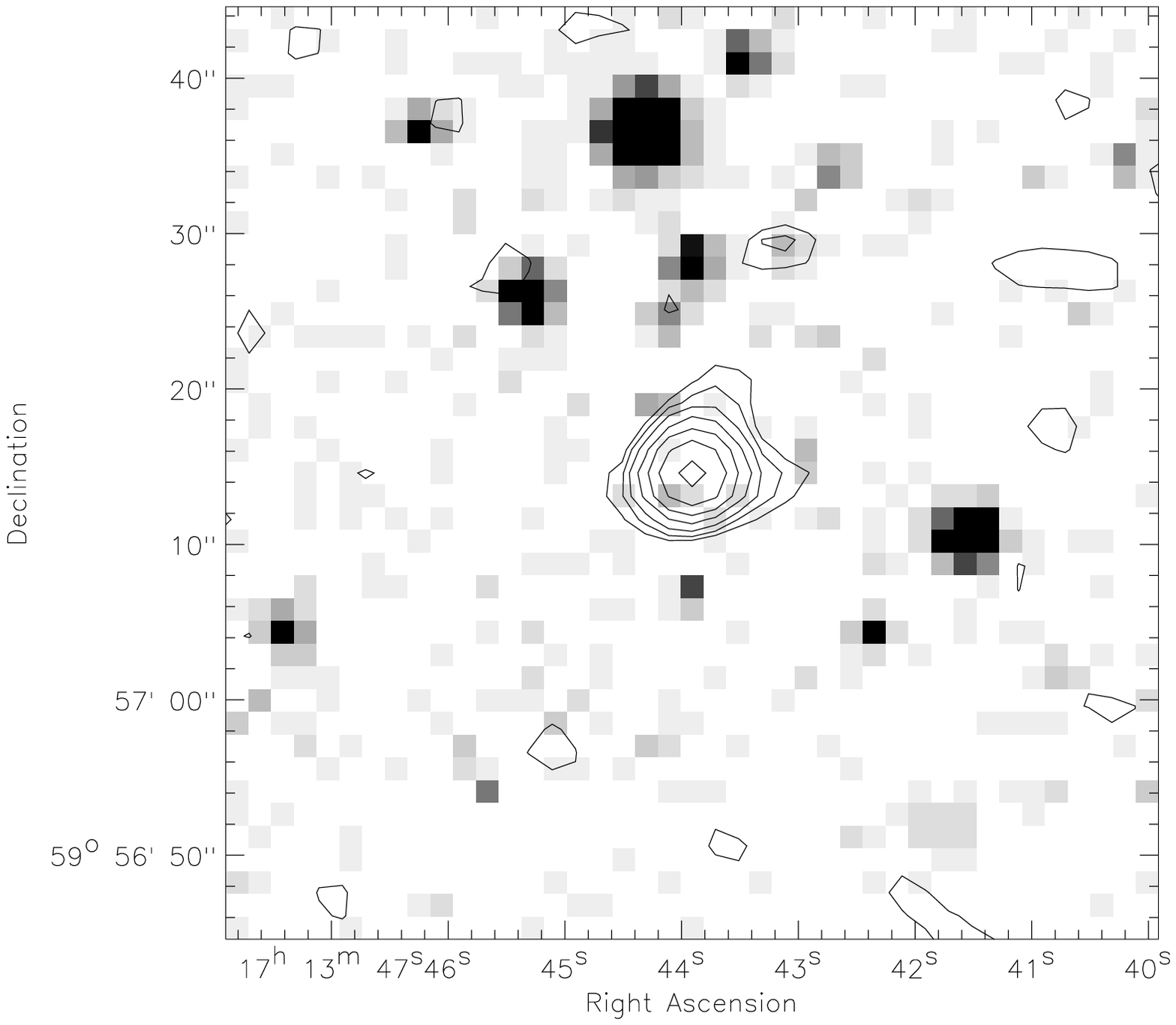,width=4.5cm,angle=0} 
}
\vspace{0.5cm} 
\hbox{
\hspace{-. cm}
\psfig{file= 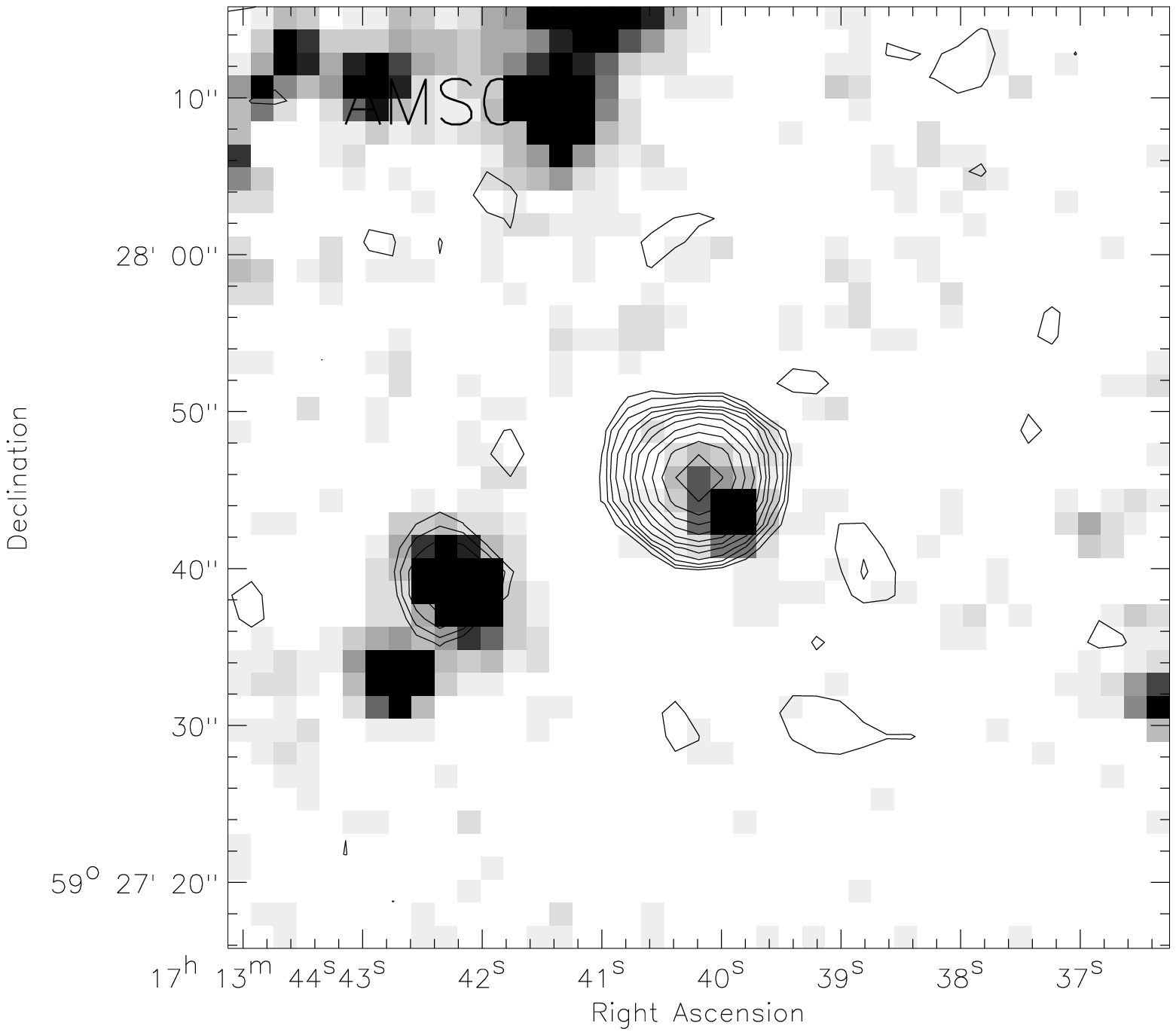,width=4.5cm,angle=0} 
\hspace{-.5cm}
\psfig{file= 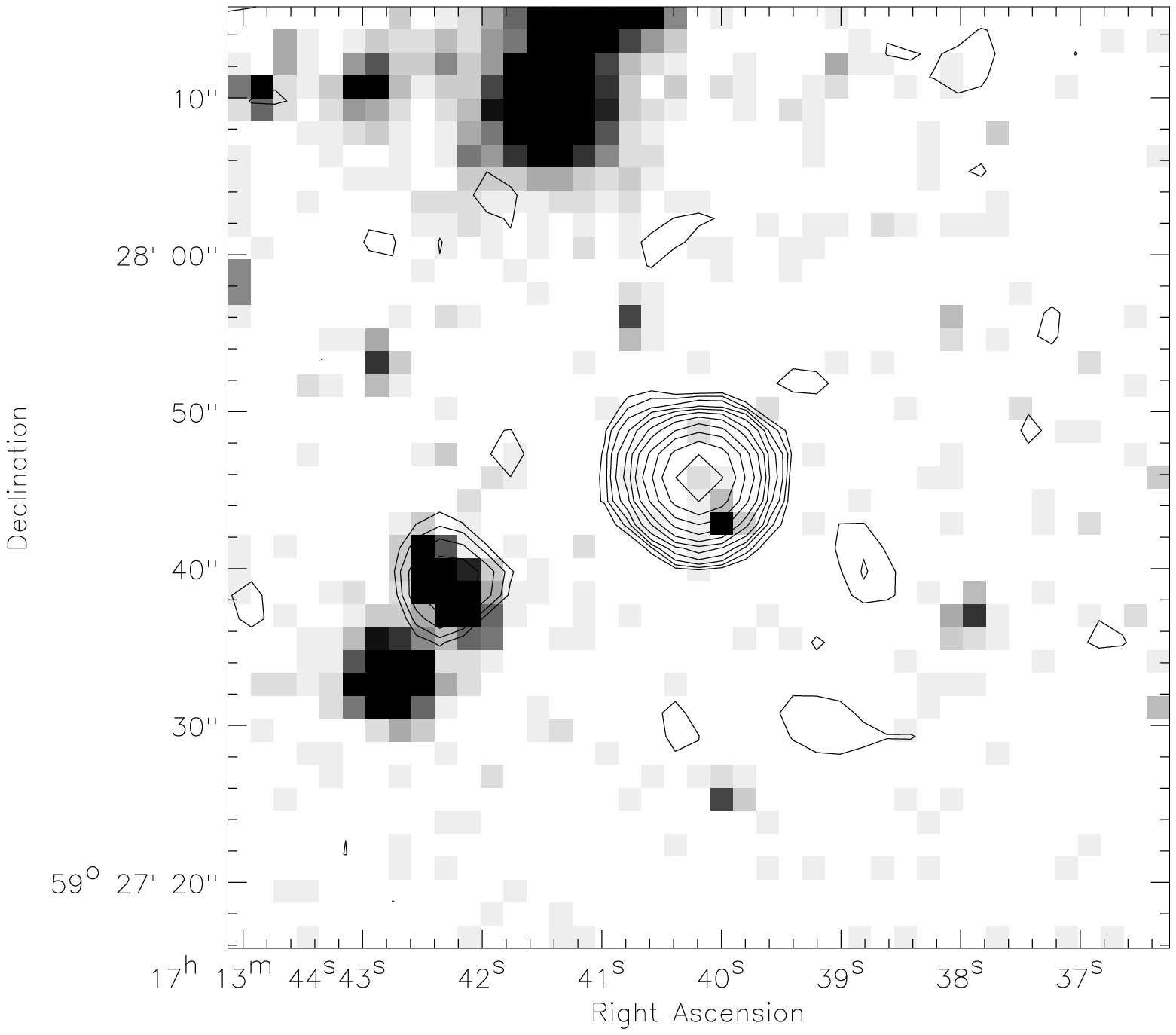,width=4.5cm,angle=0} 
\hspace{0. cm}
\psfig{file= 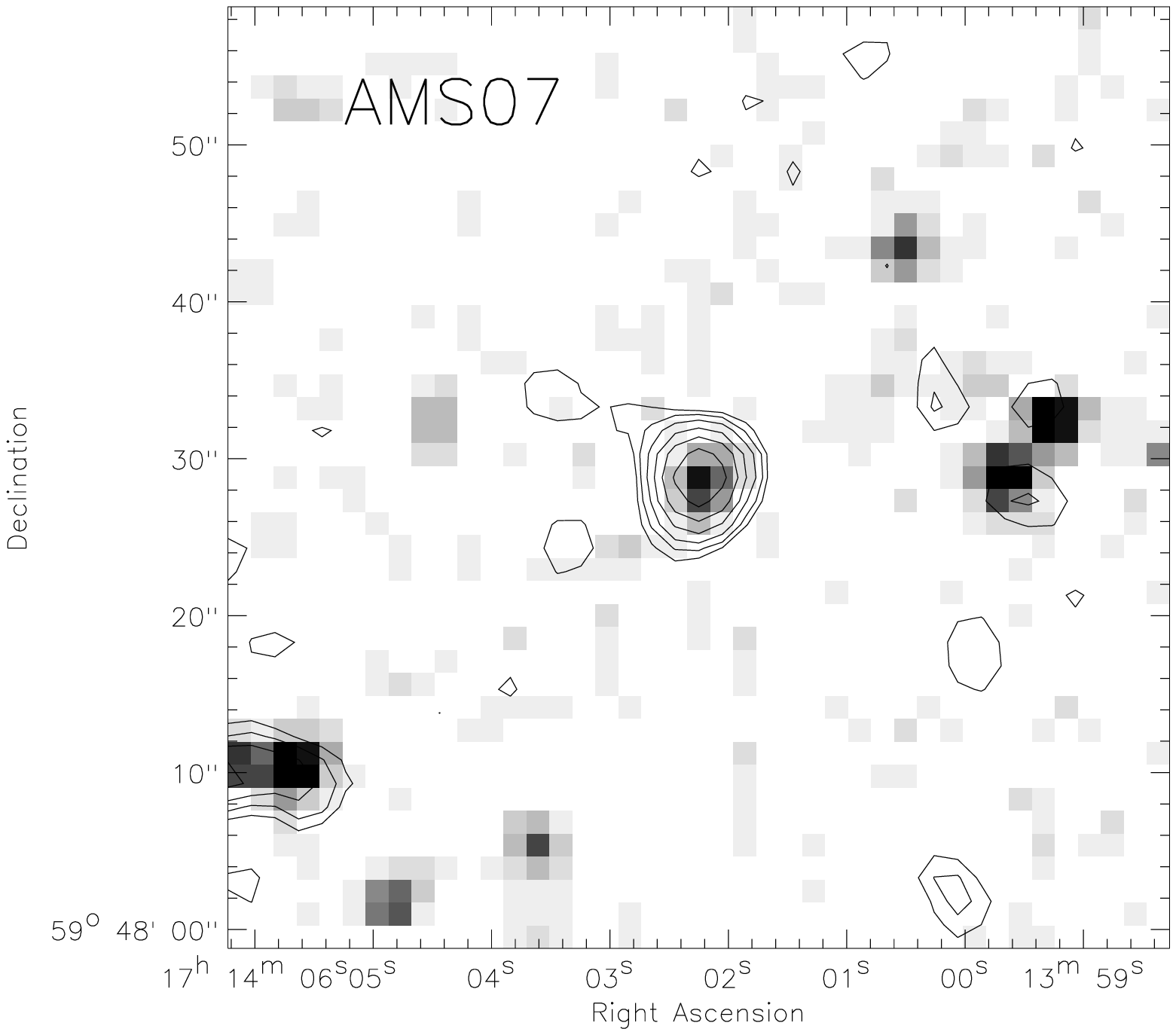,width=4.5cm,angle=0} 
\hspace{-.5cm}
\psfig{file= 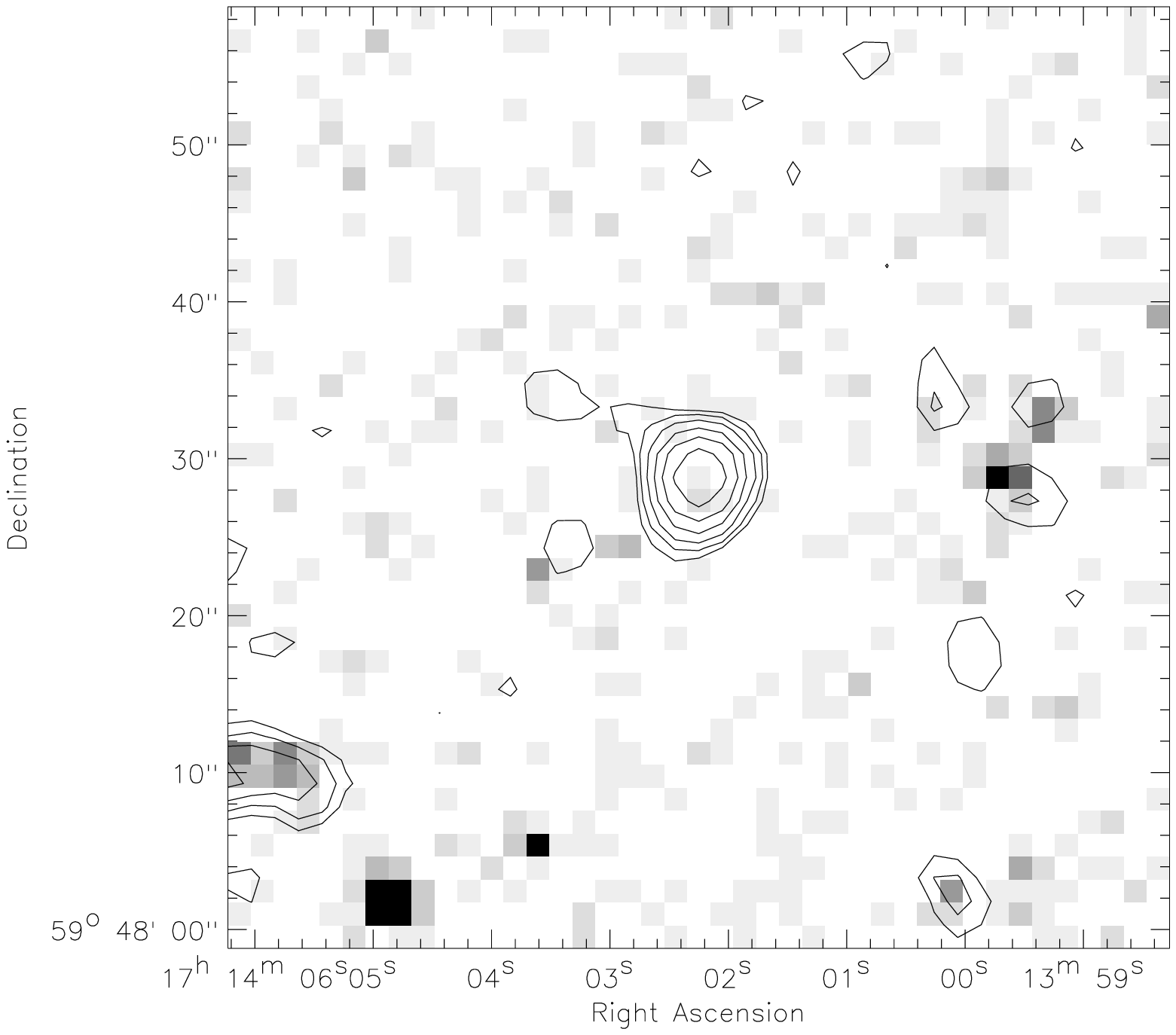,width=4.5cm,angle=0} 
}
\vspace{0.5cm} 
\hbox{
\hspace{-. cm}
\psfig{file= 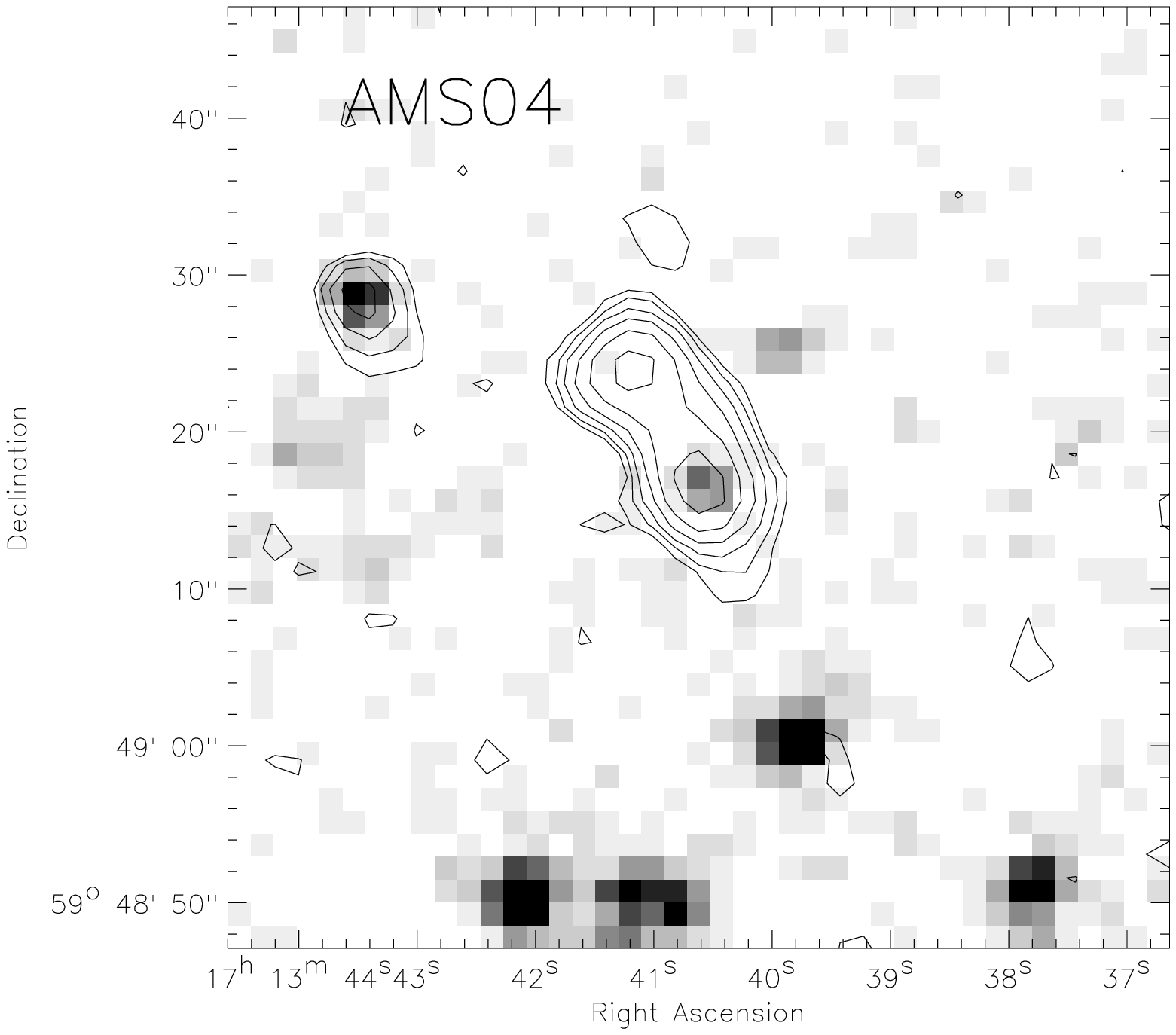,width=4.5cm,angle=0} 
\hspace{-.5cm}
\psfig{file= 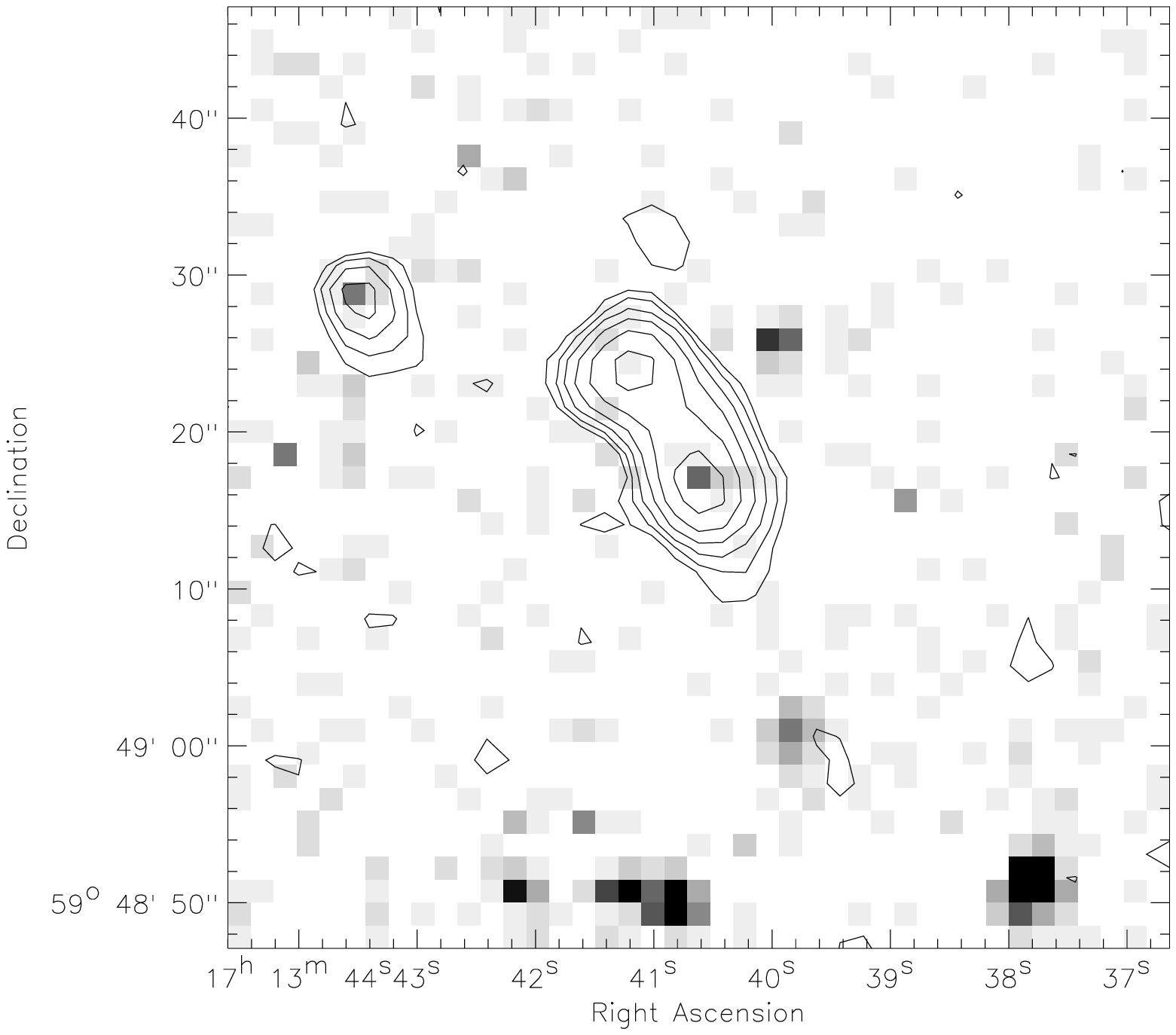,width=4.5cm,angle=0} 
\hspace{0. cm}
\psfig{file= 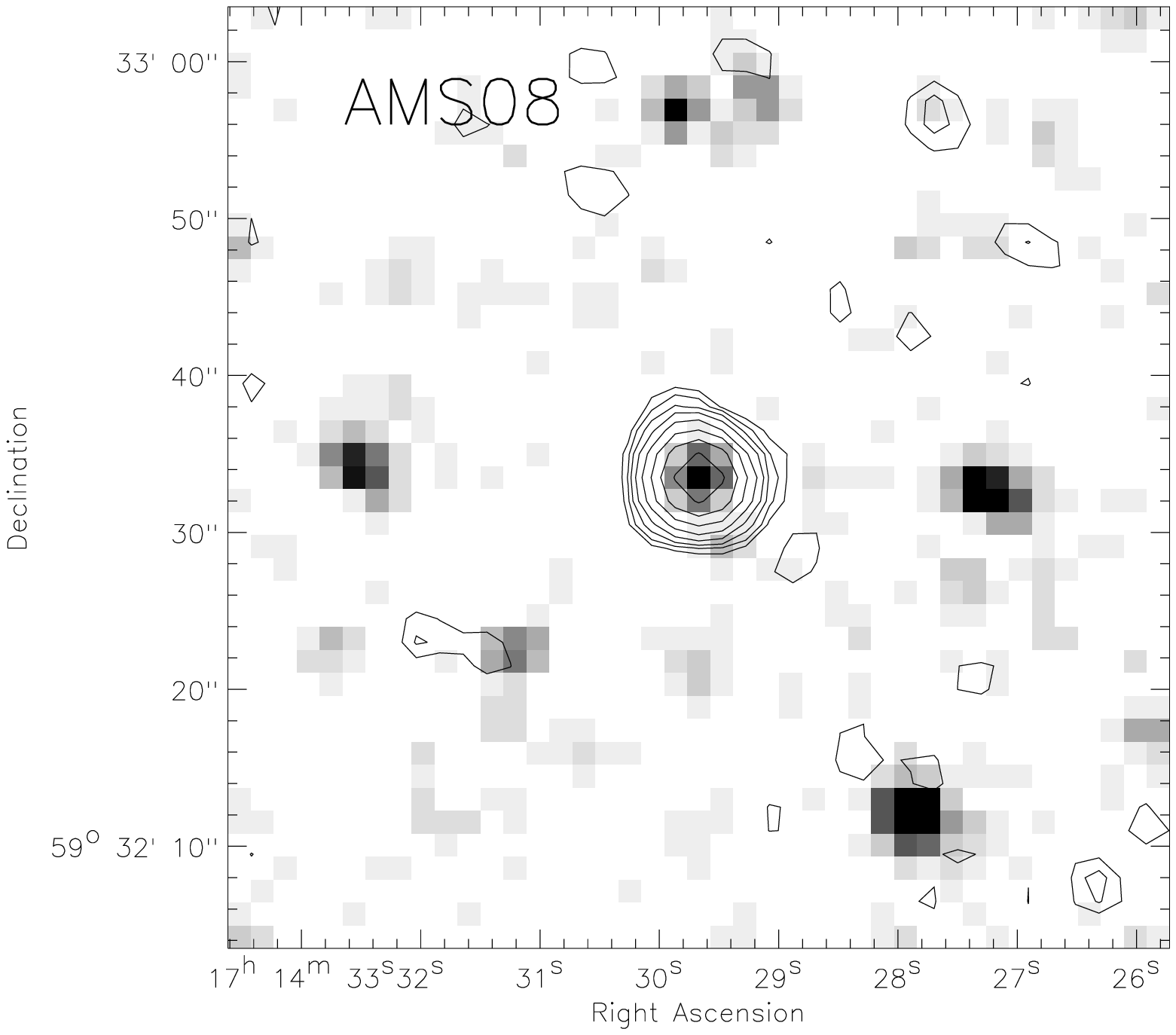,width=4.5cm,angle=0} 
\hspace{-.5cm}
\psfig{file= 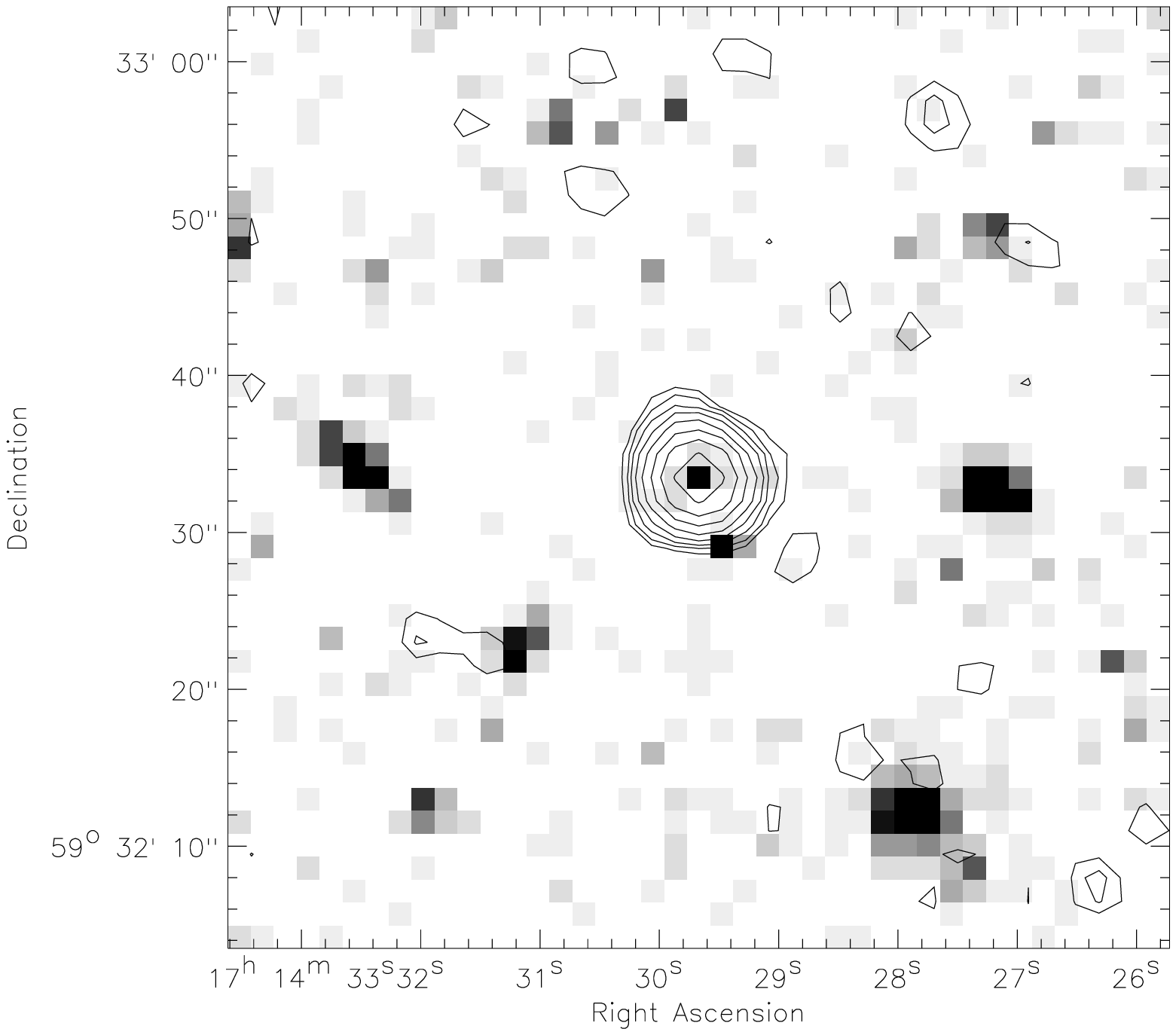,width=4.5cm,angle=0} 
}
\vspace{0.5cm} 
\caption{Radio contours overlayed over the 3.6-$\mu$m and R-band
images for the sample \citep[images from][]
{2003AJ....125.2411C,2004AJ....128....1F,2005ApJS..161...41L}. In each pair of
overlays, the 3.6-$\mu$m image is on the left (with the name of the
source) and the R-band on the right. Radio contours are plotted at
-2$\sigma$,2$\sigma$ and subsequently at 2$^{n/2}\sigma$ up to
2$^{6}\sigma$, where $\sigma$ is the rms noise of the radio image
($\approx$ 23 $\mu$Jy). The gray scale has been chosen to overcome the
high dynamical range and highlight faint objects. Faint grey pixels
have values 3-times that of the noise.}
\label{fig:overlays}
\end{center}
\end{figure*}

Thus, to find all type-2 quasars (even the Compton thick ones), one
requires a survey with a 2-10 keV flux limit of about $10^{-18}$ W
m$^{-2}$ over a reasonably large area. Therefore, the SXDF survey
(horizontal dotted line: Ueda et al. in prep.), which covers $\sim$1
deg$^{2}$ to this depth, should be able to find X-ray-selected Compton
thick quasars. Comparing to the SWIRE depth, we can see that in the
region of the SXDF, X-ray selection would be more sensitive than
24-$\mu$m selection, while still being able to find powerful type-2
quasars. 

For Compton-thick objects, the X-ray selection function is almost flat
for $0.5 \ltsimeq z \ltsimeq 2$. The redshift-distribution of
Compton-thick type-2s in a survey deep enough to find them would
therefore be dominated by the volume of the survey as a function of
redshift and the evolution of the luminosity function. This means that
X-ray surveys deep enough to be sensitive to Compton-thick type-2
quasars should find a wealth of $1 \ltsimeq z \ltsimeq 2$
objects. Coincidentally, this would be complementary to 24-$\mu$m
surveys, as the mid-infrared surveys would struggle to find heavily
obscured type-2s ($A_{\rm V} \gtsimeq 50$), especially in the range
around $z \sim 1.5$.

\addtocounter{figure}{-1}
\begin{figure*}
\hbox{
\hspace{-.0cm}
\psfig{file= 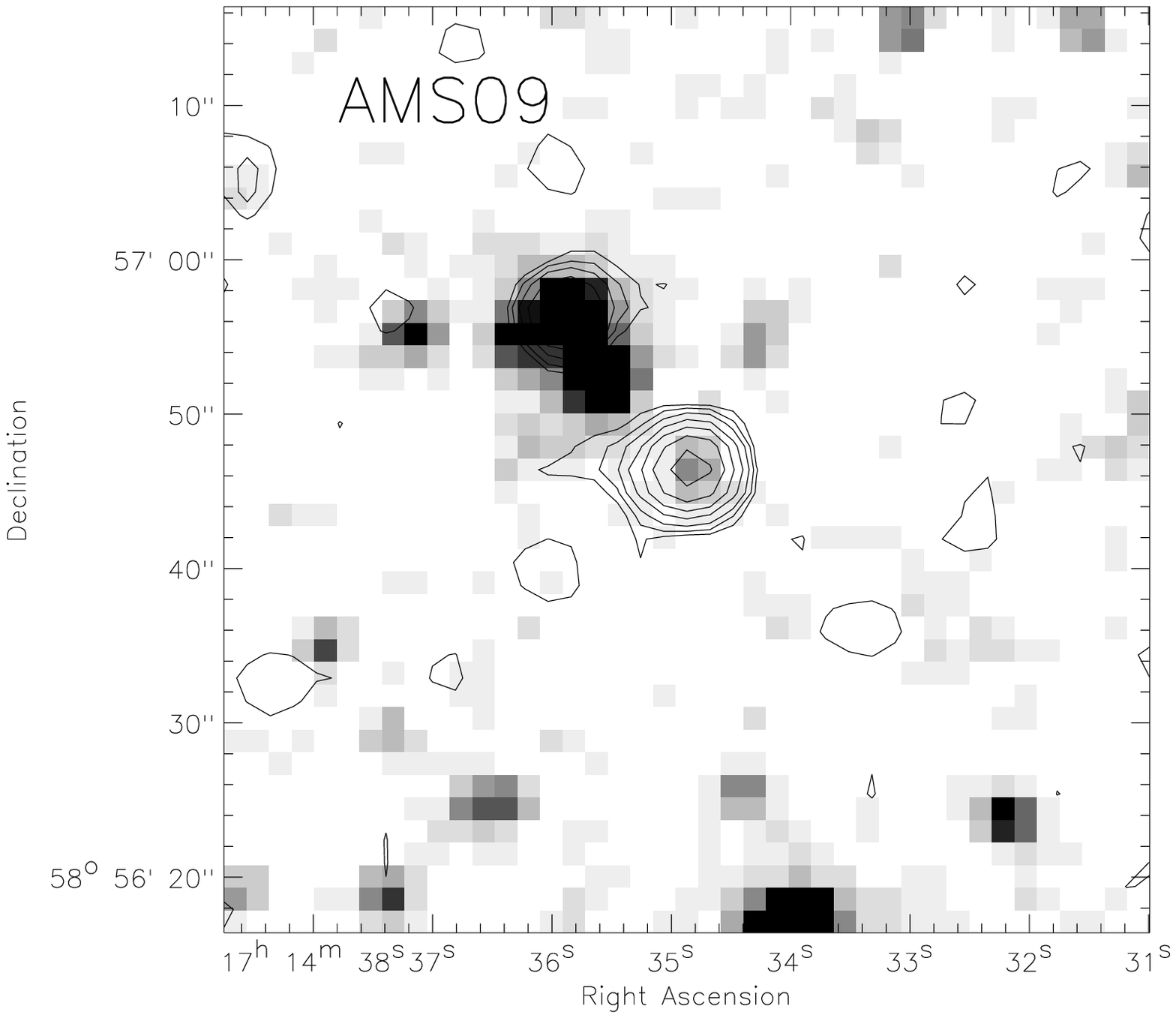,width=4.5cm,angle=0} 
\hspace{-.5cm}
\psfig{file= 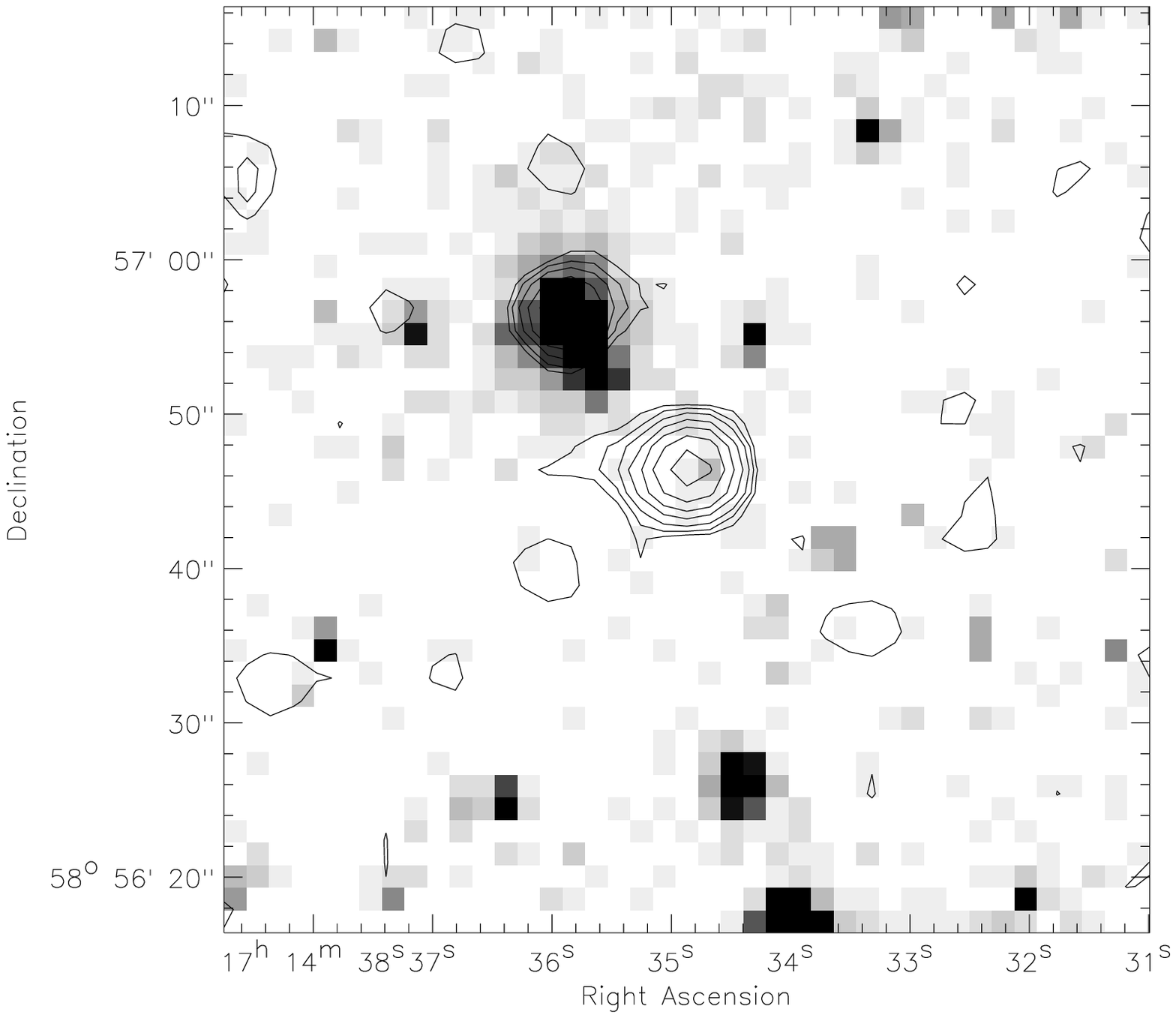,width=4.5cm,angle=0} 
\hspace{0. cm}
\psfig{file= 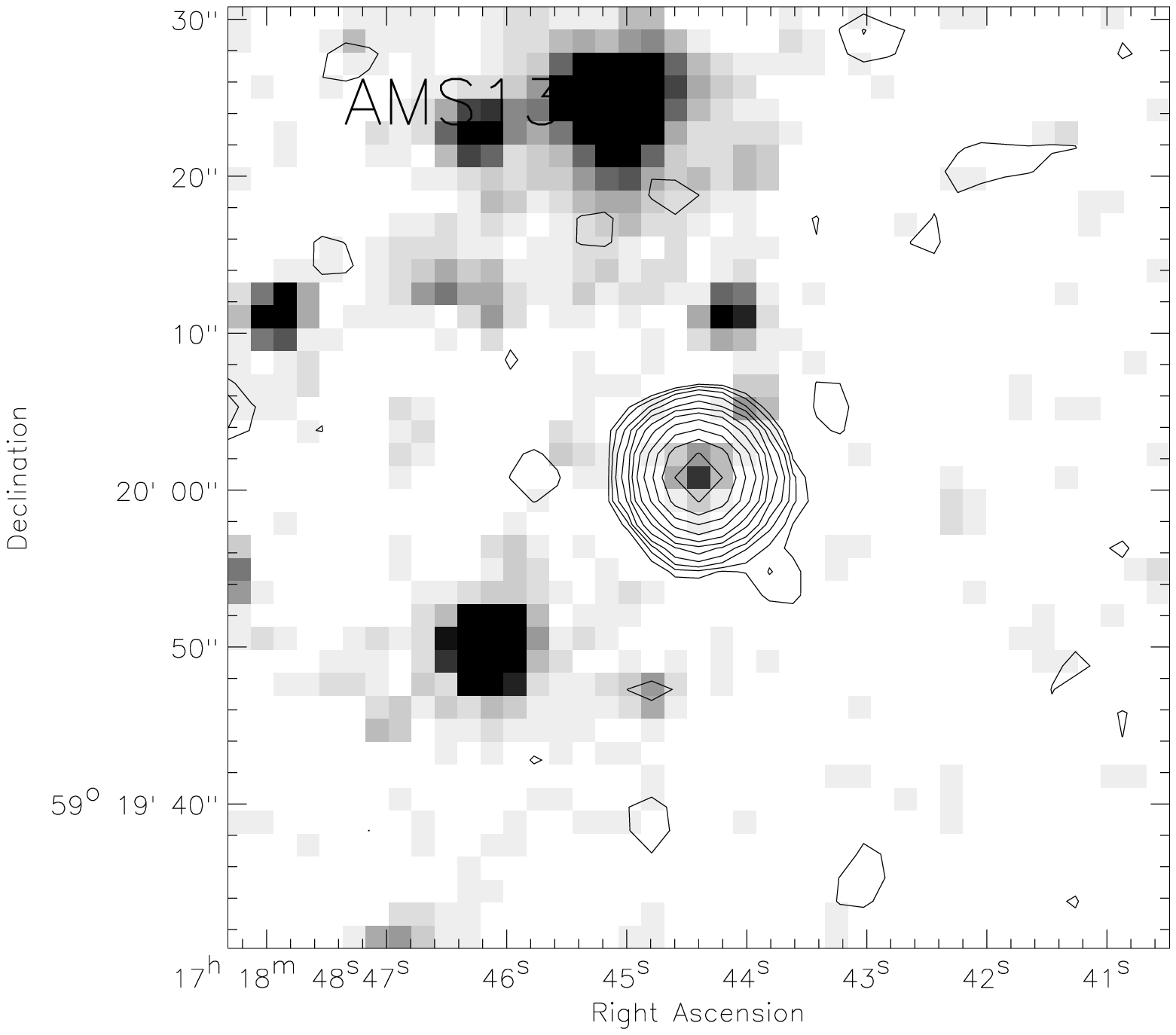,width=4.5cm,angle=0} 
\hspace{-.5cm}
\psfig{file= 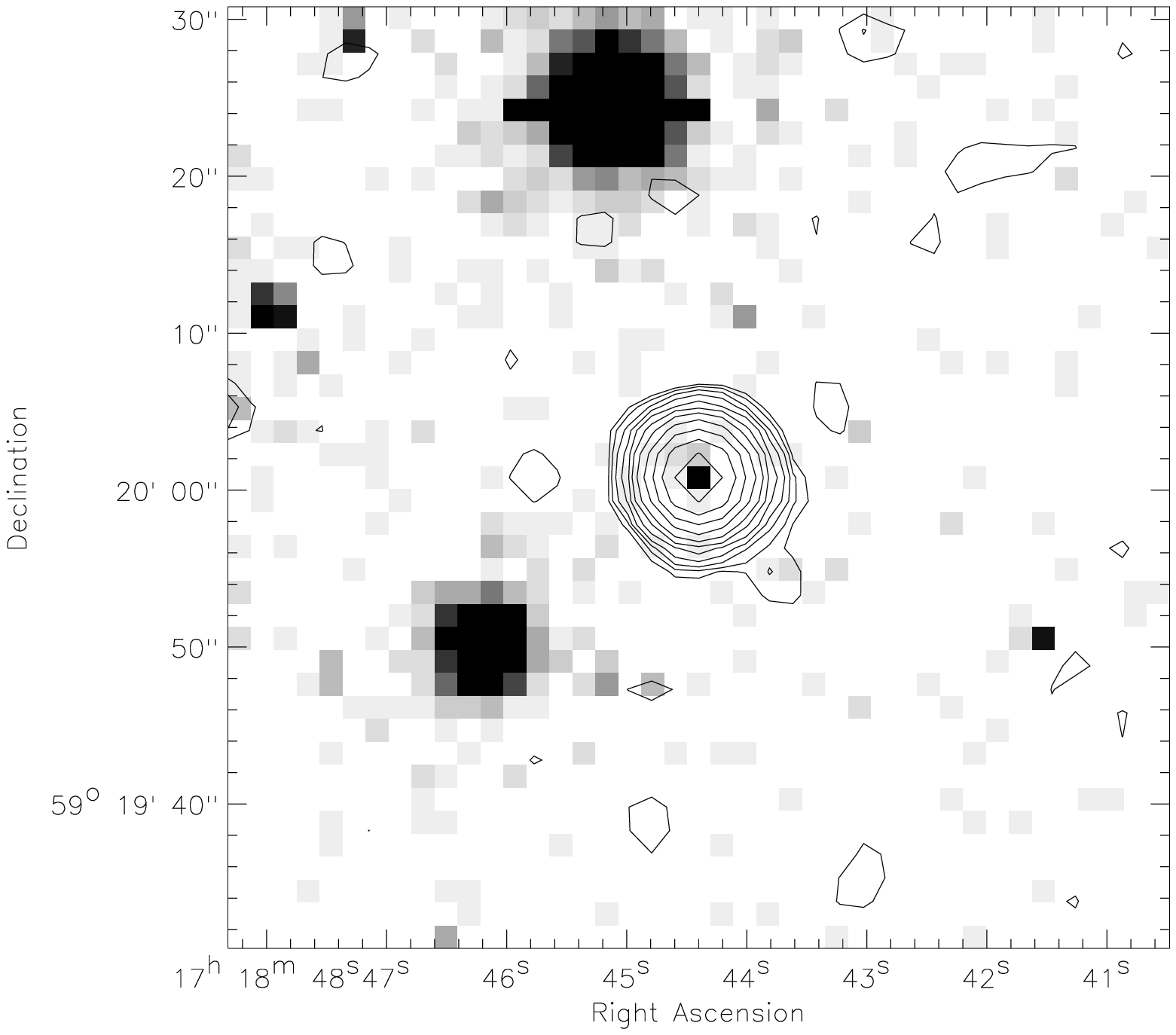,width=4.5cm,angle=0} 
}
\vspace{0.5cm} 
\hbox{
\hspace{ -.0cm}
\psfig{file= 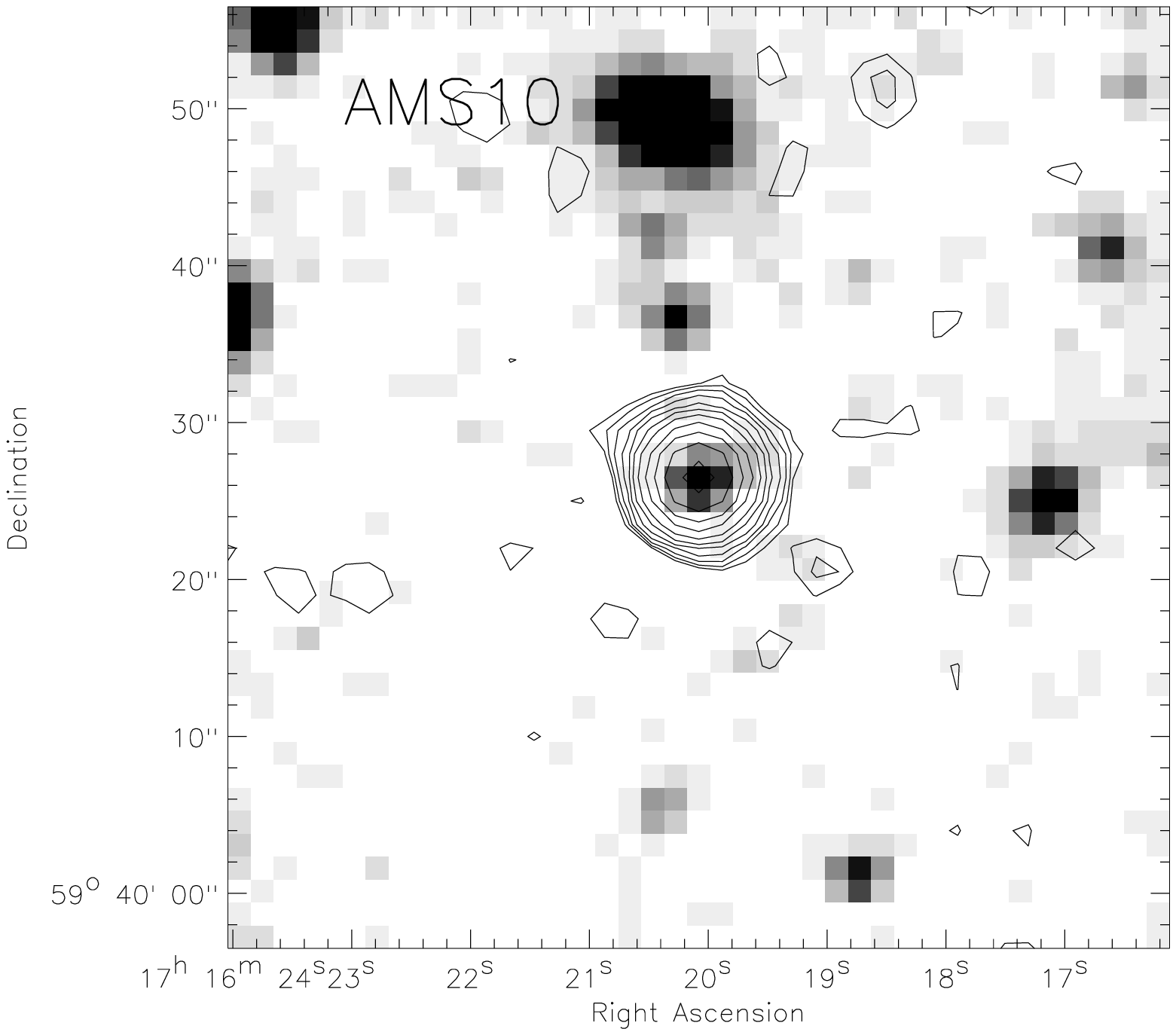,width=4.5cm,angle=0} 
\hspace{-.5cm}
\psfig{file= 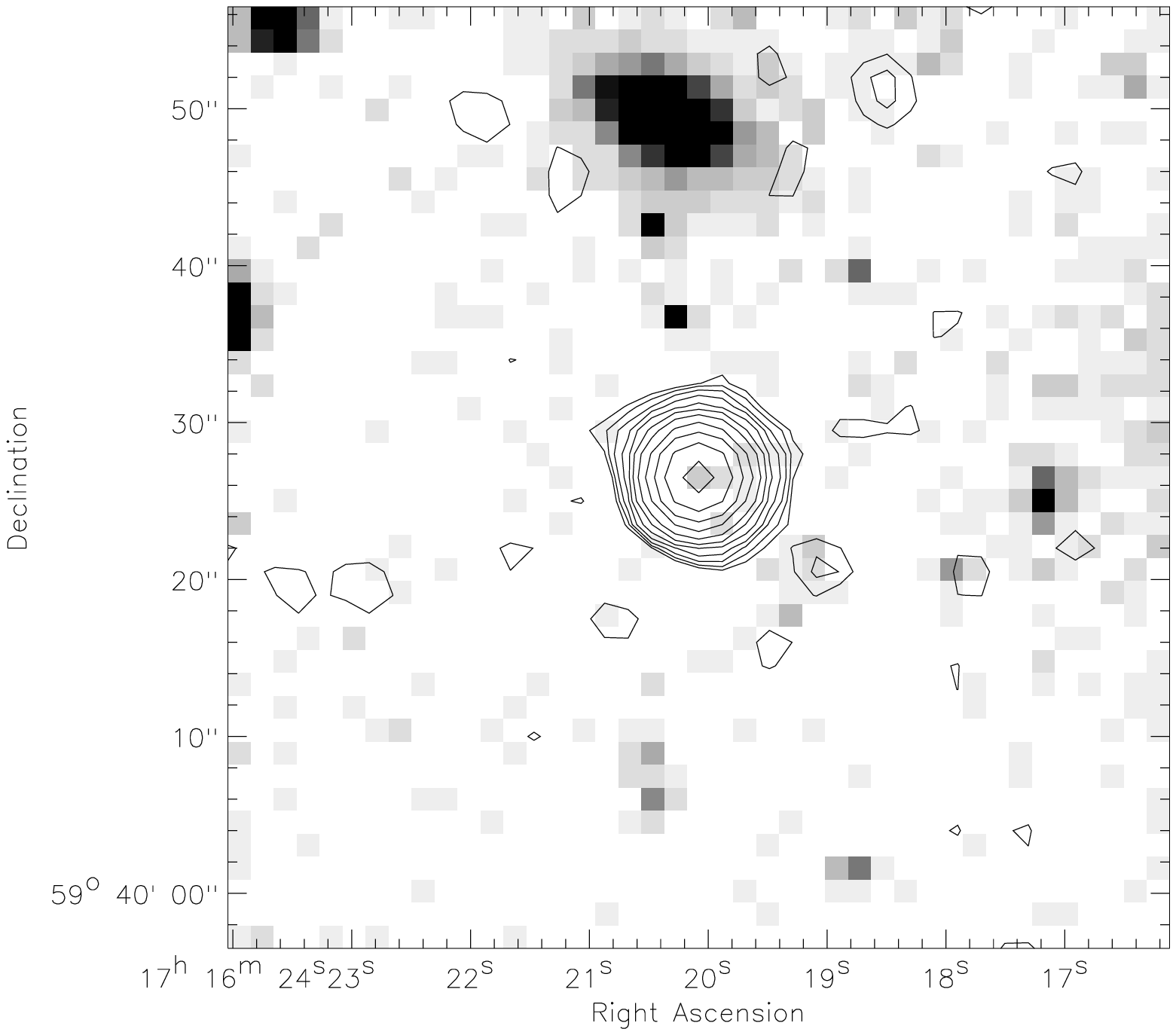,width=4.5cm,angle=0} 
\hspace{0.0cm}
\psfig{file= 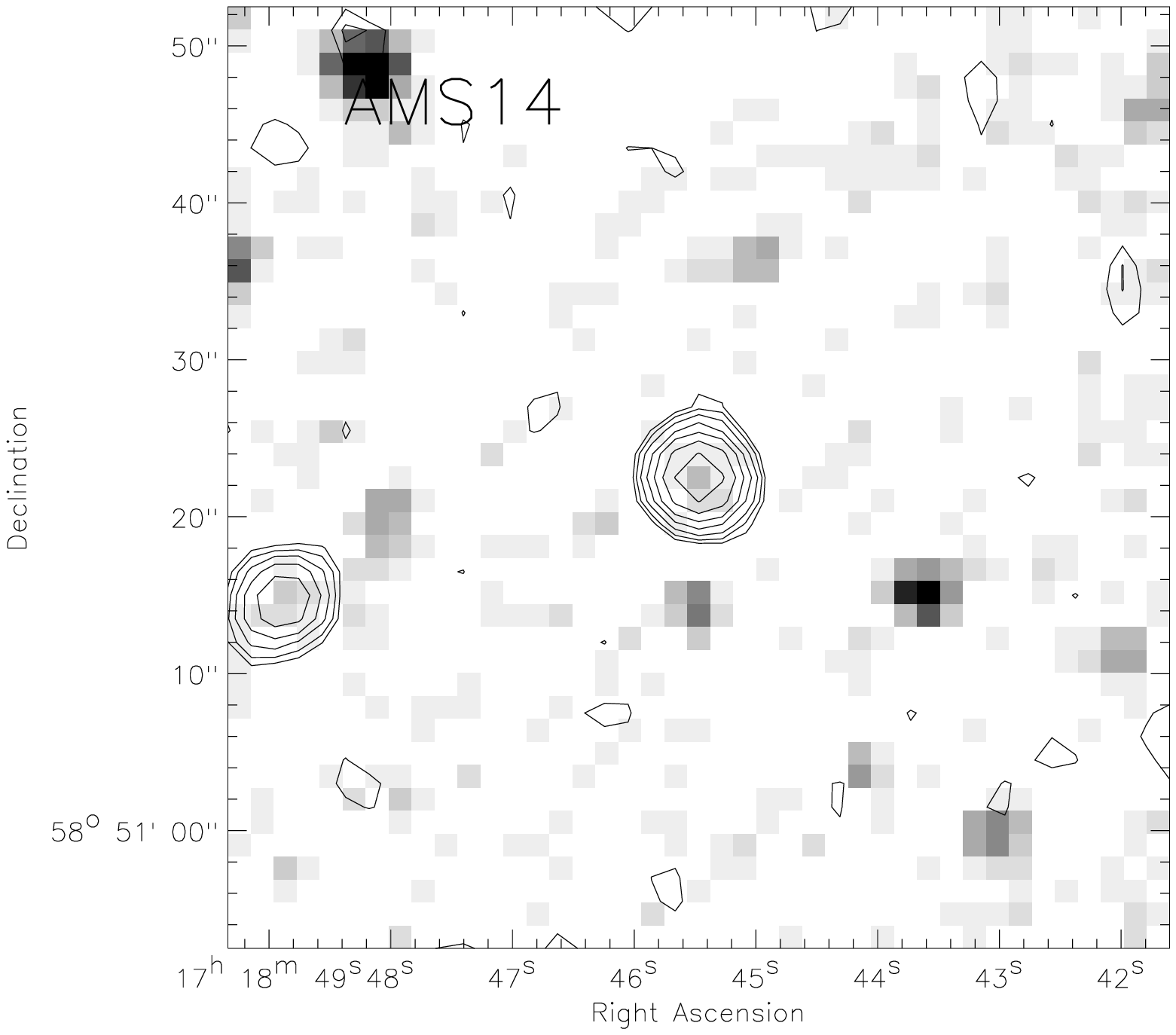,width=4.5cm,angle=0} 
\hspace{-.5cm}
\psfig{file= 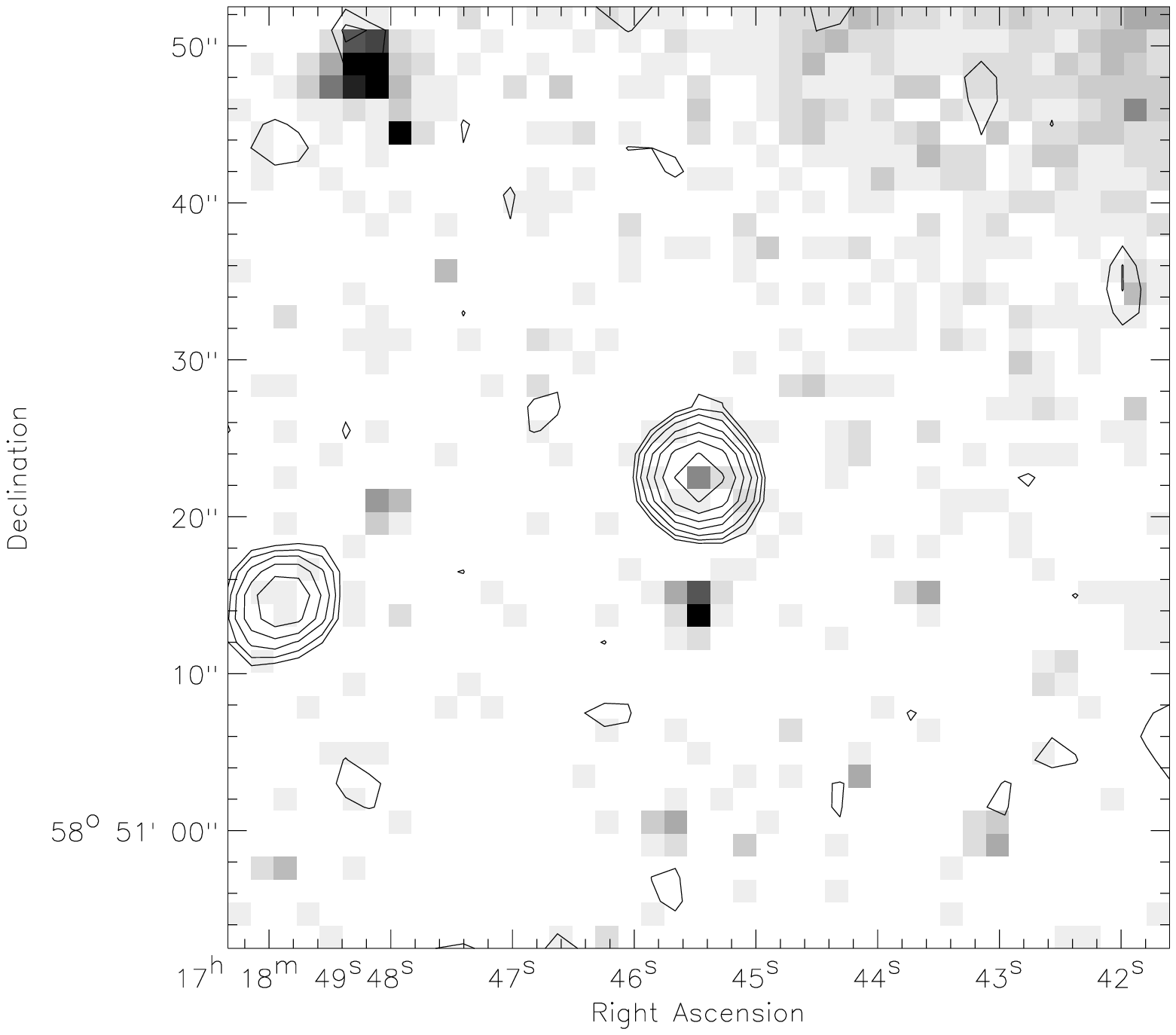,width=4.5cm,angle=0} 
}
\vspace{0.5cm} 
\hbox{
\hspace{-.0cm}
\psfig{file= 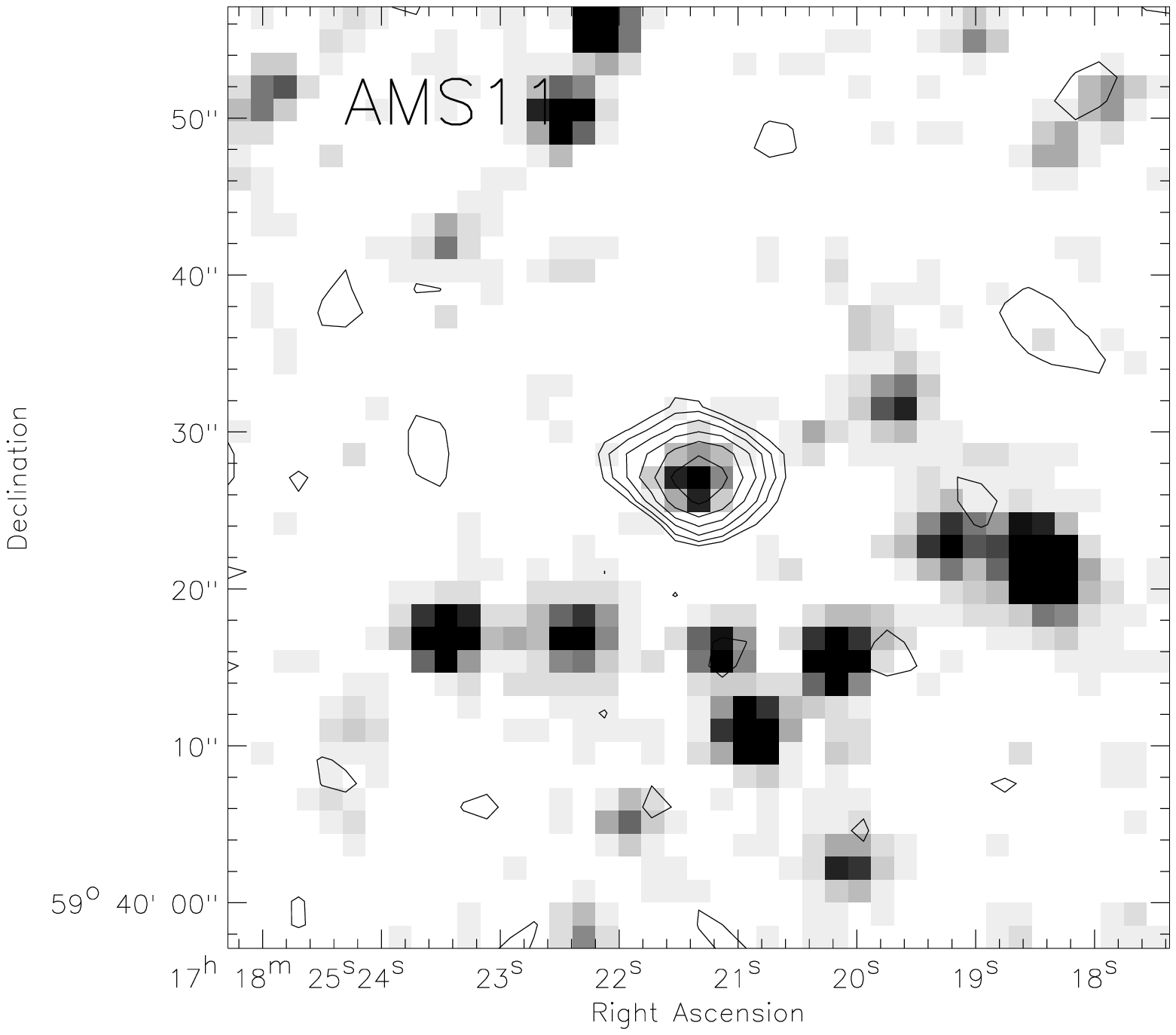,width=4.5cm,angle=0} 
\hspace{-.5cm}
\psfig{file= 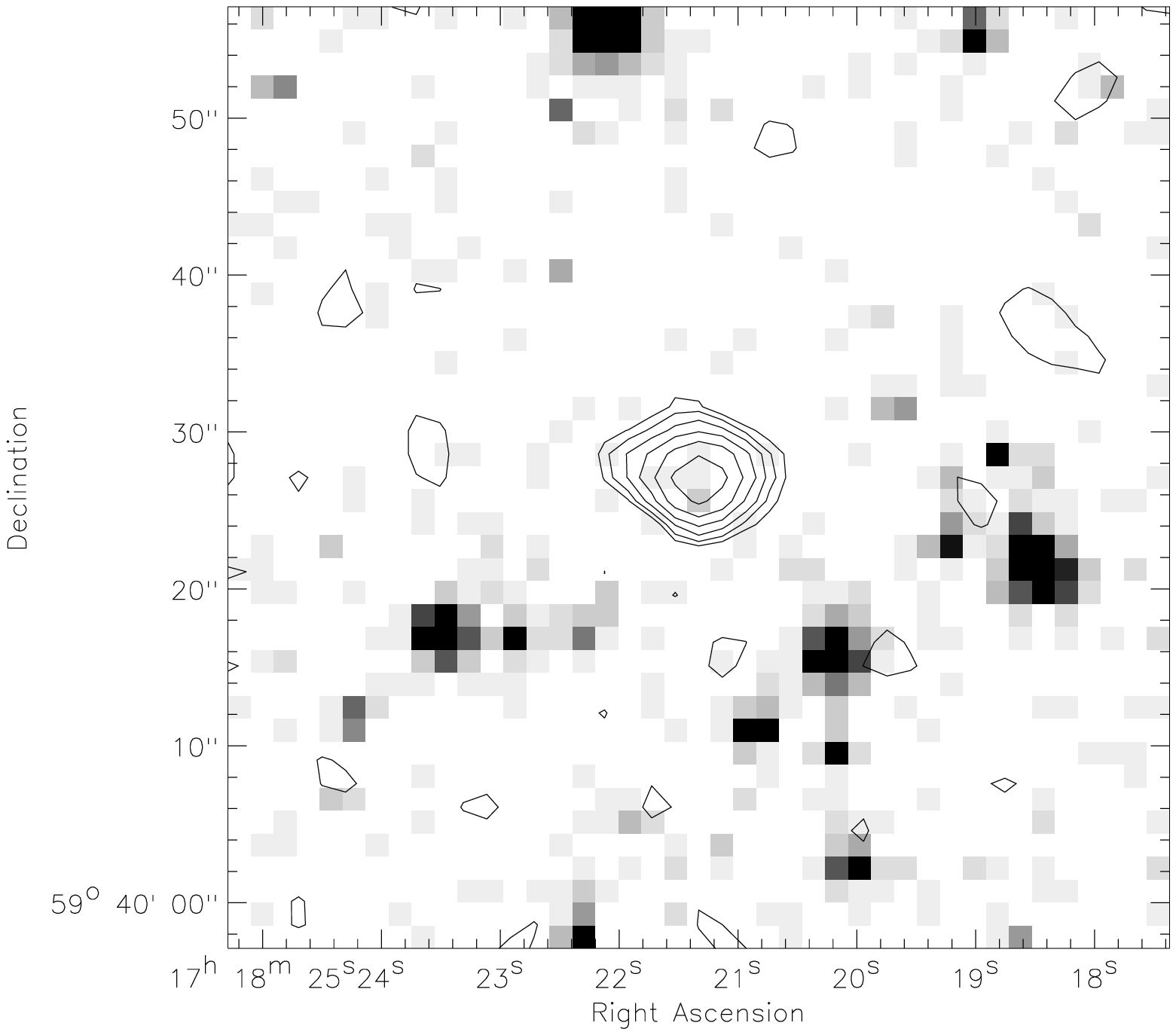,width=4.5cm,angle=0} 
\hspace{0.0cm}
\psfig{file= 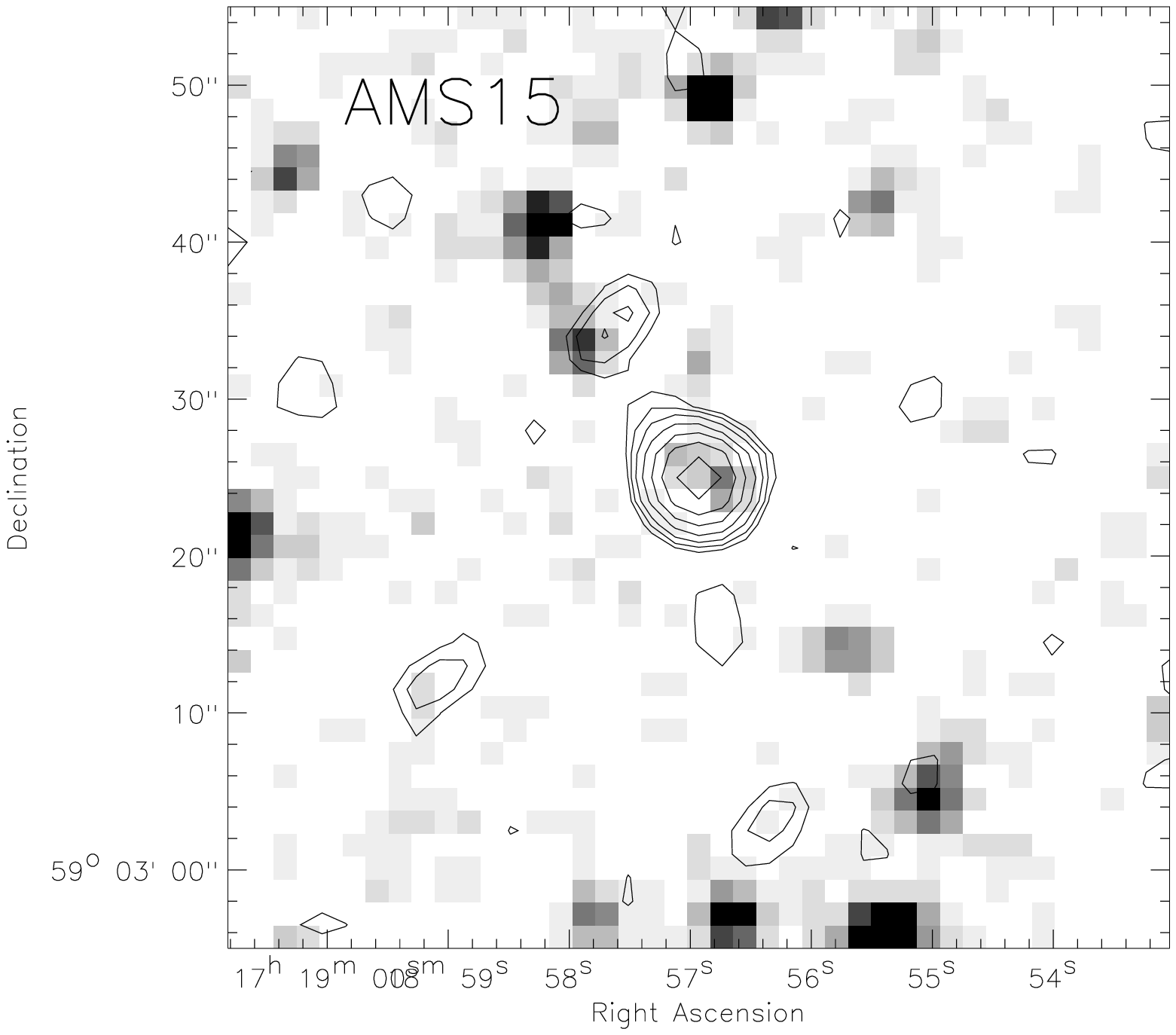,width=4.5cm,angle=0} 
\hspace{-.5cm}
\psfig{file= 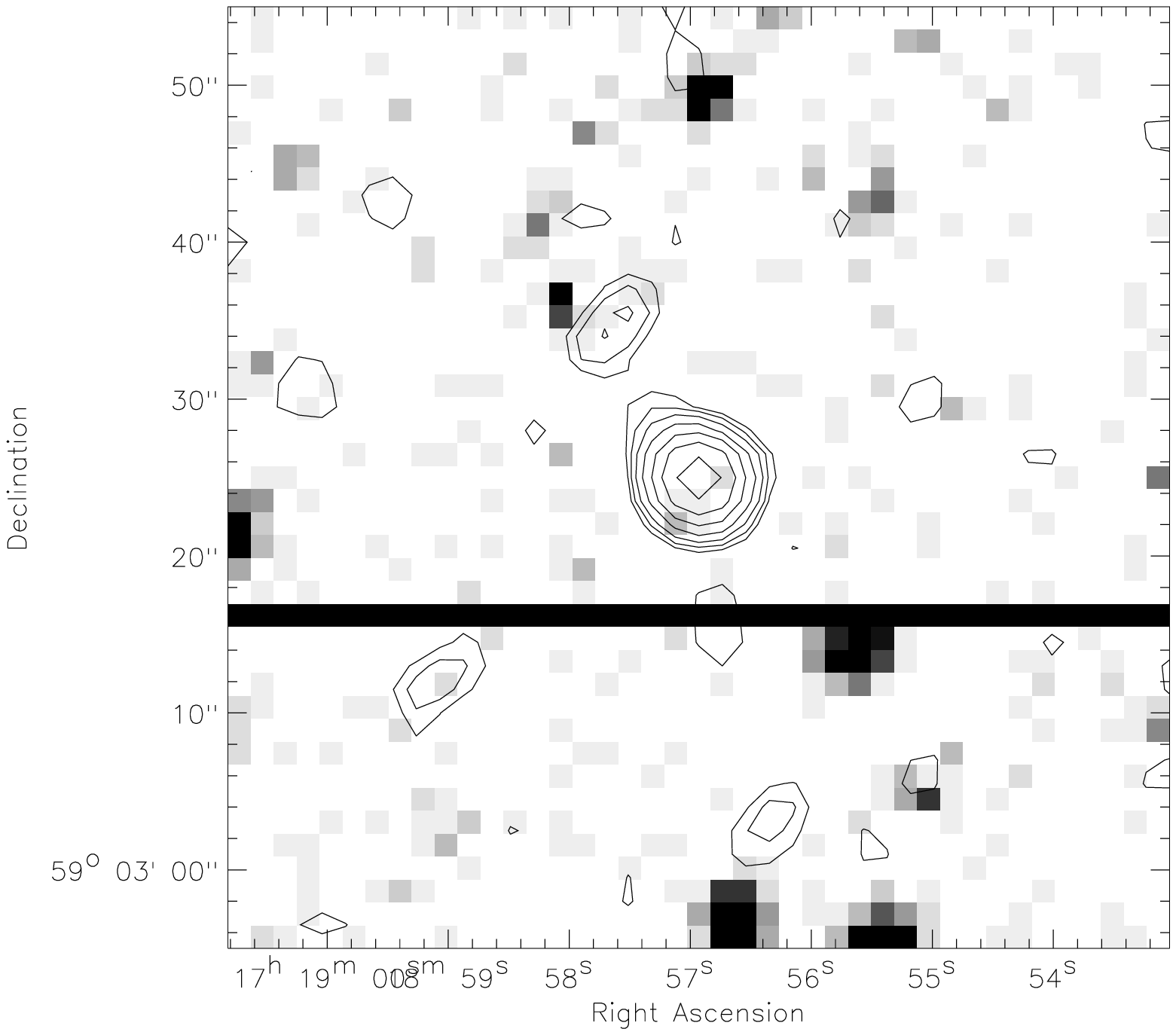,width=4.5cm,angle=0} 
}
\vspace{0.5cm} 
\hbox{
\hspace{-.0cm}
\psfig{file= 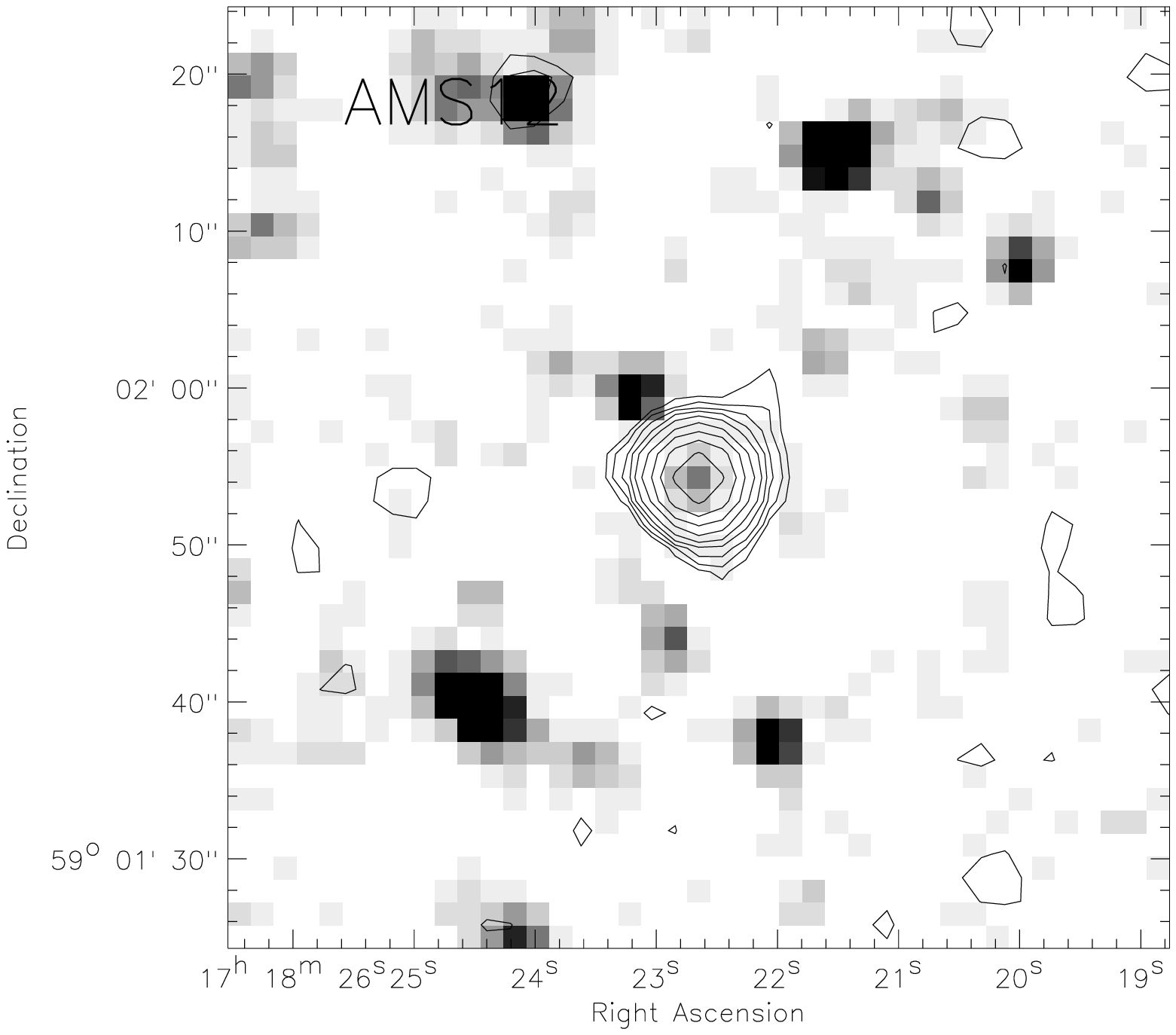,width=4.5cm,angle=0} 
\hspace{-.5cm}
\psfig{file= 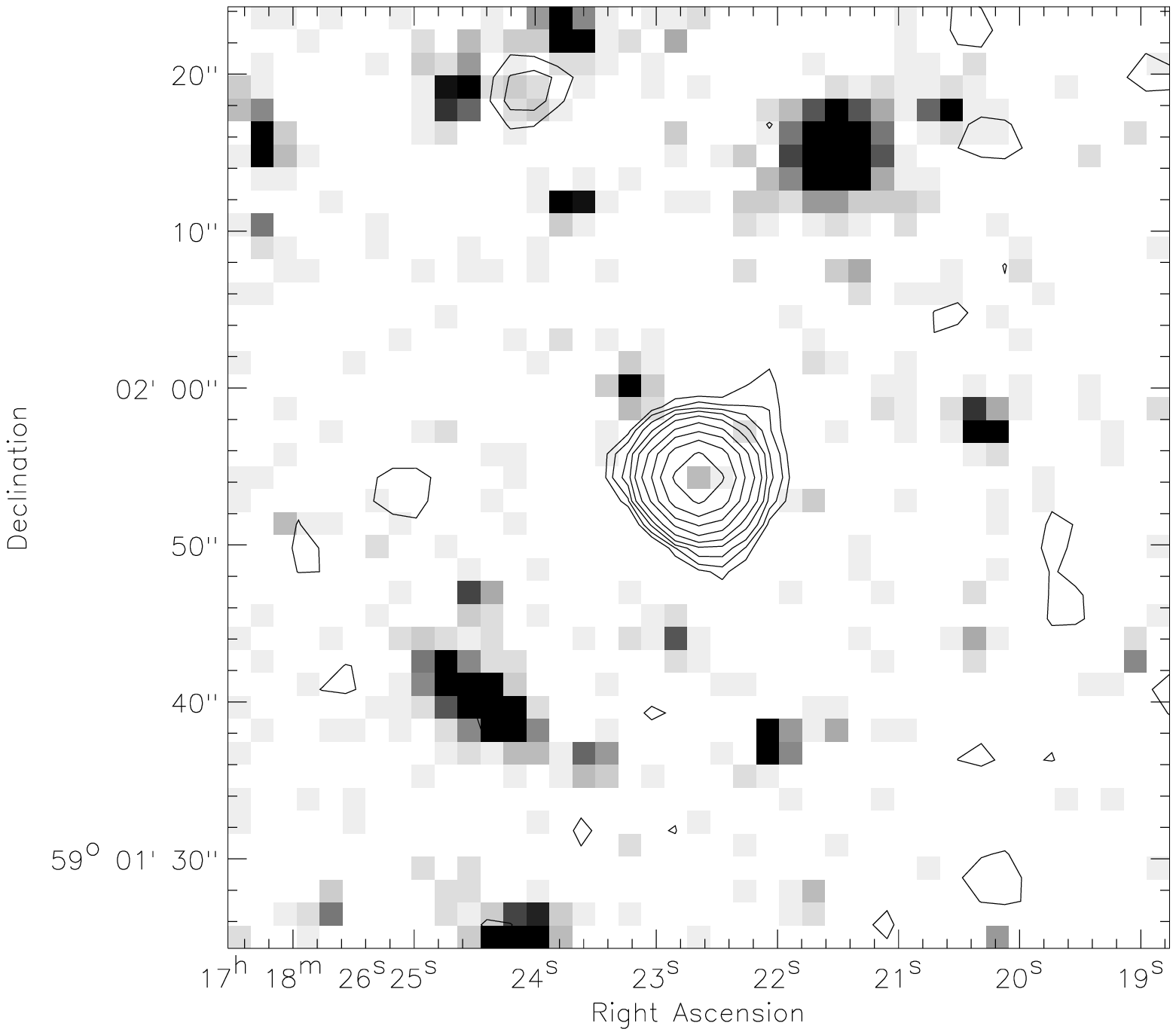,width=4.5cm,angle=0} 
\hspace{0.0cm}
\psfig{file= 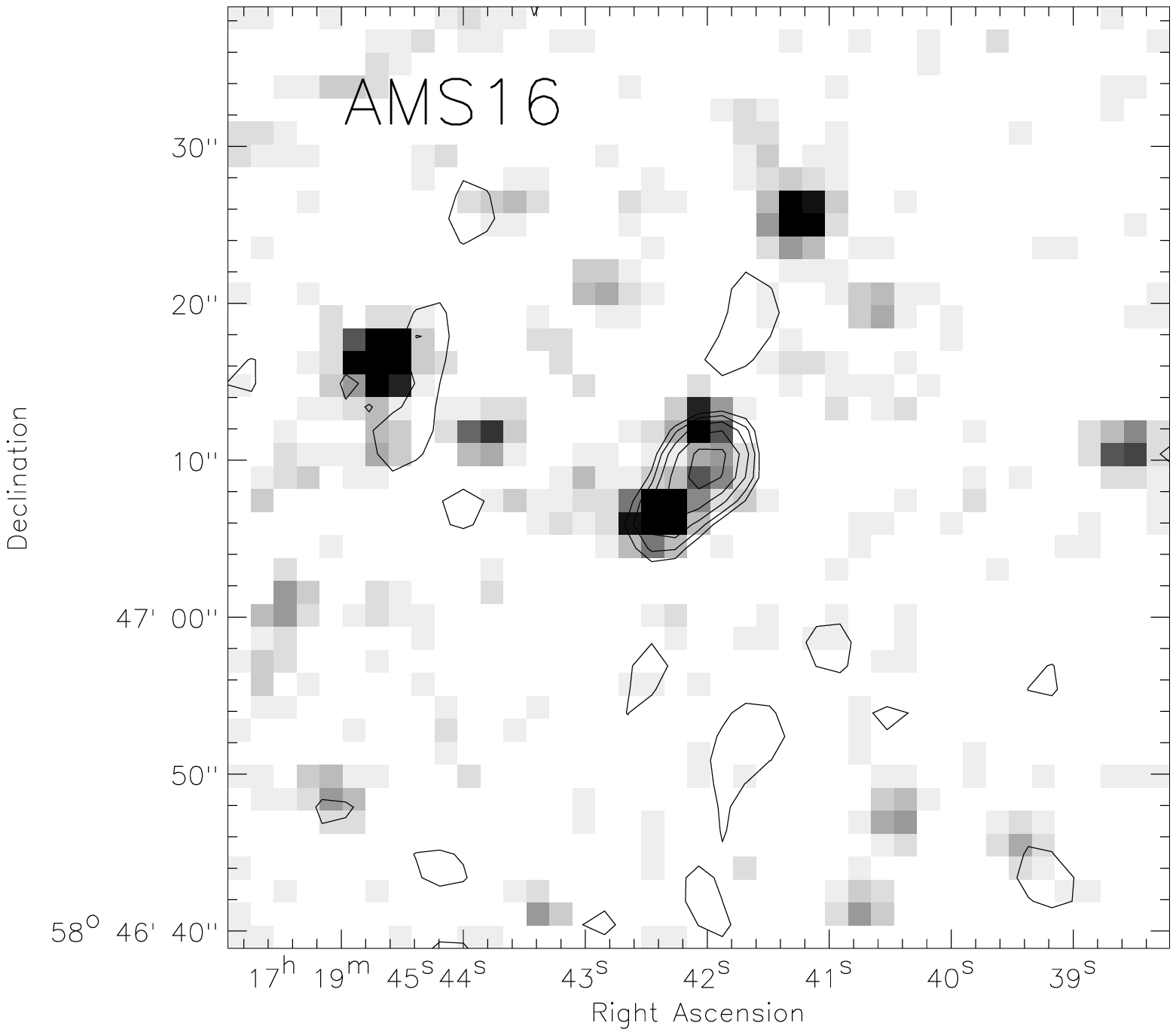,width=4.5cm,angle=0} 
\hspace{-.5cm}
\psfig{file= 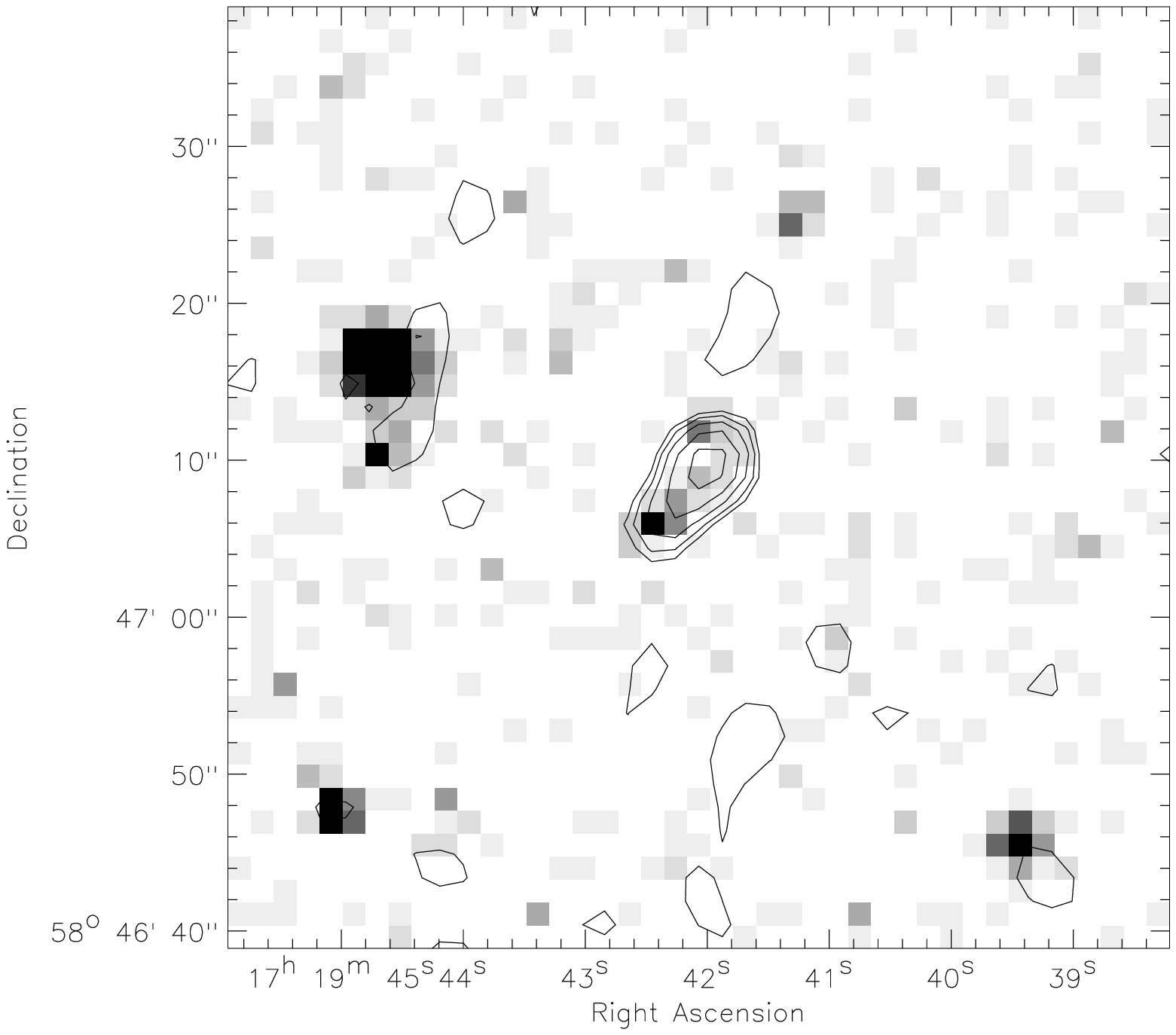,width=4.5cm,angle=0} 
}
\vspace{0.5cm} 
\caption{Continued}
\label{fig:overlaysB}
\end{figure*}

\subsection{Dust type}\label{sub:dusttype}

A second-order effect for mid-infrared selection is the type of dust
obscuring the AGN, which is particularly important in the range $1.2
\ltsimeq z \ltsimeq 1.8$ due to the varying depth of the silicate
absorption feature. Figure \ref{fig:dustcomp} shows the ratio of
transmission (see Figure caption), at observed 24-$\mu$m, for MW- and
SMC-type dust. It shows clearly that for $z \ltsimeq 3$ SMC-type dust
will cause more absorption in the mid-infrared, in particular around
9.7-$\mu$m rest-frame ($z \sim 1.5$). Thus if the dust obscuring
type-2s is more like the dust in the SMC than in the MW, then
mid-infrared selection might perform slightly worse than expected from
Figure~\ref {fig:Xraycomp}. As mentioned earlier, the dust reddening
type-1 quasars has been found to be more similar to SMC-dust, but that
does not necessarily mean that the dust present in the obscuring torus
is also SMC-type dust.

\subsection{Orientation of the quasar}\label{sub:orientation}

The analysis is mainly concerned with 24 $\mu$m selection of
objects at $z \ltsimeq 5$, so we are seeing light reprocessed by dust
in the torus at a range of temperatures, $T = 120-725$ K. We now use
the results of a radiative transfer code by \citet
{1994MNRAS.268..235G} to investigate the orientation dependence of
observed 24-$\mu$m emission. In Figure~\ref{fig:l24_lbol} we quantify
the fraction of the bolometric luminosity, $L_{\rm bol}$, seen at
24-$\mu$m as a function of redshift. For a pole-on unreddened source
(necessarily a type-1), this fraction is approximately constant
between $2 \ltsimeq z \ltsimeq 5$, while for an edge-on source (a
type-2 with $A_{\rm V} \approx 50$), the fraction is similar to that
of a type-1 and approximately constant in the range $2 \ltsimeq z
\ltsimeq 4$. At higher values of $z$, the observed 24-$\mu$m
corresponds to emitted wavelengths ($\ltsimeq 5 \mu$m) at which the
dust is becoming progresively more opaque. Around $z \sim 1.5$ the
silicate feature is seen in emission by the type-1 and absorption by
the type-2, so the observed 24-$\mu$m emission is not a very good
indicator of $L_{\rm bol}$. We can therefore conclude that 24-$\mu$m
is a good indicator of bolometric luminosity at $2 \ltsimeq z \ltsimeq
4$ and, as shown in Figure~\ref{fig:l24_lbol}, this result is not
strongly dependent on the assumed size of the torus.

\subsection{Summary}\label{sub:summary}

To summarise, we have discussed the advantages of mid-infrared
selection over hard X-ray selection. For gas-to-dust ratios similar to
those found in the Milky Way, X-ray selection is superior, and can
probe further down the luminosity function. However, many AGN have
gas-to-dust ratios significantly larger than the MW. We have seen in
this section how type-2 quasars with $N_{\rm H} = 5 \times 10^{26}$
m$^{-2} \times A_{\rm V}$ would naturally drop out of
existing large-area X-ray samples without requiring extreme optical
obscuration ($A_{\rm V} \gtsimeq 20$ would be enough). We have seen
that hard X-ray selected samples are in principle able to find even
Compton-thick type-2s, but they require a combination of depth and
area that is only just becoming available. We have also seen that
dust-type is important, but mainly in the range $1 \ltsimeq z \ltsimeq
2$ (due to the 9.7-$\mu$m silicate absorption feature being more
pronounced for certain dust types). Using the results of a radiative
transfer code, we have also seen that (observed) 24-$\mu$m luminosity
is a constant tracer of type-2 bolometric luminosity in the range $2
\ltsimeq z \ltsimeq 4$. Therefore, 24-$\mu$m selection is a natural
starting place to look for the population of high-redshift type-2
quasars missing from X-ray surveys.

\addtocounter{figure}{-1}
\begin{figure*}
\hbox{
\hspace{-.0cm}
\psfig{file= 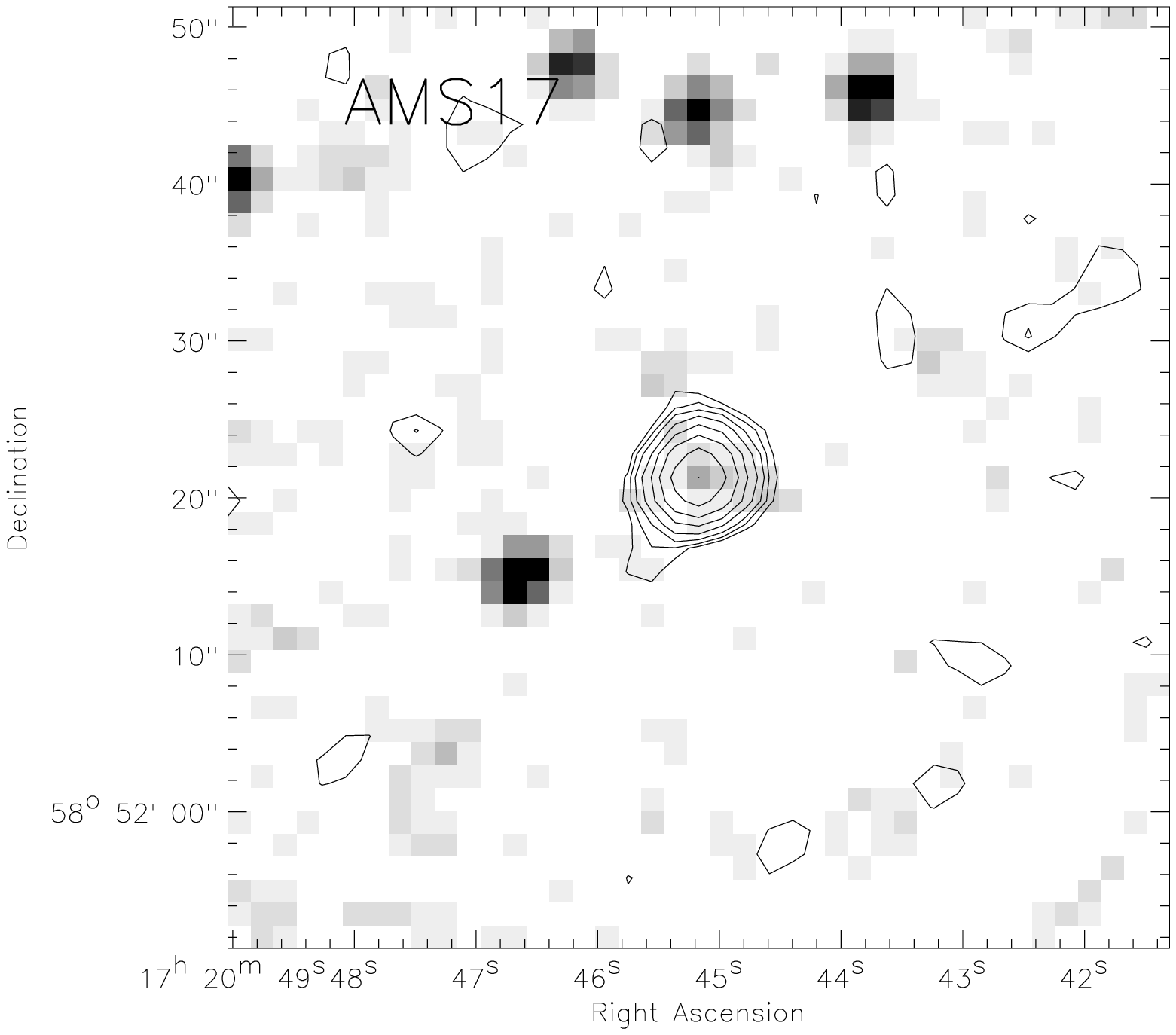,width=4.5cm,angle=0} 
\hspace{-.5cm}
\psfig{file= 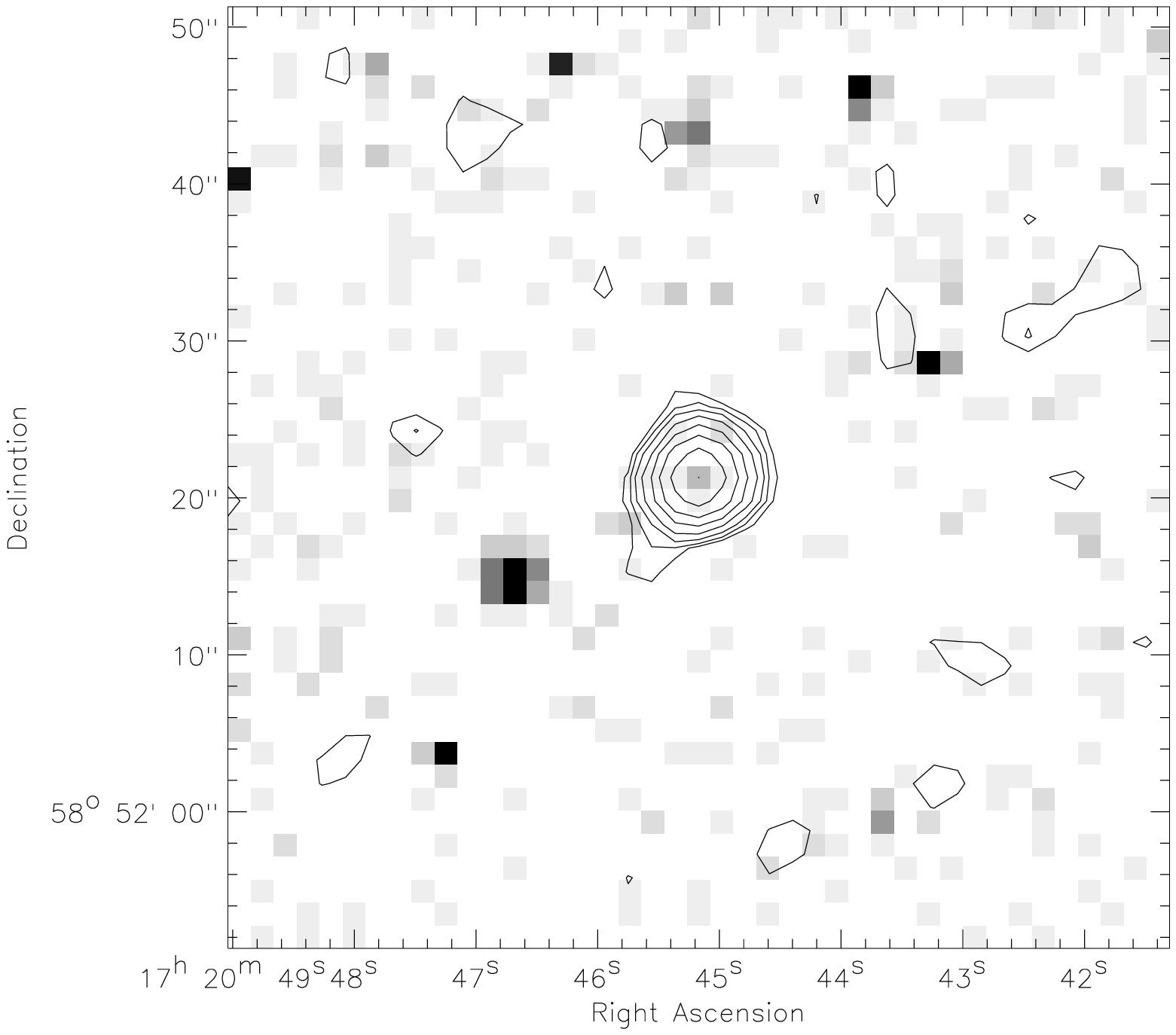,width=4.5cm,angle=0} 
\hspace{0.0cm}
\psfig{file= 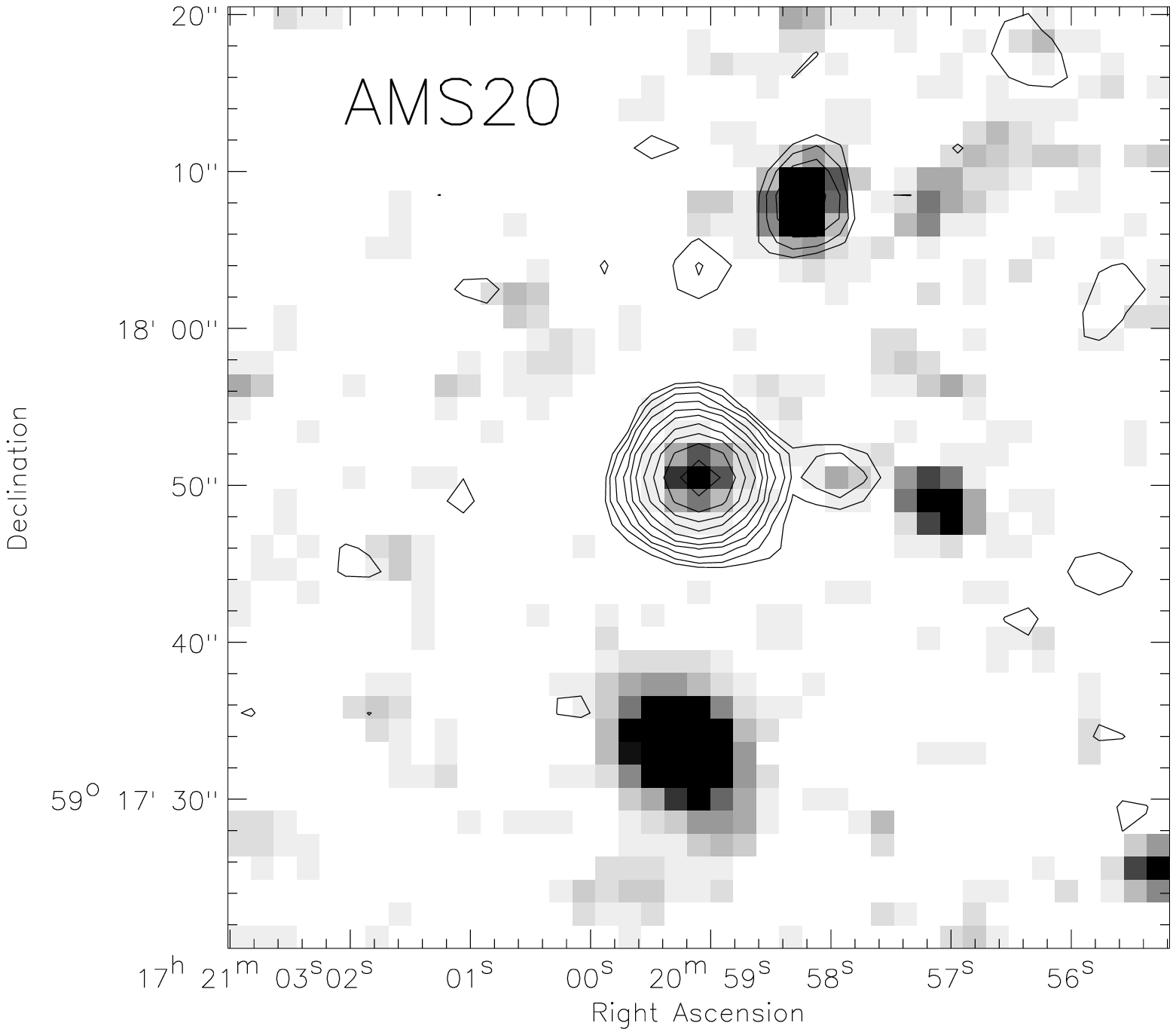,width=4.5cm,angle=0} 
\hspace{-.5cm}
\psfig{file= 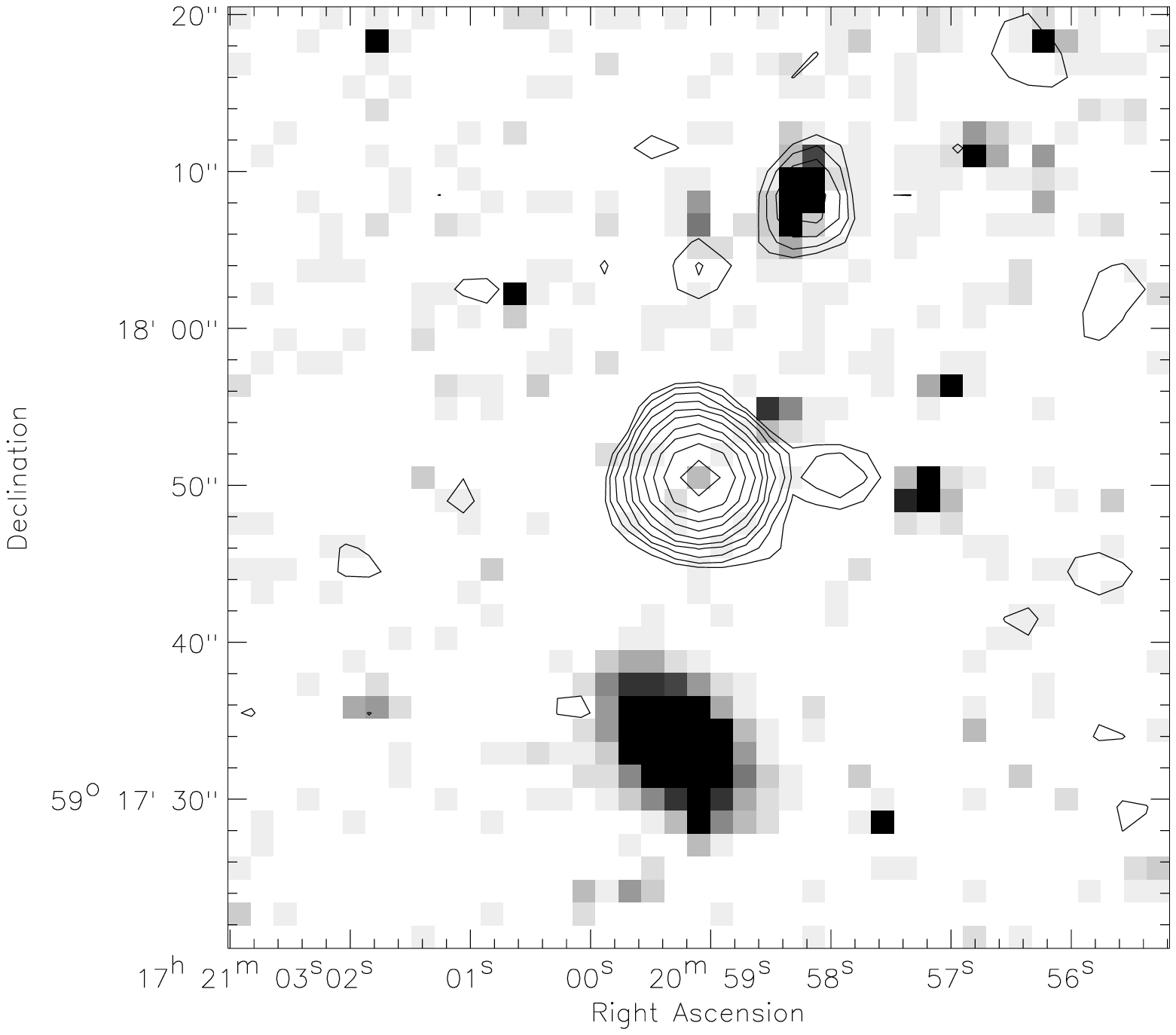,width=4.5cm,angle=0} 
}
\vspace{0.5cm} 
\hbox{
\hspace{ -.0 cm}
\psfig{file= 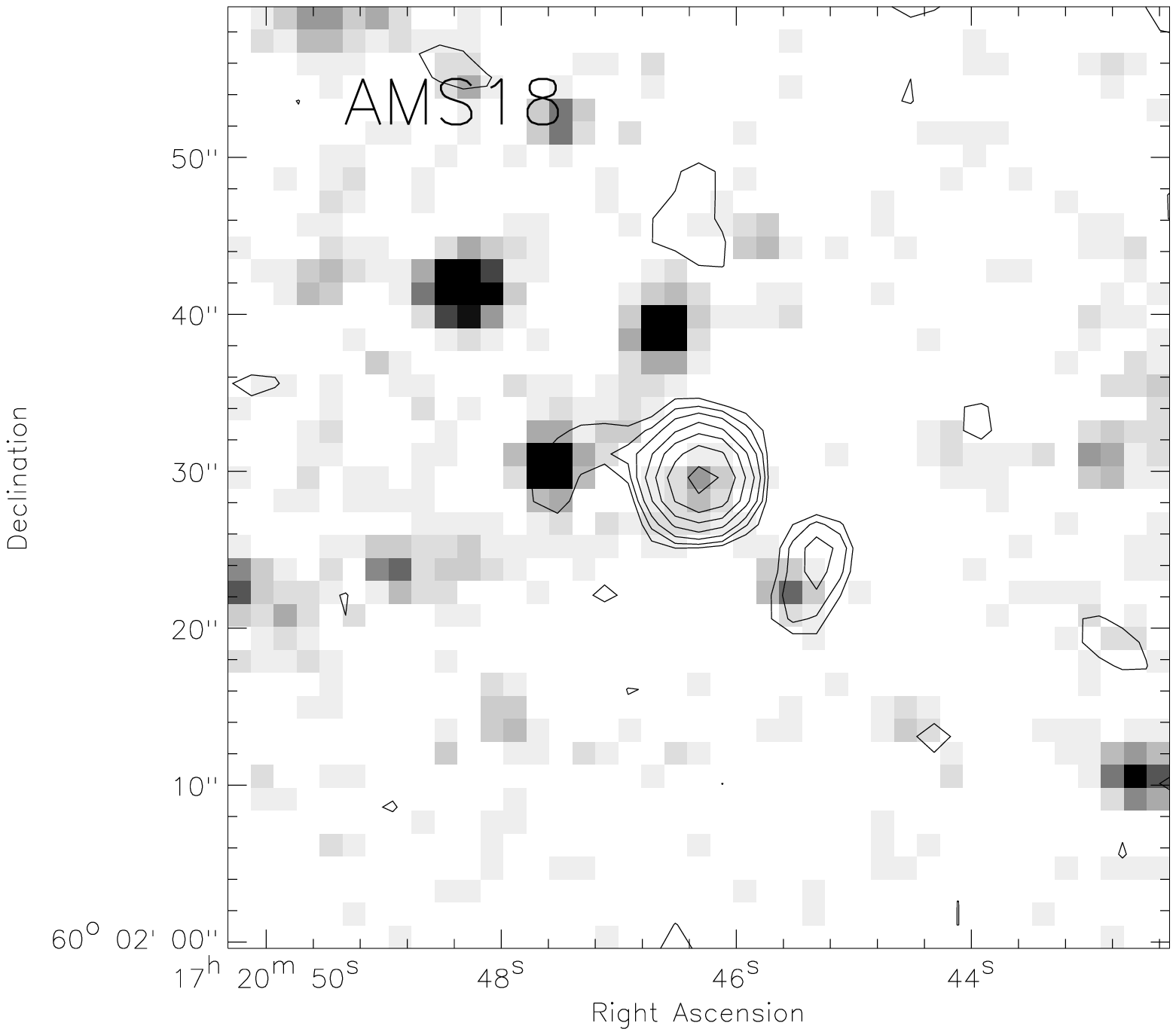,width=4.5cm,angle=0} 
\hspace{-.5cm}
\psfig{file= 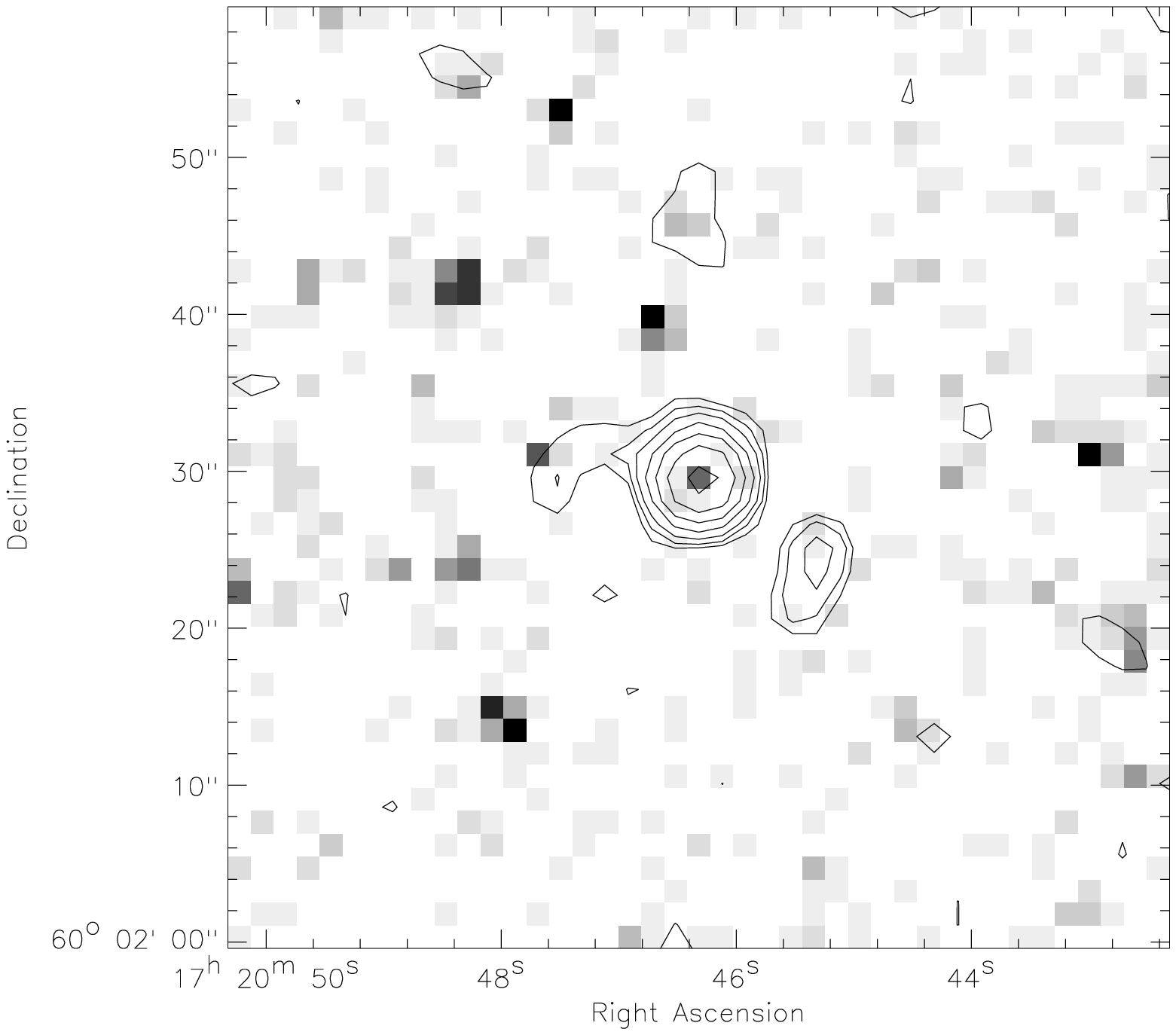,width=4.5cm,angle=0} 
\hspace{0.0cm}
\psfig{file= 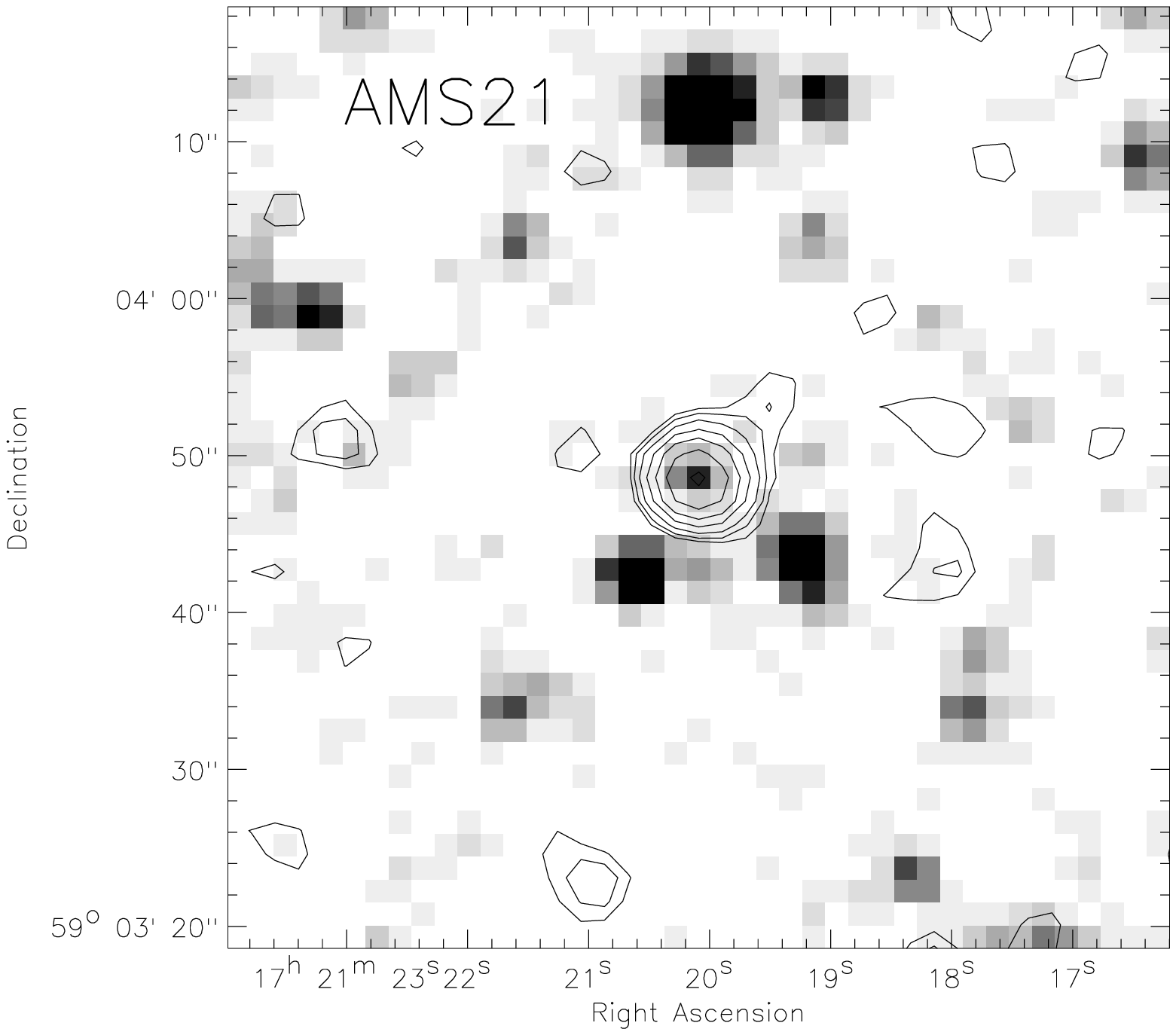,width=4.5cm,angle=0} 
\hspace{-.5cm}
\psfig{file= 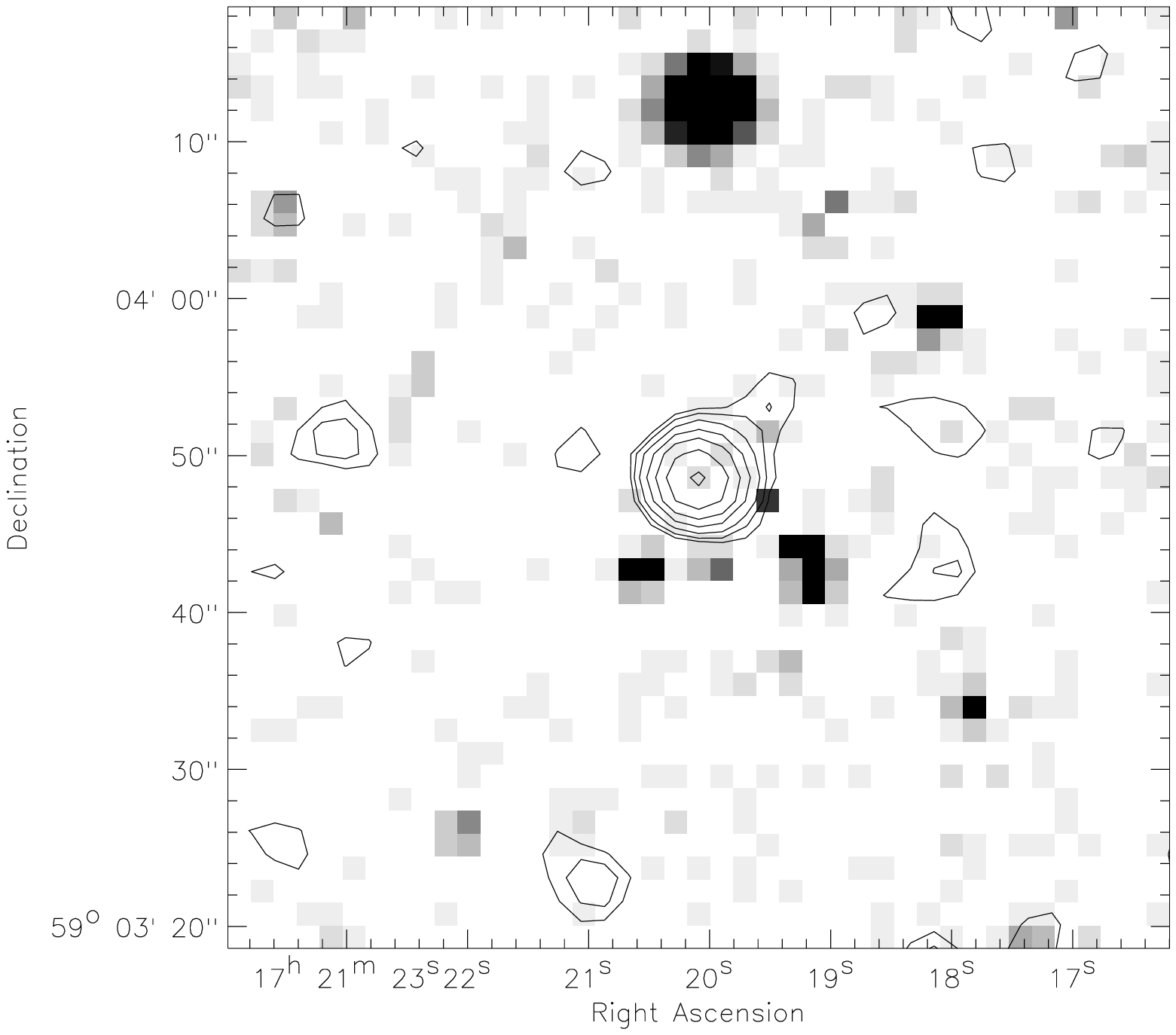,width=4.5cm,angle=0} 
}
\vspace{0.5cm} 
\hbox{
\hspace{-.0cm}
\psfig{file= 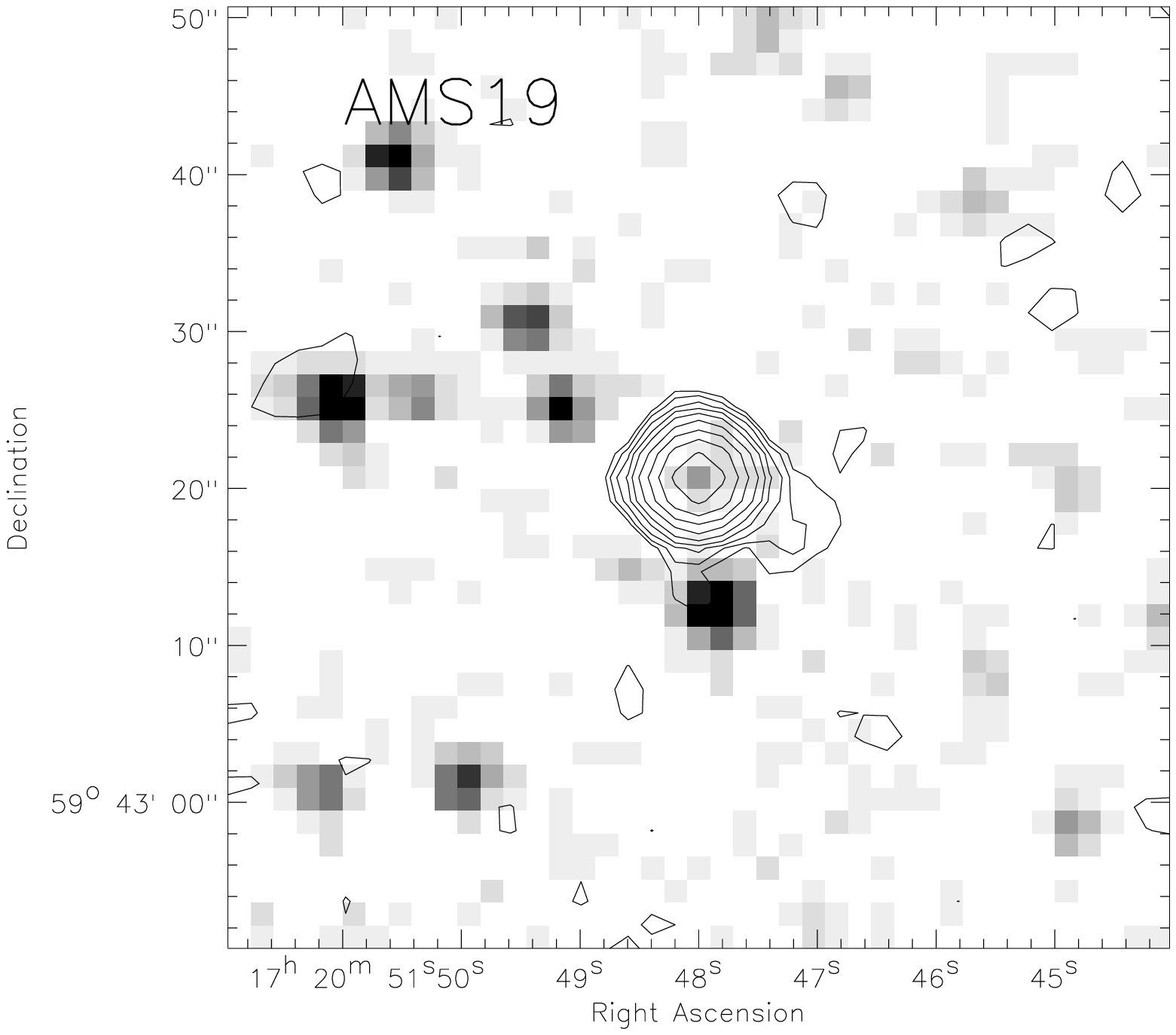,width=4.5cm,angle=0} 
\hspace{-.5cm}
\psfig{file= 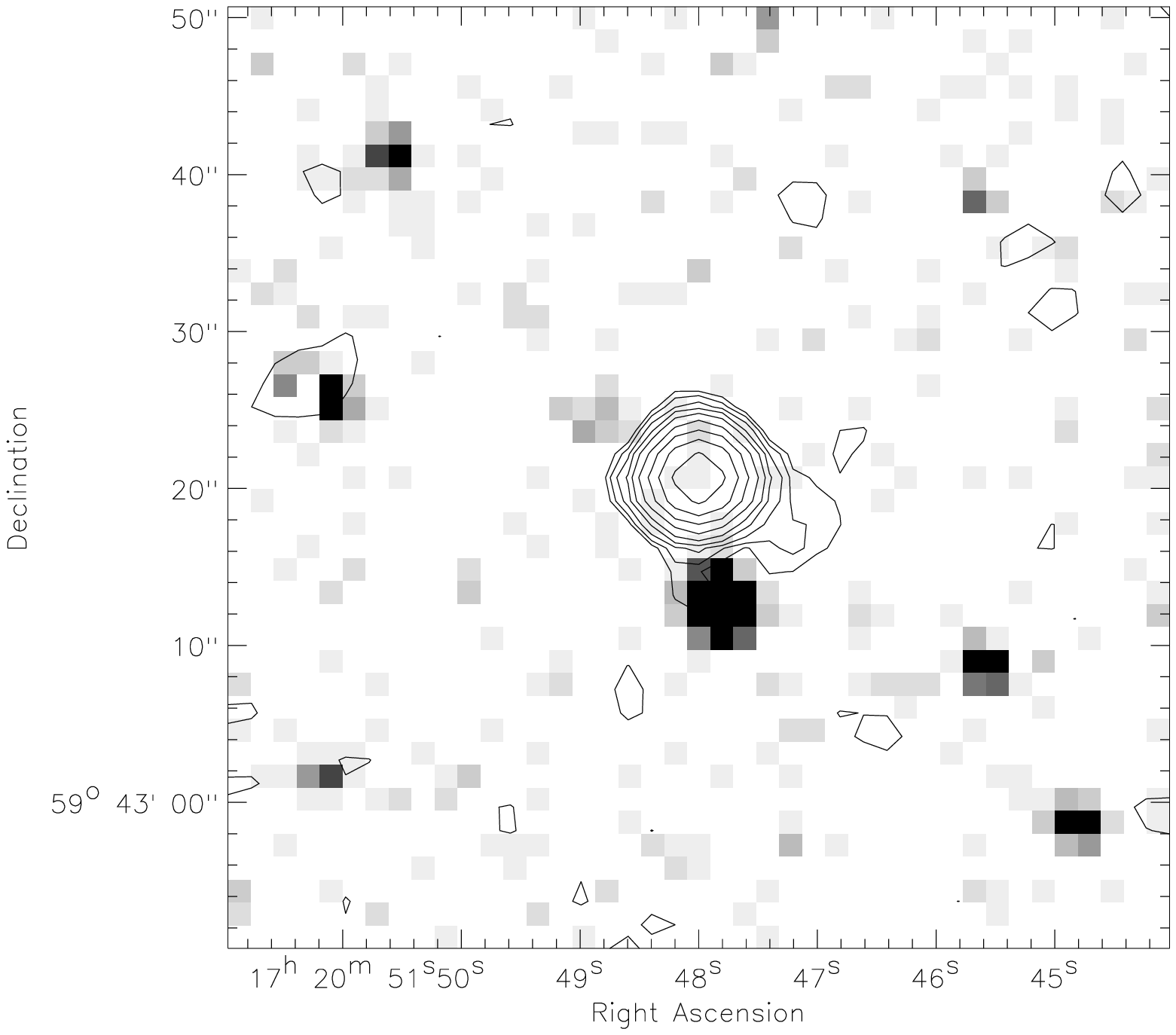,width=4.5cm,angle=0} 
\hspace{0.0cm}
}
\vspace{0.5cm} 
\caption{Continued}
\label{figure1c}
\end{figure*}

\section{Selection Criteria} \label{sec:selcrit}

Our aim was to find the elusive high redshift type-2 population around
the peak in the quasar activity, $z \gtsimeq 2$ \citep[e.g.][]
{2003A&A...408..499W}, and for this we used the following selection
criteria:

\begin{enumerate}
\item $S_{24 ~ \mu \rm m} > 300~\mu$Jy  
\item $S_{3.6~ \mu \rm m} \leq 45~\mu$Jy
\item $350 ~\mu \rm Jy \leq S_{1.4 \rm~GHz} \leq 2$ mJy
\end{enumerate}

\subsection{Datasets}\label{sub:catalogues}

The work described in this paper was all done in the Spitzer
Extragalactic First Look Survey (FLS).  Three separate catalogues were
cross-matched for the initial selection: the IRAC 3.6 $\mu$m \citep
{2005ApJS..161...41L} and the MIPS 24 $\mu$m \citep
{2004ApJS..154...66M,Fadda:2006mx} from the Spitzer First Look Survey
and the 1.4 GHz catalogue from the NRAO VLA \citep
{2003AJ....125.2411C}.  The flux limits used for 3.6 $\mu$m, 24 $\mu$m
and 1.4 GHz were 20, 300 and 100 $\mu$Jy respectively. 

Over the area with coverage in all three bands (3.8 deg$^{2}$), the
radio and 24-$\mu$m catalogues were matched, to select all sources
detected at 24-$\mu$m and within the radio flux cuts. The chosen
sources were then contrasted with the 3.6 $\mu$m catalogue, to obtain
their flux densities in this IRAC channel. However, objects were not required
to be detected at 3.6 $\mu$m to be kept in the candidate list. All
matching was done with a 2.5 arcsec tolerance to allow for  positional offsets
between catalogues.

In addition, in this paper we use tabulate R-band magnitudes from data
from the 4-m Mayall Telescope Kitt-Peak National Observatory \citep
{2004AJ....128....1F} and we also make use of a Westerbork Synthesis
Radio Telescope deep survey at 1.4 GHz in the Spitzer First Look
survey \citep {2004A&A...424..371M}. However, these two datasets were
not used to select the candidate type-2 quasars.

\subsection {Mid-infrared criteria}\label{sub:midir}

The mid-infrared criterion $S_{24~\mu \rm m} > 300~\mu$Jy was chosen
to select a flux-density-limited sample of active galaxies. At lower
redshifts, a 24-$\mu$m selected sample would be dominated by
starbursting galaxies, but combined with the 3.6-$\mu$m selection
(which ensured high-redshift objects, see below), this 24-$\mu$m flux
limit should yield powerful AGN.  The $S_{24~\mu \rm m} > 300$ $\mu$Jy
criterion is the flux-density limit from a preliminary catalogue, and
is actually a 7-$\sigma$ limit in the 24-$\mu$m final catalogue from
the Spitzer FLS data \citep {2004ApJS..154...66M,Fadda:2006mx}. At $z
= 2$, this corresponds to an emitted 8-$\mu$m luminosity $L_{8~ \mu
\rm m}= 10^{24.4}$ W Hz$^{-1}$. Assuming a typical type-1 SED \citep
{1995MNRAS.272..737R}, this is $L_{\rm B}= 10^{23.2}$ W Hz$^{-1}$ or
$M_{\rm B} = -23.8$. At $z = 2$, the break in the quasar luminosity
function, $L^{*}_{quasar}$, has a value of $M_{\rm B} = -25.7$ \citep
[] [with Pure Luminosity Evolution] {2004MNRAS.349.1397C}, so our
24-${\mu \rm m}$ selection will select quasars $\gtsimeq 0.2 ~
L^{*}_{quasar}$ at $z = 2$, and more luminous quasars at higher
redshifts. As explained in Section \ref{sec:Xray}, the selection
becomes more complicated at $z < 2$ due to the silicate absorption, so
we restrict our discussion to $z \geq 2$.  The peak of the quasar
activity occurred around $z \geq 2$ and by targeting the quasars
around the ``break'' we are sensitive to the part of the population
that contributes most of the luminosity density.

Quasars are considered as being type-2 if $A_{\rm V} \gtsimeq 5$
\citep {1999MNRAS.306..828S}, and such an amount of extinction will
make the observed near-infrared emission of type-2s much fainter than
that of type-1s. The 3.6-$\mu \rm m$ criterion was therefore chosen to
remove naked (type-1) quasars as well as lower redshift ($z \ltsimeq
1.4$) type-2s.  At $z \geq 2$ the detected 3.6-$\mu$m flux density
corresponds to light emitted at $\lambda \leq$ 1.2 $\mu$m so dust
extinction ensures that type-2 quasars are much fainter than type-1
quasars, even for a moderate $A_{V}$.  Indeed, the $S_{3.6 \mu \rm m}$
emission is likely to be dominated by starlight for $A_{V} \gtsimeq
10$, and since light at 3.6-$\mu$m is dominated by the old stellar
population, there will be an $S_{3.6 \mu \rm m}-z$ correlation,
analogous to the $K-z$ relation for radio galaxies \citep
{2001MNRAS.326.1585J,2003MNRAS.339..173W}. A typical host galaxy for a
$z = 2$ radio-quiet quasar is the progenitor of a $\sim 2L^{*}_{gal}$
\citep {2001MNRAS.326.1533K} galaxy in the local Universe, so we adapt
the $K-z$ relation (which corresponds to $\sim3L^{*}_{gal}$ hosts for
radio galaxies) to $\sim2L^{*}_{gal}$ hosts at 3.6-$\mu$m.  To do
this, we use the models of \citet {2003MNRAS.344.1000B} to form an
elliptical galaxy at $z = 10$, assign it a (stellar) mass and allow it
to evolve passively. This allows us to predict the 3.6 $\mu$m flux as
a function of redshift, for an elliptical galaxy of this given mass,
and hence obtain a rough photometric redshift. The stellar mass of the
galaxy was set to $3.8 \times 10^{11} M_{\odot}$, and in \citet
{2005Natur.436..666M} we quoted this as corresponding to
$\sim2L^{*}_{gal}$.  If we assume a Salpeter IMF
\citep{1955ApJ...121..161S} and take $M_{\rm K} = -24.3$ from the
fitting of Schechter functions by \citep {2001MNRAS.326..255C}, then
our assumed stellar mass translates to a $2.6L^{*}_{gal}$ galaxy. This
is the host (elliptical) galaxy luminosity assumed to obtain the rough
photometric redshifts in \citet {2005Natur.436..666M} and in this
paper. The criterion $S_{3.6~ \mu \rm m} \leq 45~\mu$Jy corresponds to
a limiting photometric redshift $z_{\rm phot} \gtsimeq 1.4$. This was
chosen to target $z \sim 2$ type-2 quasars, while allowing for scatter
in the crude photometric redshift estimation and filtering out type-1
quasars and low-redshift contaminants like radio galaxies.

The combination of these criteria is illustrated in Figure
\ref{fig:sed}. It shows the observed spectral energy distributions
of two $z=2$ quasars \citep [model from][]{1995MNRAS.272..737R}, a
type-1 and a type-2 \citep [the latter obscured by   $A_{\rm V}=5$, dust model
from][] {1992ApJ...395..130P}. A 2.6$L^{*}$ elliptical galaxy from
\citet {2003MNRAS.344.1000B} is alos plotted at $z=1.4$, showing how the
3.6-$\mu$m criterion serves also as a crude photometric redshift cut.

\subsection {Radio criterion}

The radio selection criteria is added to ensure that the candidates
are quasars rather than starburst galaxies. The 3.6-$\mu$m - 24-$\mu$m
``colour'' demanded by our criteria can be achieved by lower redshift
($z < 1$) ultra luminous infrared galaxies (ULIRGs). For example, NGC
7714 would satisify these criteria if it were at $z = 0.25-0.4$ \citep
[see Figure \ref{fig:sed}, SED from][] {2004ApJS..154..188B} and the
more luminous ULIRG Arp 220 if it where at $z = 0.5-0.7$.  In the
absence of AGN radio activity, the radio flux densities of such
objects would follow the far-IR/radio correlation \citep
{1992ARA&A..30..575C}, and hence would have values of $S_{1.4~\rm GH}
\sim 50$ (for ULIRGs at $z \sim 0.3$) and $500~\mu$Jy (for more
powerful ULIRGs at $z \sim 0.6$).

To avoid such starburst contaminants, we chose a lower limit on
$S_{1.4~\rm GHz}$ which also excludes higher-redshift
(submillimetre-selected) starburst galaxies without the benefit of
strong gravitational lensing \citep[e.g.][] {2005ApJ...622..772C}.
For a $z \sim 0.6$ ULIRG to have $S_{1.4~\rm GH} \geq 350~\mu$Jy, it
would require a far-infrared luminosity of $L_{\rm FIR} \gtsimeq
1\times10^{12}$ $L_{\odot}$. Objects with this far-infrared luminosity
are rare in space-density \citep {1996ARA&A..34..749S}, and the
comoving volume probed at $z \sim 0.6$ by 3.8 deg$^{2}$ is such that
$\ltsimeq 1$ contaminants would be expected in our sample.

An upper limit on $S_{1.4~\rm GHz}$ was also used to filter out the
radio-loud objects, whose extended jets might complicate
interpretation. The radio flux-density cut introduced here will select
the radio-bright end of the radio-quiet quasar population \citep
{2003MNRAS.346..447C}, so these objects can be described as
radio-intermediate. In addition, the $S_{3.6 \mu \rm m}$ criterion
includes objects without IRAC detections which we expect to have the
highest redshifts. Since the 24-$\mu$m positions are accurate to $\sim
1$ arcsec, the FLS radio positions \citep {2003AJ....125.2411C},
accurate to $\sim 0.5$ arcsec, were also better for spectrosocopic
follow-up.

\begin{table*}
\begin{center}
\begin{tabular}{lllllll}
\hline
\hline
Name & RA & Dec & Date & Airmass & PA & Slit \\
   & (J2000)  &  &  Observed &  &/deg & /arcsec \\
\hline
AMS01 & 17 13 11.17 & +59 55 51.5  & 15/07/04 & 1.17 & 103 & 1.5 \\ 
AMS02 & 17 13 15.88 & +60 02 34.2  & 14/07/04 & 1.37 & 149 & 2.0 \\
AMS03 & 17 13 40.19 & +59 27 45.8  & 14/07/04 & 1.17 &114 & 1.5\\
AMS04 & 17 13 40.62 & +59 49 17.1  & 14/04/05 & 1.18 & 36 & 2.0 \\
AMS05 & 17 13 42.77 & +59 39 20.2  & 14/07/04 & 1.47/1.56$^{[1]}$ & 98 & 2.0 \\
AMS06 & 17 13 43.91 & +59 57 14.6  & 13/07/04 & 1.19 & 120 & 2.0 \\
AMS07 & 17 14 02.25 & +59 48 28.8  & 16/07/04 & 1.22 & 19 & 1.5 \\
AMS08 & 17 14 29.67 & +59 32 33.5  & 15/07/04 & 1.32 & 128 & 1.5 \\
AMS09 & 17 14 34.87 & +58 56 46.4  & 15/07/04 & 1.17 & 111 & 1.5 \\
AMS10 & 17 16 20.08 & +59 40 26.5  & 14/07/04 & 1.20 & 114  & 1.5 \\
AMS11 & 17 18 21.33 & +59 40 27.1  & 16/07/04 & 1.19 & 76 & 1.0 \\
AMS12 & 17 18 22.65 & +59 01 54.3  & 15/07/04 &1.16 &150 & 1.5 \\
AMS13 & 17 18 44.40 & +59 20 00.8  & 14/04/05 & 1.16 & 186 & 2.0 \\
AMS14 & 17 18 45.47 & +58 51 22.5  & 13/07/04 & 1.20 & 105 & 2.0 \\
AMS15 & 17 18 56.93 & +59 03 25.0  & 15/04/05$^{[2]}$ & - & - & - \\
AMS16 & 17 19 42.07 & +58 47 08.9  & 16/07/04 & 1.25 & 141 & 2.0\\
AMS17 & 17 20 45.17 & +58 52 21.3  & 15/04/05 & 1.27 & 117 & 1.5\\
AMS18 & 17 20 46.32 & +60 02 29.6  & 13/07/04 & 1.40 & 133 & 2.0 \\
AMS19 & 17 20 48.00 & +59 43 20.7  & 14-15/04/05 & 1.17/1.17 & 170/185 & 2.0\\
AMS20 & 17 20 59.10 & +59 17 50.5  & 14/07/04 & 1.22 & 160 & 2.0 \\
AMS21 & 17 21 20.09 & +59 03 48.6  & 15/04/05 & 1.20 & 84 & 1.5 \\
\hline
\hline
\end{tabular}
\caption{ \noindent Summary of the spectroscopic observations. Most of
the objects were observed in July 2004, and the spectroscopy was
completed in April 2005. The objects were all observed with airmass $<
1.5$ (except AMS05), with good seeing (in the range 0.7-1.0
arcsec). As described in Section \ref{sec:obs}, the position angles
(PAs) of the slit were generally chosen to go through any other radio
source in the field, or any bright 3.6 $\mu$m source, to help the
astrometry and identification. In some cases neither were available,
or the airmass was high, so the PA was chosen to match the parallactic
angle and therefore minimise dispersion of the blue
light. $^{[1]}$AMS05 was observed twice on the same night, at
different airmasses. $^{[2]}$AMS15 was targetted in the April 2005,
but the position used was slightly wrong ($\sim 0.8$ arcsec) and
therefore we do not consider it to have been observed.}
\label{tab:observations}
\end{center}
\end{table*}

The two simple infrared criteria together with radio selection 
were chosen over an IRAC colour requirement because the latter
criterion would have selected type-2s with a narrow range in $A_{\rm
V}$ and $z$. Our three criteria represent a simple way of selecting
type-2 quasars and allow the
modelling necessary to estimate the quasar fraction at high redshift
\citep {2005Natur.436..666M}.

After these selection criteria, we were left with 21 candidate
high-redshift type-2 quasars (see Table \ref{tab:sample}), which were
followed up spectroscopically (as described in Section \ref
{sec:obs}).  Note that all our candidates were originally selected
from the preliminary 3.6 and 24 $\mu$m catalogues, but have had their
fluxes slightly revised and the flux densities quoted are the final
ones. This explains why two candidates are slightly outside the
selection boundaries. We also note that although the R-band (Vega)
magnitudes are not part of our selection criteria, they are consistent
with optically obscured quasars in high-redshift host galaxies. At $z
= 1.4$, our model elliptical galaxy would have $m_{\rm R}= 23.3$.

\section{Observations and Data Reduction}\label{sec:obs}

The 21 candidates were observed with both arms of the ISIS instrument
at the William Herschel Telescope in July 2004 and April 2005 (see
Table~1). Low-resolution long slit spectroscopy
was performed across the entire visible band (3200~\AA~ to 9300~\AA)
using the gratings R158B and R158R, and the EEV12 and Marconi
detectors. The 6100~\AA~ dichroic was used in July 2004, while in
April 2005 we used the new 5300~\AA~ dichroic.  Using slit widths of
1.5-2.0 arcsec, we obtained spectral resolutions of 12-16~\AA~ in both
arms. Bias, flat, arc, sky-flat and standard-star frames were all taken
at the beginning and end of the night to ensure reasonable
callibration.

\begin{table*}
\begin{center}
\begin{tabular}{lllrrrrrrl}
\hline
\hline
Name & RA & Dec & $S_{24\mu \rm m}$ &  $S_{1.4 \rm GHz}$ &  $S_{3.6\mu \rm m}$&  $S_{4.5\mu \rm m}$  &  $m_{\rm R}$ &
$z_{\rm phot}$ & $z_{\rm spec}$  \\

& (J2000)  & &  / $\mu $Jy & / $\mu $Jy & / $\mu $Jy &  /$\mu $Jy & & \\
\hline
AMS01 & 17 13 11.17 & +59 55 51.5  & 536 & 490 &25.0 &30.0 & 25.1 &2.1 &\\
AMS02 & 17 13 15.88 & +60 02 34.2  & 294 & 1184 & 44.5 & 61.0 & 25.3 &1.4 &\\
AMS03 & 17 13 40.19 & +59 27 45.8  & 500 & 1986 & 16.0 & 28.0 & 24.4 & 3.1 & 2.698$^{[1]}$\\
AMS04 & 17 13 40.62 & +59 49 17.1  & 828 & 536 & 18.0 &22.0 & 22.96 &2.8 &1.782 \\
AMS05 & 17 13 42.77 & +59 39 20.2  & 1769 & 1038 & 34.7 & 61.4 & 25.2 & 1.7 & 2.017\\
AMS06 & 17 13 43.91 & +59 57 14.6  & 969 & 444 & $<$20 & $<$25 & $\geq$25.3 &$\geq$2.5 & 1.76$^{[2]}$ \\
AMS07 & 17 14 02.25 & +59 48 28.8  & 503 & 354 & 37.9 & 47.8 & 24.7 & 1.5 &\\
AMS08 & 17 14 29.67 & +59 32 33.5  & 792 & 655 & 41.6 & 46.3 & 22.48 & 1.4 & 1.979\\
AMS09 & 17 14 34.87 & +58 56 46.4  & 685 & 426 & 25.2 & $<$25 &$\geq$25.3 & 2.1 &\\
AMS10 & 17 16 20.08 & +59 40 26.5  & 338 & 1645 & 39.2 & 44.7 & 24.1 & 1.5 &\\
AMS11 & 17 18 21.33 & +59 40 27.1  & 442 & 356 & 32.4 & 55.6 & 24.7 & 1.7 &\\
AMS12 & 17 18 22.65 & +59 01 54.3  & 518 & 946 & 25.24 & $<$25 & 23.77 & 2.0 & 2.767\\
AMS13 & 17 18 44.40 & +59 20 00.8  & 4196 & 1888 & 24.7 & 49.1 & 22.89 & 2.1 & 1.974$^{[3]}$\\
AMS14 & 17 18 45.47 & +58 51 22.5  & 937 & 469 & 8.0 & 15.0 & 23.67 & 4.6 & 1.794 \\
AMS15 & 17 18 56.93 & +59 03 25.0  & 371 & 440 & 16.0 & 16.0 & $\geq$25.3 & 3.0 & Not Observed$^{[4]}$\\
AMS16 & 17 19 42.07 & +58 47 08.9  & 788 & 390 & 18.0 & 21.0 & 23.07 & 2.8 & 4.169\\
AMS17 & 17 20 45.17 & +58 52 21.3  & 1134 & 615& 10.0 & 15.0 & 23.80 & 3.9 & 3.137$^{[1]}$\\
AMS18 & 17 20 46.32 & +60 02 29.6  & 925 & 390 &  $<$20  & $<$25 & 23.53 & $\geq$2.5 & 1.017 \\
AMS19 & 17 20 48.00 & +59 43 20.7  & 1433 & 822 &  $<$20 & 26.9 & $\geq$25.3 & $\geq$2.5 & 2.25$^{[2]}$\\
AMS20 & 17 20 59.10 & +59 17 50.5  & 492 & 1268 & 45.2 & 64.9 & 23.80 & 1.4 & $^{[5]}$ \\
AMS21 & 17 21 20.09 & +59 03 48.6  & 720 & 449 & 25.2 & 38.6 & 24.2 & 2.1 &  \\
\hline
\hline
\end{tabular}
\caption{ \noindent Basic data on the 21 type-2 quasars in our sample.
The J2000.0 positions are from the Spitzer FLS radio catalogue
\citep{2003AJ....125.2411C}. The MIPS 24-$\mu$m flux density $S_{24\mu
\rm m}$ is obtained by point spread function (PSF) fitting
\citep{2004ApJS..154...66M} as all objects (except AMS16) are point
sources at the $\sim 6$ arcsecond resolution of the MIPS observations,
with positional errors $\sim 1$ arcsecond. It has a typical error of
$\pm 10-15$\%. The 1.4-GHz flux density is the peak value (in $\mu$Jy
per beam) from the radio catalogue.  The IRAC 3.6- and 4.5-$\mu$m flux
densities are measured in 5-arcsec-diameter apertures
\citep{2005ApJS..161...41L} and have typical errors of $\pm 10$\%.
The R-band magnitudes are measured using a 3-arcsecond diameter
aperture from the images of the 4-m Mayall Telescope Kitt-Peak
National Observatory \citep {2004AJ....128....1F}, with typical errors
of 0.03 for objects with magnitude 22-23, 0.10 for objects with 23-24
and 0.25 mags for those with magnitudes of 25.  $^{[1]}$ These objects
have redshifts identified from a single line, which we identified as
Lyman-$\alpha$: in AMS03, from the amount of structure in the line,
and in AMS17 from the fainter continuum blueward of the line. $^{[2]}$
These two objects have redshifts from Spitzer-IRS spectroscopy from
\citet{2006ApJ...638..613W}, who give two values for the redshifts. Here we
show the mean of the two values.  $^{[3]}$ This object was observed
with Spitzer-IRS by \citet {2005ApJ...628..604Y} who measured a
redshift of $z = 2.1 \pm 0.1$ in good agreement with our
value. $^{[4]}$ As explained in Table~1, AMS15
has not really been observed spectroscopically.$^{[5]}$ AMS20 Shows faint red continuum
but no emission lines. Note there were errors in the redshifts of
AMS14 and AMS18 presented in Table~1 of \citet {2005Natur.436..666M}
and these have been corrected here, together with minor corrections to
several other redshifts. }
\label{tab:sample}
\end{center}
\end{table*}

The slit was placed on the radio positions, without optical
identification, by ``blindly'' offsetting from a nearby guide star
\citep[e.g. ] []{1990MNRAS.243P..14R}. Whenever another radio source
was in the field, the slit angle was chosen to go through the source
as well as our candidate. When no other radio sources were available,
bright sources from the IRAC 3.6-$\mu$m catalogue were placed in the
slit.  This ensured multiple objects were detected in the slit,
improving the astrometry and identification of our sources in the 2-D
spectra. At high airmasses ($\sim 1.4-1.5$) we observed the objects
with slit position angles (PAs) close to the parallactic angle, to minimise
atmospheric dispersion of the blue light.  Each object was observed for 30 minutes
with a 1-2 arcsec slit width, with one 1800 second exposure in the blue arm and
two 900 second exposures in the red arm.

\begin{figure*}
\begin{center}
\vspace{5cm}
\hbox{
\hspace{1.0cm}
\psfig{file= 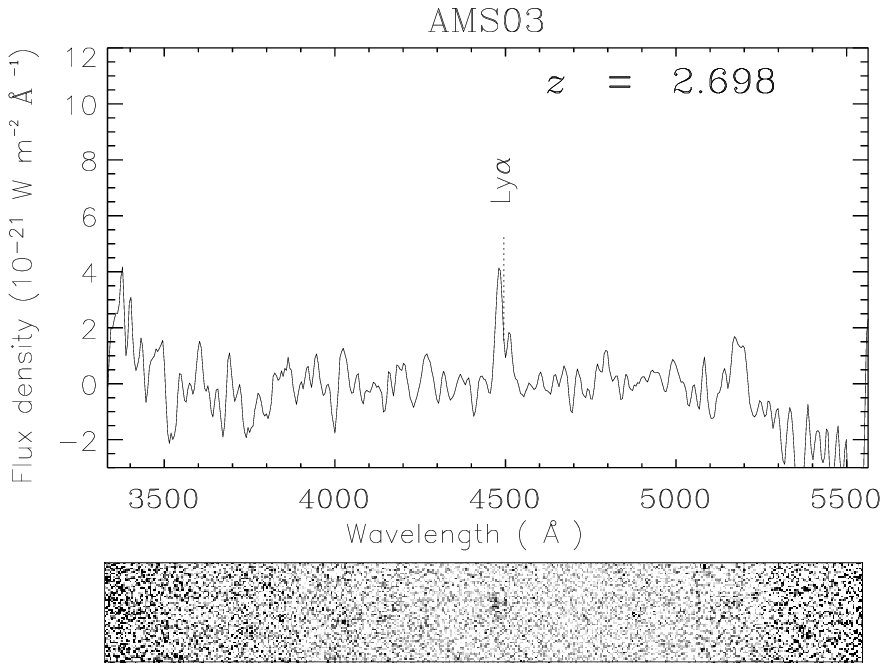,width=6.5cm,angle=0} 
\hspace{2.0cm}
\psfig{file= 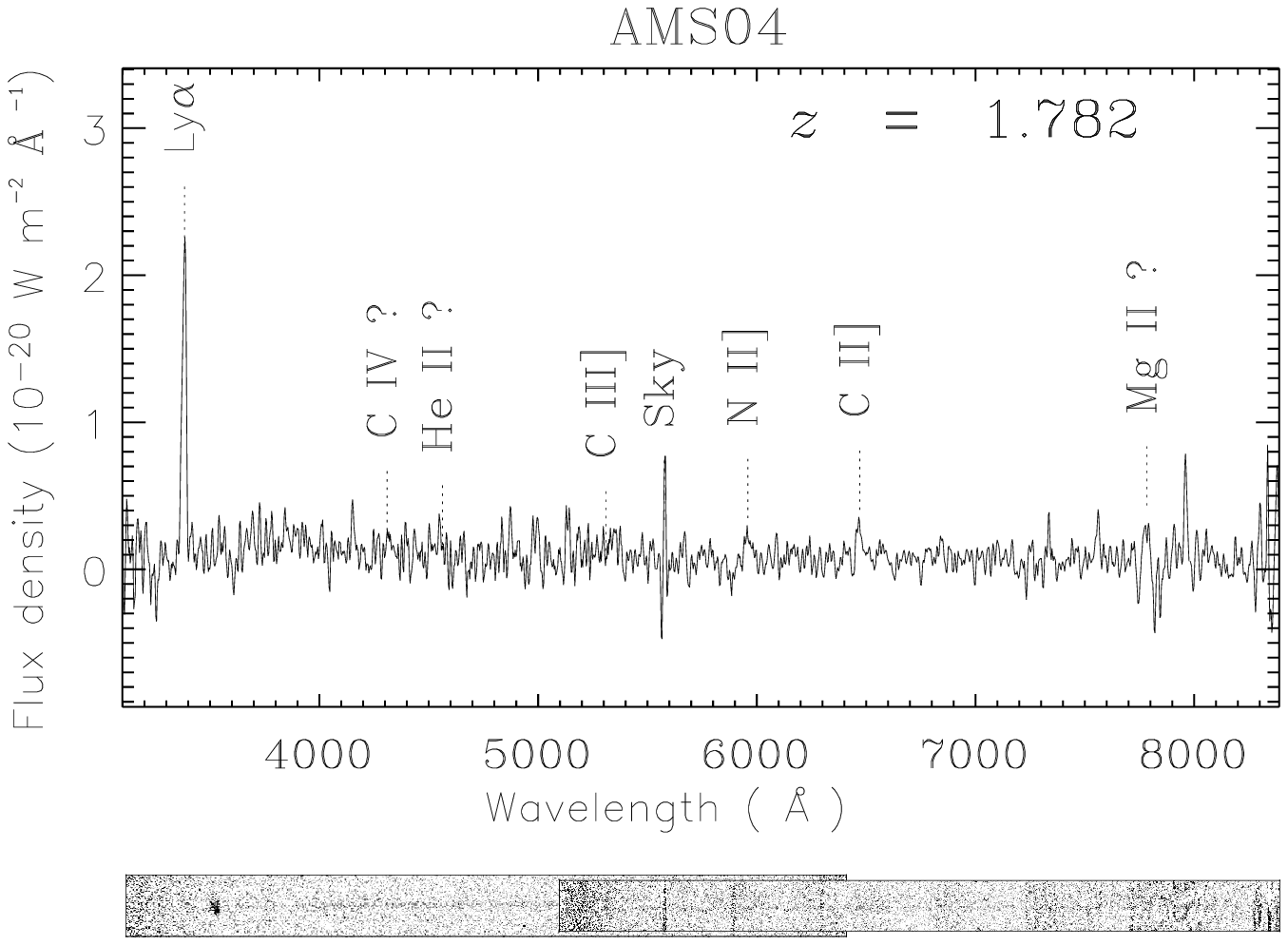,width=6.5cm,angle=0} 
}
\vspace{5.5cm}
\hbox{
\hspace{1.0cm}
\psfig{file= 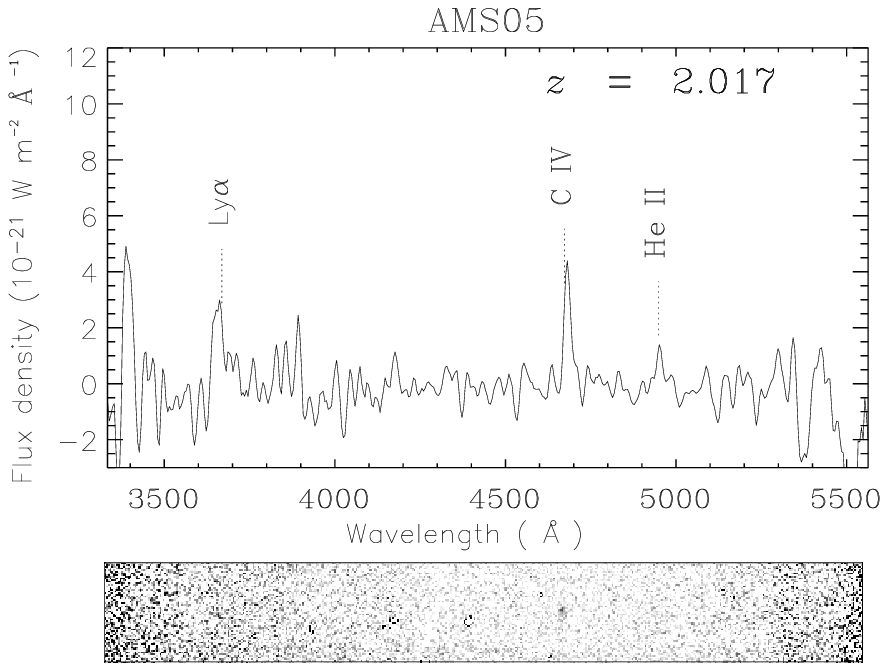,width=6.5cm,angle=0} 
\hspace{2.0cm}
\psfig{file= 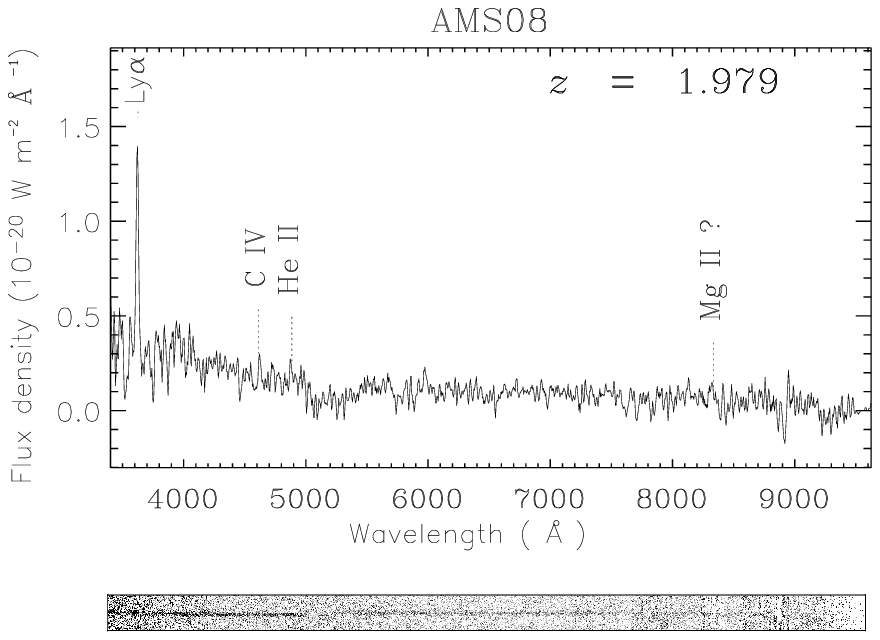,width=6.5cm,angle=0} 
}
\vspace{5.5cm}
\hbox{
\hspace{1.0cm}
\psfig{file= 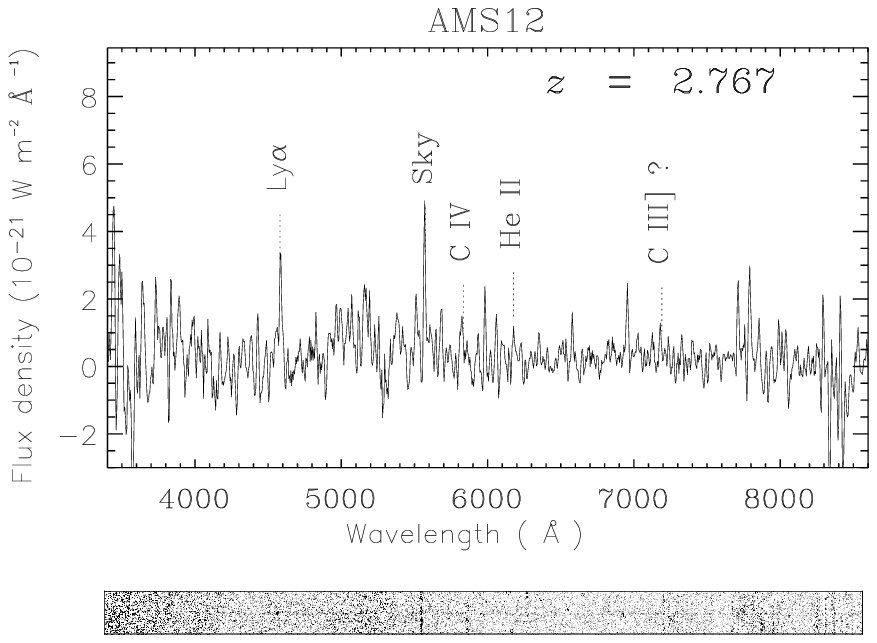,width=6.5cm,angle=0} 
\hspace{2.0cm}
\psfig{file= 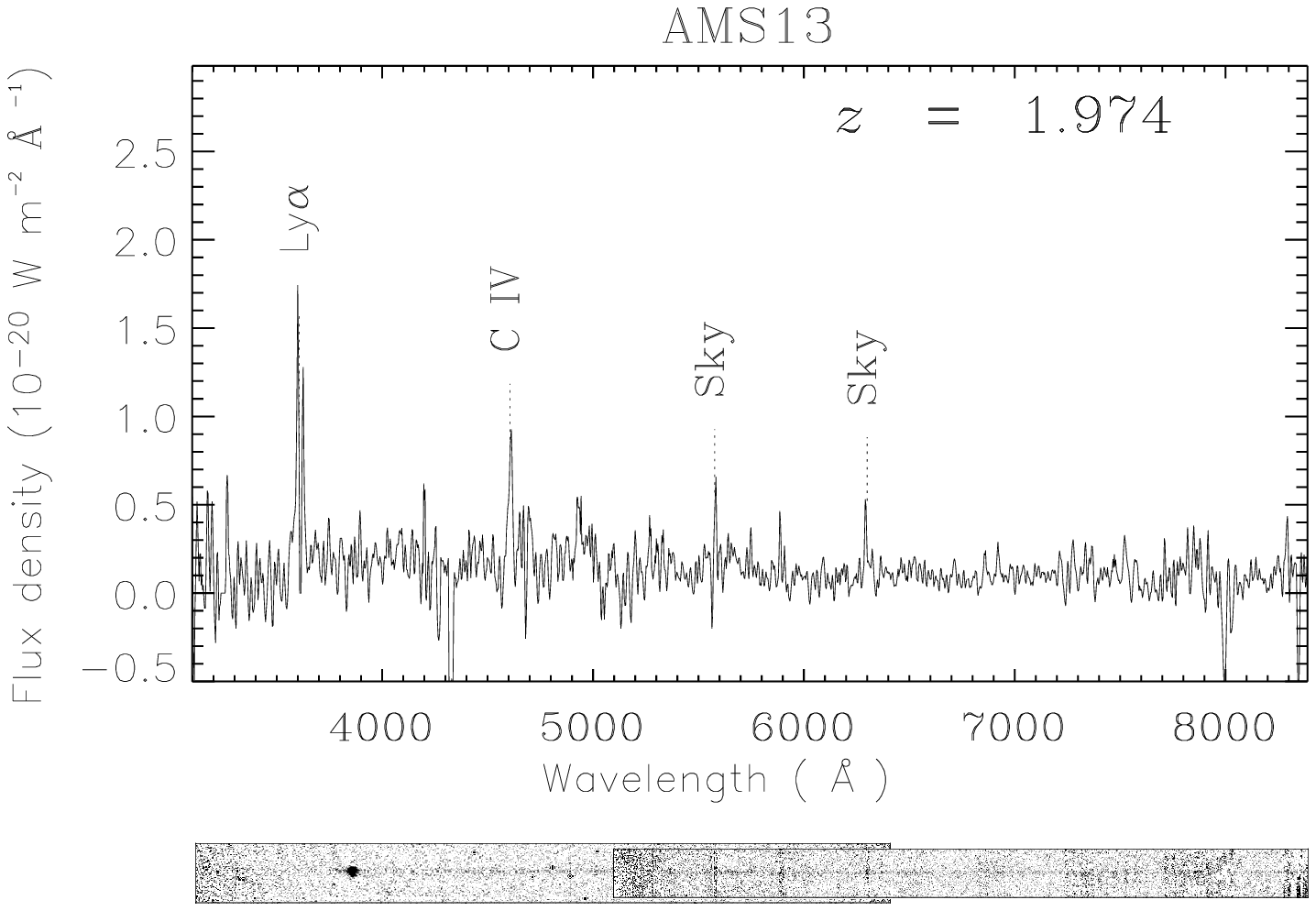,width=6.5cm,angle=0} 
}

\caption{Optical spectra of objects in the sample which showed emission lines. The emission lines from which
 the redshifts were obtained are labelled, while tentative
 identifications of lines which have not used for the redshifts are
 labelled with a question mark.  The corresponding two-dimensional
 spectra are shown underneath the one-dimensional ones as a visual
 help to find the emission lines. They have approximately the same
 scale in the horizontal direction as the one-dimensional spectra. The
 grey scale has been chosen to bring out low signal-to-noise lines and
 can lead to visual artefacts such as the strange change in continuum
 strength of AMS08. For the two objects observed in April 2005 (AMS04
 and AMS13), the pixel scale in the dispersion axis was not the same
 for the blue and red arms, and so the two-dimensional spectra are not
 combined, just overlayed on each other. Particularly strong residual
 sky lines have been labelled in the one-dimensional spectra and are
 clearly visible in the two-dimesional ones. In AMS04, the sky lines
 are close to the N~II] and C~II] lines, making these two lines hard
 to detect in the grey scale image. The reader is assured that these
 two lines are detected and are not due to the residual sky lines. }
\label{fig:spectra}
\end{center}
\end{figure*}

The spectra were all reduced using the {\sc iraf} package {\sc
twodspec} {\sc longslit}. The wavelength calibration was performed
using CuAr+CuNe lamps and flux callibration was obtained from
appropriate spectrophotometric stars. The two red frames for each
object were reduced separately and then combined to remove cosmic
rays. The 1D spectra were extracted separately for the blue and the
combined red frames, using the task {\sc apsum}. The resulting one-and
two-dimensional spectra are plotted in Figure~\ref{fig:spectra}.

\noindent{\bf AMS01} The 3.6 $\mu$m image of this object shows it to be
visible, although possibly confused with a galaxy just to the North (N)
of it (see Figure \ref {fig:overlays}). However, in the R-band image 
we see our target is distinguishable as an extremely faint galaxy. The
two images clearly show  the very red colour of AMS01. There was
another radio source of similar flux East (E) of AMS01 and so a
position angle (PA) of 103 degrees was chosen to go through both
sources. The spectrum of AMS01 was completely blank in a 30-minute exposure..

\noindent{\bf AMS02} In this case, our target is clearly visible in the
IRAC image, but is only just  detected  in the R-band. The PA  was chosen 
to go through another radio source, 45 arcsec SE of AMS02. The 
spectrum of our target was once again completely blank in a 30-minute exposure.

\noindent{\bf AMS03} Here we see another possible source of confusion
in the 3.6 $\mu$m image, but the R-band image shows our source to be
very faint but distinct from the brighter source SW of it. There is a
nearby radio source which is confused in the WSRT images (see
Section~\ref{sec:radio}). Here the PA was chosen to go through this
other radio source 18 arcsec E-SE of AMS03. The two-dimensional
spectrum shows a faint, double, spatially-extended line ($\sim$ 1.6
arcsec once deconvolved from the seeing), which we identify as
Ly~$\alpha$ (the redward line is not N V). The spectrum shows quite a
lot of structure (see Figure~\ref{fig:spectra}): there are two
Ly-$\alpha$ lines, one at 4481~\AA~ and the other at 4512~\AA. The
redward line has a full-width half maximum (FWHM) of $\sim$
500~km~s$^{-1}$, while the blueward line has a hint of absorption in
the spatial centre, and has a greater FWHM ($\sim$ 1400~km~s$^{-1}$)
but still narrow enough to qualify as a type-2 quasar. A possible
interpretation is that the type-2 quasar resides in a Ly-$\alpha$
halo, and there is a large difference in systemic velocity for the
quasar and the halo (or the halo is collapsing, leading to an
additional redshift).  Another possibility is associated absorption - 
this is present in small radio sources \citep {1997A&A...317..358V}.

\noindent{\bf AMS04} This object is very faint at 3.6 $\mu$m (18
$\mu$Jy) but reasonably bright in R-band ($m_{\rm R}=$23). The radio
emission around AMS04 is extended towards the NE. There is another
distant radio source, identified at both 3.6 $\mu$m and R-band further
to the east. This is possibly associated with another extremely faint
galaxy about 10 arcsec away, although the radio and optical positions
of the adjacent radio source and the faint galaxy do not match
perfectly. We placed the slit to go through both AMS04 and the other
radio source. AMS04 shows strong continuum, a spatially extended (and
bright) Ly~$\alpha$ as well as C~III], N~II] and C~II] lines but nothing was seen associated with the extended radio emission. The
Ly~$\alpha$ line has an angular size of $\sim 5.5$ arcsec. This object
has a redshift of 1.782, with all the redshifts from individual lines
agreeing well (see Table~\ref {tab:emli1}).  The fact that this object
is not that faint in R-band is due to the continuum and narrow-lines
at the red end of the spectrum.

\noindent{\bf AMS05} Here we see again a source which is relatively
bright at 3.6 $\mu$m (34 $\mu$Jy) compared to R-band ($m_{\rm
R}=$25.2). The IRAC image is quite crowded and there is the danger of
confusing sources, but we can see in the R-band that the radio
position falls on a faint galaxy.  This object showed an interesting
spectrum: C~IV is actually as strong as Ly~$\alpha$, and He~II is also
detected, but there is no continuum in the blue or red arm. All lines
are unresolved, and the weakness of the Ly~$\alpha$ may be due to
it falling in the blue end of the spectrum while having been observed
at a high airmass (see Table~1) and away from the parallactic angle.

\noindent{\bf AMS06} This target is just about visible in the 3.6
$\mu$m images, but it is not bright enough to make it to the IRAC
catalogue. It is undetected in the R-band. A PA of 120\degree was
chosen to go through the ``semi-major axis'' of the marginally-resolved radio source. The
spectrum was blank at the location of the target source. This could be
due to the entire galaxy being very dusty and therefore faint in the
optical and even the infrared (see the Section \ref{sec:discussion}) or it could
be a very high redshift source. Recently a redshift of 1.76 has been
obtained from silicate absorption and PAH emission with Spitzer IRS
\citep {2006ApJ...638..613W}, confirming this is a very dusty system at
high redshift.

\noindent{\bf AMS07} AMS07 is quite a red object (within our
sample) in that the 3.6 $\mu$m is relatively bright, whereas the
R-band is very faint. There is another radio source 56 arcsec SW of
the target, so the slit (with PA 19\degree) was made to go through
both radio sources. The spectrum was blank at the target position.

\addtocounter{figure}{-1}
\begin{figure*}
\vspace{5cm}
\hbox{
\hspace{1.0cm}
\psfig{file= 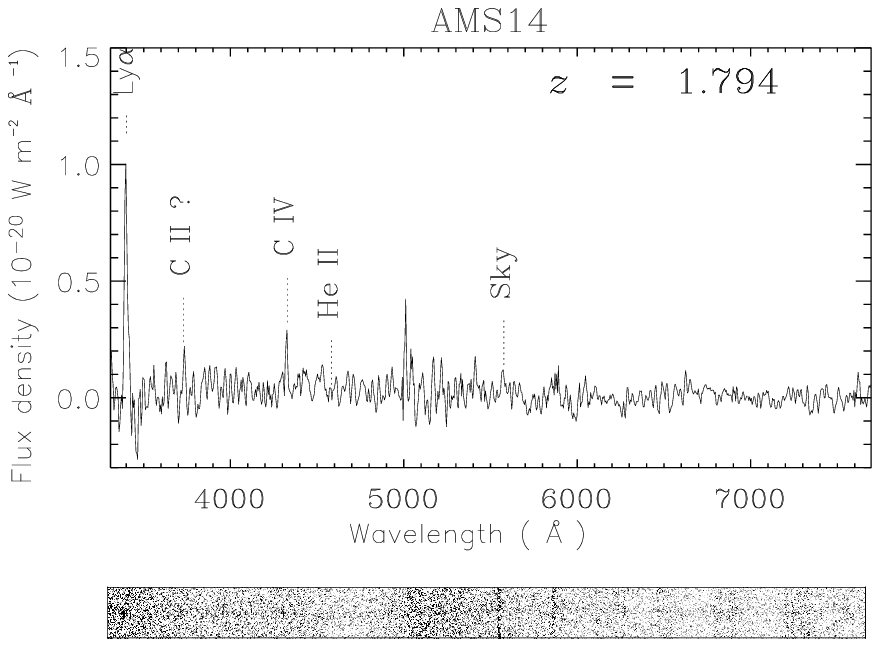,width=6.5cm,angle=0} 
\hspace{3.0cm}
\psfig{file= 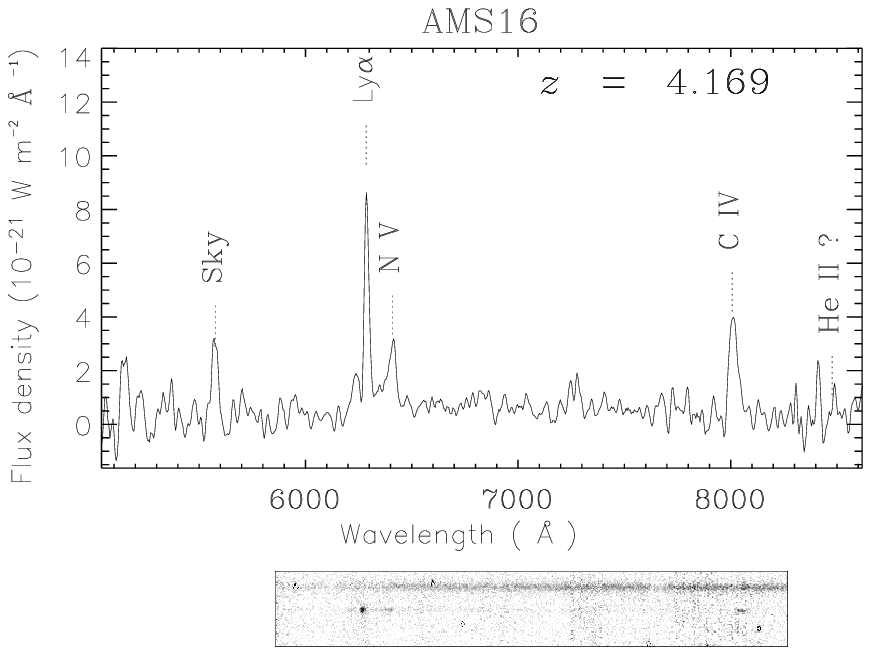,width=4.5cm,angle=0} 
}
\vspace{5.5cm}
\hbox{
\hspace{1.0cm}
\psfig{file= 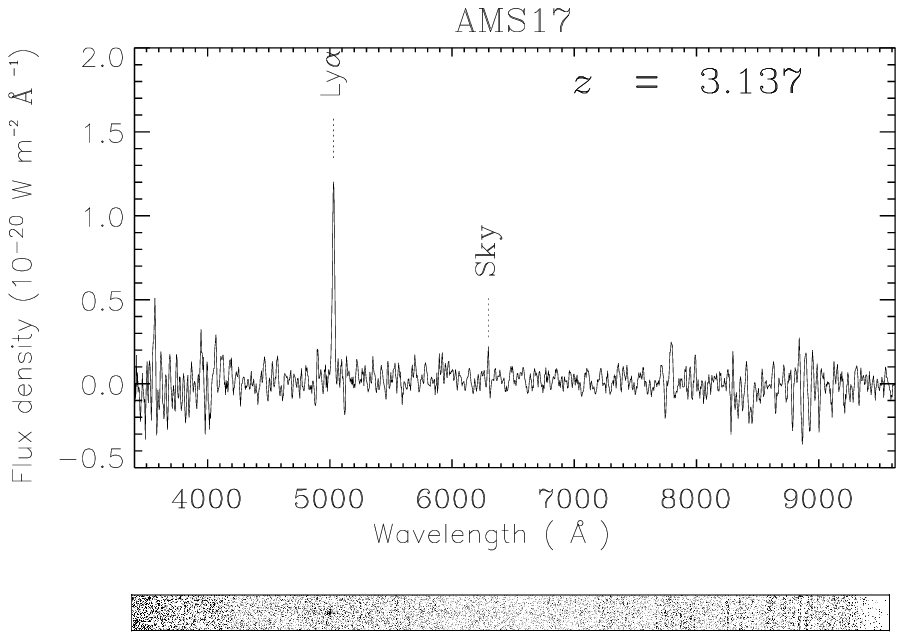,width=6.5cm,angle=0} 
\hspace{2.0cm}
\psfig{file= 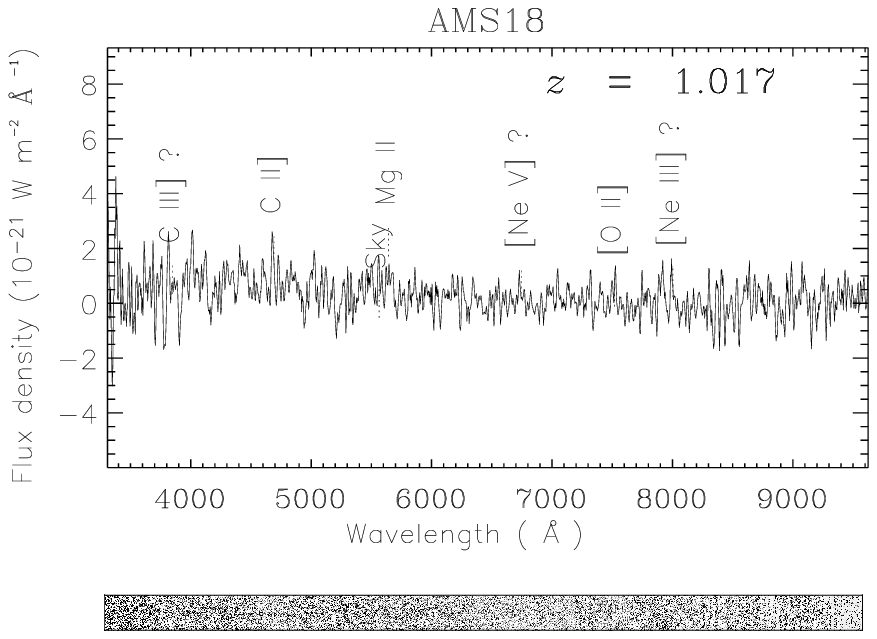,width=6.5cm,angle=0} 
}

\caption{Contined.}
\label{fig:spectraB}
\end{figure*}

\noindent{\bf AMS08} Here we have a relatively bright source for our
sample, both in the IRAC and R bands. A faint (slightly extended)
radio source, 18 arcsec SE of our source, was visible in the radio
images. The slit was made to go through both this source and our
target. AMS08 appears in the spectrum, showing Ly~$\alpha$ ($\sim$ 3.7
arcsec wide after deconvolving from the seeing), C~IV and He~II and blue continuum.

\noindent{\bf AMS09} This is a faint source at 3.6 $\mu$m and is
undetected in R-band.  The 3.6 $\mu$m image shows only one infrared ID
for our radio source, but the R-band image shows there might be a
serious confusion issue, as there are several faint sources near the
radio position. There were two other radio sources in the field, one close to the N (visible in Figure~\ref{fig:overlays})
and one E. We chose to place the slit through the eastern radio source
(41 arcsec away) and the PA angle (119 deg) is actually appropriate
to try and catch several of the possible optical IDs. There is barely a hint
of continuum from AMS09 in the red two-dimensional spectrum and no hint of the other
source.

\noindent{\bf AMS10} AMS10 is a relatively bright source in the
IRAC bands, and has $m_{\rm R}=$24.1. The 3.6 $\mu$m source is
actually confusing two R-band sources, but the radio position is
clearly associated to only one of the R-band galaxies.  There were no
other radio sources in the field, so we chose the PA to match the
parallactic angle. The 2D spectrum of AMS10 is blank.

\begin{table*}
\footnotesize
\begin{center}
\begin{tabular}{lrlcccrcc}
\hline\hline
\mc{1}{c}{Name} &\mc{1}{c}{$z$} &\mc{1}{l}{Line} &\mc{1}{c}{$\lambda_{\mathrm rest}$} &\mc{1}{c}{$\lambda_{\mathrm obs}$} &\mc{1}{c}{FWHM}&\mc{1}{c}{Flux} &\mc{1}{c}{log$_{10}L_{\rm line}$}   &\mc{1}{l}{Error in flux} \\  
\mc{1}{c}{} &\mc{1}{c}{ } &\mc{1}{c}{ }    &\mc{1}{c}{/\AA}  &\mc{1}{c}{(\AA)} &\mc{1}{c}{/km s$^{-1}$}  &\mc{1}{c}{/W m$^{-2}$} &\mc{1}{c}{/W} &\mc{1}{c}{/\% }\\
\hline\hline
AMS03 & 2.698 $\pm$ 0.009 &  Ly~$\alpha$            & 1216  & 4481    &  1400   & $1.22\times10^{-19}$    &35.86  &  20  \\ 
&  &  Ly~$\alpha$            & 1216  & 4512    &   550  & $3.69\times10^{-20}$    &35.34 &  18  \\ \hline
AMS04      & 1.782 $\pm$ 0.001  &  Ly~$\alpha$            & 1216  & 3382   &  1800  & $5.01\times10^{-19}$      &36.04  & 12  \\
&    &  C~III]             & 1909  & 5311  & 1450  & $1.02\times10^{-19}$    &35.35 & 21 \\ 
&    &  N~II]             & 2142  & 5963& 2550  & $1.16\times10^{-19}$    &35.40 & 20 \\ 
&    &  C~II]             & 2326  & 6467   & 1250   & $8.59\times10^{-20}$    & 35.27 & 28 \\ \hline
AMS05      & 2.017 $\pm$ 0.006  &  Ly~$\alpha$            & 1216  & 3658     &  1250   & $1.32\times10^{-19}$   &35.59  &38 \\  
&   &  C~IV     & 1549  & 4681    &  1000   & $1.16\times10^{-19}$   &35.53& 18 \\  
&    &  He~II             & 1640  & 4952   & 500   & $3.16\times10^{-20}$   &34.97 & 41 \\ \hline
AMS08      & 1.979 $\pm$ 0.004  &  Ly~$\alpha$            & 1216  & 3622   &  1900   & $5.05\times10^{-19}$   &36.15  & 15  \\  
&   &  C~IV    & 1549  & 4622   &  1400   & $4.49\times10^{-20}$    & 35.10 &  23  \\
&   &  He~II     & 1640  & 4877   &  900   & $2.30\times10^{-20}$    &34.81 &  35  \\ \hline
AMS12      &  2.767 $\pm$ 0.005 &  Ly~$\alpha$            & 1216  & 4586     &  650   & $4.41\times10^{-19}$  & 36.45  & 17 \\ 
             &    &  C~IV             & 1549  & 5824   & 1150   & $3.50 \times10^{-20}$  & 35.35& 25 \\ 
&    &  He~II             & 1640  & 6181   & 1100   & $2.22\times10^{-20}$  & 35.15 & 26 \\ \hline
AMS13      & 1.974 $\pm$ 0.003  &  Ly~$\alpha$            & 1216  & 3613     &  1650   & $9.88\times10^{-19}$  &36.44  & 12 \\  
&   &  C~IV     & 1549  & 4611    &  1300   & $1.67\times10^{-19}$  &35.67& 20  \\\hline 
AMS14      &  1.794 $\pm$ 0.001  & Ly~$\alpha$              & 1216  & 3398   & 1500  & $3.69\times10^{-19}$    & 35.91 & 18 \\ 
&    & C~IV             & 1549  & 4326   & 600  & $5.61\times10^{-20}$   & 34.92 & 21 \\ \hline
AMS16      &  4.169 $\pm$ 0.002 &  Ly~$\alpha$            & 1216  & 6288   &  1050   & $3.19\times10^{-19}$  & 36.73   & 12 \\ 
  &   &  N~V     & 1240  & 6406    &  2050   & $1.56\times10^{-19}$  & 36.42   &28 \\
             &    &  C~IV             & 1549  & 8009   & 2050   & $2.69\times10^{-19}$ & 36.66  & 16 \\ \hline
AMS17      & 3.137 $\pm$ 0.002 &  Ly~$\alpha$            & 1216  & 5031     & 1200   & $2.89\times10^{-19}$ & 36.40  & 16 \\ \hline
AMS18      &  1.017 $\pm$ 0.004  &  C~II]             & 2326  & 4680   & 1400   & $8.19\times10^{-20}$   & 34.65 & 25  \\ 
&    &  Mg~II             & 2798  & 5652  & 3500  & $1.14\times10^{-19}$   &34.79  & 44 \\ 
&    &  [O~II]             & 3727  & 7527   & 2200   & $1.12\times10^{-19}$   &34.79 & 27 \\ 
\hline\hline         
\end{tabular}
{\caption{Table of emission line properties of the 10 objects with
spectroscopic redshifts from our optical spectroscopy. The lines were
fitted with gaussians to find $\lambda_{\mathrm obs}$, FWHM, and the
flux. For objects with several lines, the error quoted is the square
root of the average variance. For objects with only one line, the
error is assumed to be dominated by the fitting of a gaussian to the
line. This in turn is dominated by the spectral resolution and the
seeing. In two cases (AMS14 and AMS18) an error in wavelength
calibration led to incorrect identification of lines and redshifts in
\citet{2005Natur.436..666M}. The error in flux (and luminosity) is the
sum in quadrature of the photometric error (typically 10\%), the error
from the slit (typically 5\%) and the uncertainty in the continuum.
The FWHM is deconvolved by subtracting in quadrature the seeing from
the observed size of the emission line. For lines with errors $>20$\%
the measurements of the FWHM are not reliable, and the presence of
broad lines (e.g. Mg~II in AMS18) is therefore not significant. }
{\label{tab:emli1} }} \normalsize
\end{center}              
\end{table*}

\noindent{\bf AMS11} This is another quite red source, relatively bright at 3.6 $\mu$m and faint in R-band. There are no other radio sources near but the radio
emission  seems to be slightly extended in the E direction, so
we chose the PA so the slit would go through it. AMS11 was not
detected by ISIS.

\noindent{\bf AMS12} Here we have a faint source in all bands. The PA
was chosen to go through a radio source 50 arcsec SE of our
target. The spectrum of AMS12 shows several emission lines:
Ly~$\alpha$ (spatially unresolved), C~IV and He~II (see
Table~\ref{tab:emli1}) and faint continuum.

\noindent{\bf AMS13} AMS13 is a point-like radio source, and is quite faint at
3.6 $\mu$m, so we matched the PA to the parallactic to minimise
dispersion of the blue light (even though the airmass, 1.16, was not
particularly high).  In the 2D spectrum, however, it is very bright,
showing spatially extended Ly~$\alpha$ and C~IV lines ($\sim$ 4.3 and 3.5 arcsec wide respectively) as well as
continuum. The continuum explains why it is not that faint in the
R-band. The optical spectra show AMS13 is at a redshift of 1.974, and
\citet {2005ApJ...628..604Y} find a redshift of 2.1 $\pm$ 0.1 from the silicate absorption feature in their Spitzer-IRS spectroscopy, in
agreement with our redshift.

\noindent{\bf AMS14} This source is extremely faint in the IRAC bands
(8 $\mu$Jy at 3.6 $\mu$m) making us expect this to be a very high
redshift source.  A PA of 105\degree was chosen to go through another faint
radio source 10 arcsec E of our target. The spectrum, however, shows
a very strong (and spatially resolved) Ly~$\alpha$ line ($\sim$ 1.6
arcsec wide) as well as C~IV at a redshift of $z = 1.794$. A mistake
in an early wavelength calibration led us to mis-identify the lines and the
redshift reported in \citet {2005Natur.436..666M} is incorrect.

\noindent{\bf AMS15} This is another very faint source at 3.6 $\mu$m
and undetected in R-band. This is a good candidate for a high-redshift
type-2 quasar. However we used a slightly wrong position (the one
published in \citet{2005Natur.436..666M} is off by $\sim$0.8 arcsec)
for follow-up spectroscopy. For this reason we do not consider this
object to have been observed.

\begin{figure}
\centerline{\psfig{file=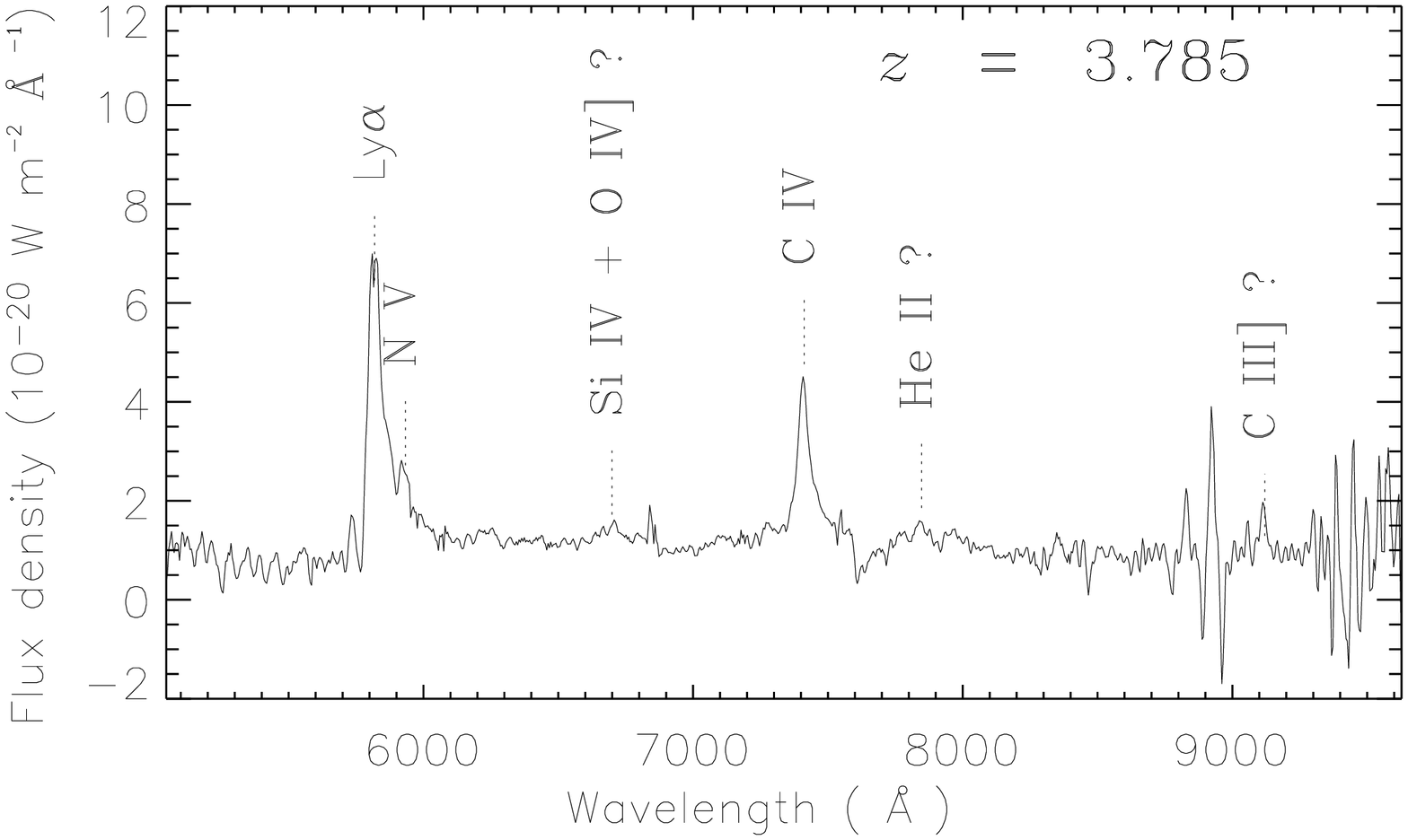,width=5.2cm,angle=90}}
\caption{The spectrum of the type-1 quasar observed at the same time as AMS16.}
\label{fig:type1}
\end{figure}

\noindent{\bf AMS16} AMS16 is very faint in the IRAC bands, although
it is not extremely faint in R-band ($m_{\rm R} = 23.07$). It has an
extended radio structure, whose central position matches the optical
and IR positions. However, the extended emission matches very well two
galaxies next to the source, and in the 24-$\mu$m image the three
sources are confused, and have a similar structure to the radio
emission. There is the possibility that the two foreground galaxies
might be lensing AMS16 (see Figure \ref{fig:overlays}). Two factors
therefore influenced the choice of PA in this case. First, the radio
emission around AMS16 is extended in the SE to NW direction
(diagonally, possibly due to the two galaxies). In addition, there is
a radio loud ($S_{1.4~\rm GHz}=$21.6 mJy) type-1 quasar about an
arcmin NW. By good fortune, the same angle is required to get the slit
through the `semi-major axis' of the extended radio emission, the
brighter foreground galaxy and the type-1 quasar. We found Ly~$\alpha$
($\sim$ 2.7 arcsec), N~V ($\sim$ 1 arcsec) and C~IV (spatially
unresolved) in AMS16, indicating a redshift of $z = 4.169$. The
emission lines of AMS16 are all strong, which explains why it is not
that faint in the R-band. To the NW of AMS16, 58 arcsec away, we found
a radio-loud type-1 quasar, at redshift $z = 3.785$ (see Figure
\ref{fig:type1}).

\noindent{\bf AMS17} This is another faint source in the IRAC bands,
suggesting a high redshift, although the R-band is not particularly
faint for this sample, and the images show the optical/IR positions 
match perfectly the centre of the radio emission, with no possible
sources of confusion nearby. The slit PA was chosen to go through
another object, about 59 arcsec SE of our target. AMS17 is indeed at
high-redshift, as it shows one narrow line in the blue spectrum (which we
identify as Ly~$\alpha$ at $z = 3.137$) and faint continuum redwards
of the line. The Ly~$\alpha$ line is extended $\sim$ 1.6 arcsec wide in the spatial
direction, once the seeing is deconvolved.

\noindent{\bf AMS18} Here is a source which is undetected in the IRAC
catalogues, which suggests a high redshift. The PA was chosen to go
through a bright radio source 57 arcsec South-East. The 2D spectrum
shows AMS18 has several faint emission lines, which we identify as
C~II], Mg~II and [O~II] at $z = 1.017$, so it is at a lower redshift
than first expected from the crude photometric redshift. Note the
weakness of the [O~II] line, which suggests a faint AGN rather than
something similar to a radio galaxy. The same wavelength calibration
error discussed for AMS14 led to the redshift of AMS18 being incorrect
in \citet {2005Natur.436..666M}. The emission lines are those of an
AGN, so this is probably a less-luminous quasar possibly hosted by a
small galaxy, which is how it made it through our selection
criteria. It is however, difficult to judge whether the lines are
spatially extended, due to the low SNR.

\noindent{\bf AMS19} The PA was chosen to be identical to the
parallactic angle, as it is very faint in the near-infrared and the radio
emission of AMS19 is point-like. This could mean a high redshift, so
we tried to maximise the blue light received. Indeed, in the
two-dimensional AMS19 does not even show faint continuum. However it
was also observed by \citet {2006ApJ...638..613W} with Spitzer IRS and is
found to be at $z = 2.25$ from it's PAH and Silicate features.

\noindent{\bf AMS20} This is a relatively bright source for our
sample, and so is not expected to lie at a particularly high
redshift. The slit was placed at an angle so it would go through two
fainter radio sources NW of AMS20 (19 and 55 arcsec away). The
spectrum of AMS20 shows faint red continuum but no lines to allow us
to obtain a redshift. Further study is required to determine whether
this is a very bright high-redshift type-2, or more similar to AMS18,
a faint, lower-redshift object. The lack of lines suggests $z \sim 1.3-1.6$.

\noindent{\bf AMS21} The last source in our sample is relatively faint in at 3.6 $\mu$m and R-band. In this case, the PA was chosen to go through a
fainter radio source about 25 arcsec E of the target. However, neither
AMS21 or the other radio source have been detected in the optical spectrum.

\section{Type-2 Spectra}\label{sec:spectra}

Approximately half of the objects (10 out of 21) showed narrow
emission lines and sometimes continuum (see Figure
\ref{fig:spectra}). Ly~$\alpha$ with full-width half maximum (FWHM) $
\ltsimeq 2000~\rm~km~s^{-1}$ was found in nine objects (with a
redshift range $1.78 \leq z \leq 4.17$) and high excitation lines such
as CIV (1549~\AA) or HeII (1640~\AA) were also found in several
objects (see Table \ref{tab:emli1}).  In two cases (AMS03 and AMS17)
we only found one emission line, which was taken to be Ly~$\alpha$ by
virtue of its extreme equivalent width, blue-absorbed line profile and
lack of other lines supportive of alternative identifications. The
Ly~$\alpha$ line of AMS03 is particularly spatially-extended.  One
object, AMS18, proved to be at $z = 1.02$, lower than expected from
our selection criteria. AMS18 is still an AGN, so there is still no
evidence of pure-starburst contamination. Only one object showed (very
faint) continuum and no lines: AMS20. As mentioned in
Section~\ref{sec:obs}, it will require more study to determine whether
this object is similar to AMS18, is at $1.3 \ltsimeq z \ltsimeq 1.7$,
or is at a higher-redshift object. If this object were a contaminant,
like a $z \sim 0.5$ ULIRG, it would have shown stronger continuum and
the [O~II] 3727~\AA~line. It is important to insist on how faint our
sources are: all objects, even those with emission lines in the R-band
or showing some continuum, have $\rm m_{\rm R} \geq 22.5$ (Vega), and
most have $\rm m_{\rm R} \geq 23$. The rest of the objects show blank
spectra, confirming that there are no low-redshift contaminants in our
sample. A 2.6$L^{*}$ elliptical at $z = 1.4$ would have a continuum
flux density of $1\times 10^{-21}$ W m$^{-2}$ \AA$^{-1}$, at the limit
of our sensitivity. AMS20, with faint red continuum, is therefore
consistent with having an elliptical host galaxy, and being somewhere
in the spectroscopic desert.

The nature of the blank-spectrum objects is discussed further in
Section~\ref{sec:discussion}, but we see that their blank optical spectra are
consistent with host ellipticals at $z \gtsimeq 1.4$. If 24-$\mu$m
flux density is a good  tracer of narrow emission line strength,
then we expect all our sources to have detectable Ly~$\alpha$
lines if they lie at $z \geq 1.7$. In addition, the C~IV (1549\AA) line is detected in six out
of the nine objects with Ly~$\alpha$, and in the composite spectrum
(see Section \ref{sec:composite}) it has 36\% of the flux of
Ly~$\alpha$. Hence, some objects in the spectroscopic desert should
also have detectable C~IV lines.  The two blank objects brightest at
24-$\mu$m have redshifts from Spitzer-IRS in the same range as the
rest of the optical spectra, where we should have seen
Ly~$\alpha$. This suggests that at least some of the blank objects are
at redshifts high enough to have narrow emission lines in the optical band, and that their spectra are
blank for another reason. This reason could be that they have weaker
emission lines, but they would have to be significantly weaker than
those of the narrow-line objects, despite having similar 24-$\mu$m
fluxes. Another alternative, is that the narrow emission line region
is itself obscured, as has proven  to be the case from the Spitzer spectroscopy of  AMS06 and AMS19.

\section{Crude Photometric Redshifts}\label{sec:photoz}

As discussed in Section~\ref{sec:selcrit}, for $A_{\rm V} \gtsimeq 10$
the 3.6-$\mu$m flux density should be dominated by light from the host
galaxy's old stellar population. We estimated crude ``photometric''
redshifts for the entire sample by assuming the host galaxies to be
the progenitors of present day massive elliptical galaxies. These
galaxies were taken from the models of \citet{2003MNRAS.344.1000B},
formed at $z = 10$ in a single starburst, with a Salpeter IMF \citep
{1955ApJ...121..161S} and a stellar mass of $3.8 \times 10^{11}
M_{\odot}$. Such host galaxies would become 2.6 $L^{*}$ galaxies in
the local $K$-band luminosity function
\citep{2001MNRAS.326..255C}. The justification for such an assumption
is that radio-loud galaxies show a $K-z$ Hubble diagram consistent
with such massive host galaxies \citep[$\sim 3L^{*}$][]
{2001MNRAS.326.1585J,2003MNRAS.339..173W} and evidence points to the
host galaxies of $z \sim 2$ radio-quiet quasars being $\sim 2L^{*}$
ellipticals \citep {2001MNRAS.326.1533K}.

The scatter in true host luminosity means that individually, the
photometric redshifts are not necessarily close to the spectroscopic
redshift, but they agree on average for 10 of the 12 objects. There
are two cases (AMS14 and AMS18) in which the photometric redshift is
clearly overestimated. There are two possible explanations for this: a
host galaxy which is smaller than the progenitors of $\sim2-3L^{*}$
ellipticals, and so our mass is an overestimate, or a star-forming
host-galaxy, in which case our elliptical galaxy template is not
appropriate. For the objects that have spectroscopy, the mean and
median values of $z_{\rm phot}$ are 2.75 and 2.65 respectively while
the mean and median spectroscopic redshifts are 2.33 and 2.00 (2.29
and 2.00 if the Spitzer-IRS are included): this represents reasonable
agreement for such a simple photometric redshift estimation and gives
us some confidence that we can estimate how many objects that remained
unobserved or had blank optical spectra are at $z \geq 2$. To bring
the median $z_{\rm phot}$ to the value of the median $z_{\rm spec}$
one would have to change the mass of the elliptical to $2.9 \times
10^{11} M_{\odot}$, which would correspond to the progenitor of a
$\sim 2L^{*}$ elliptical galaxy instead of $\sim 2.6L^{*}$. This would
also bring the mean $z_{\rm phot}$ close to the mean $z_{\rm spec}$.
The fact that, for the objects with emission lines, the
``photometric'' redshifts are roughly in agreement with the
spectroscopic ones, suggests the host galaxies are the progenitors of
massive ellipticals ($\gtsimeq 2L^{*}$).

A possible source of error for our photometric redshifts would be the
contribution of quasar light to $S_{3.6 \mu \rm m}$, together with a
large host galaxy. This could plausibly make us miss some $z\sim2$
type-2 quasars, which would only make the results of \citet
{2005Natur.436..666M} more striking. Another source of error is if the
host galaxy is not the progenitor of an elliptical, but a starburst
(this possibility is explored in Section~\ref{sec:discussion}). The
old stellar population would then be less luminous, and $z_{phot}$
would be overestimated. The lack of lower-redshift contaminants in our
spectra gives us confidence that this is not a serious problem.

\section{Composite Type-2 Spectrum}\label{sec:composite}

Following \citet {2001MNRAS.322..523R}, we proceeded to create a
composite type-2 quasar spectrum by shifting the individual spectra  to
the rest-frame with the {\sc iraf} task {\sc newredshift}. The
individual spectra were all scaled by their Ly~$\alpha$ line flux and
then averaged bin-by-bin with a sigma-clipping rejection algorithm
with the clip set to $\pm 1.5\sigma$. The object at $z = 1.02$ was
therefore not included as there was no way of scaling it
consistently. The composite of 9 objects is shown in Figure~\ref{composite} and
shows strong C~IV (1549~\AA) and He~II (1640~\AA) as well as
Ly~$\alpha$ (1216~\AA) lines. C~II] (2326~\AA) and Mg~II (2798~\AA)
are also detected.

\begin{figure}
\centerline{\psfig{file=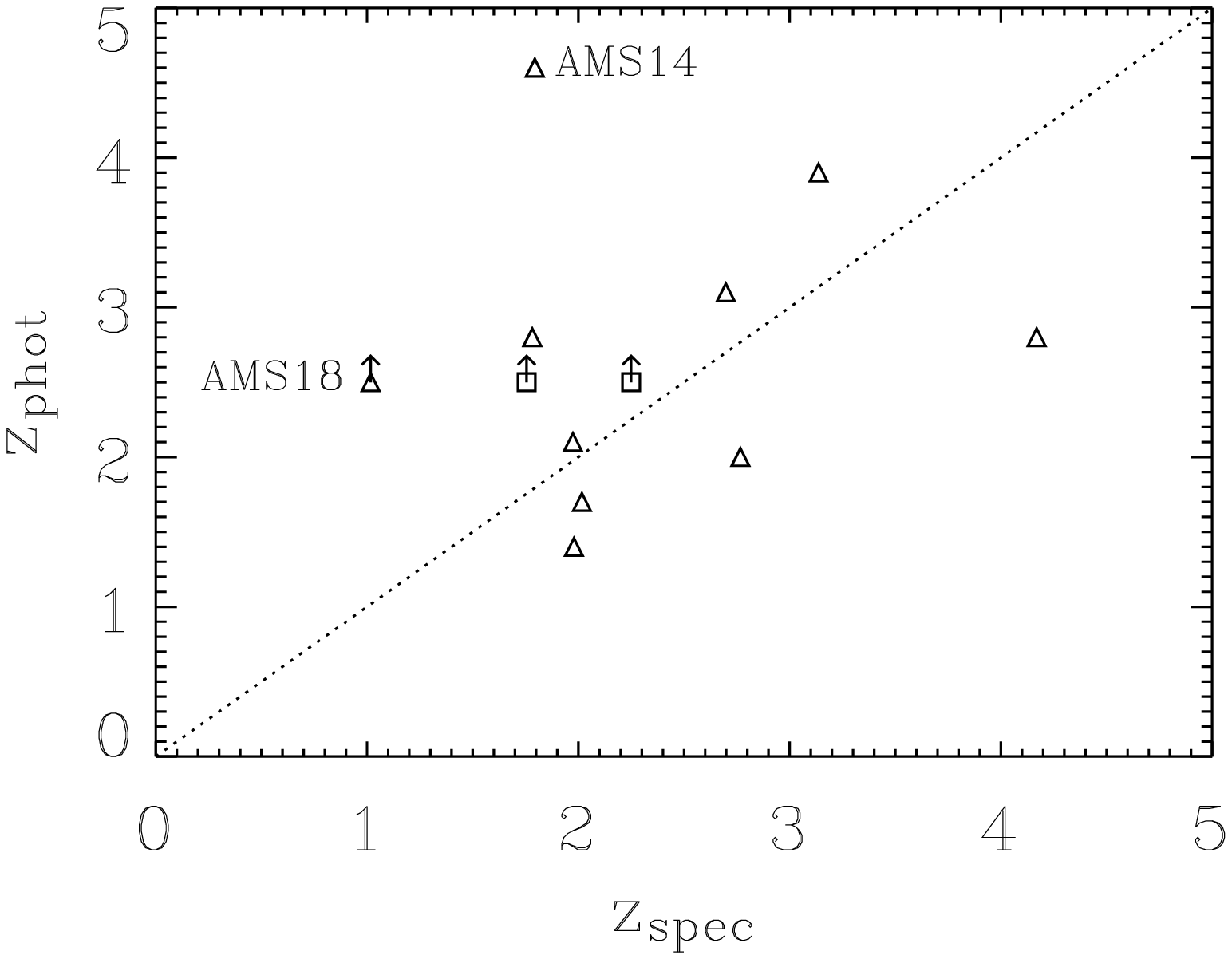,width=8.5cm,angle=0}}
\caption{Comparison between crude photometric and spectroscopic
redshift for the objects with redshifts. Triangles are objects with
narrow-lines in our spectroscopy, squares are objects with redshifts
from Spitzer-IRS only. We see a large scatter, and
individually the photometric redshifts are of no use. On the whole, the agreement is reasonable except for two objects (AMS14 and AMS18).  This means that the
objects with spectroscopic redshifts in our sample are consistent with
having massive elliptical galaxies as hosts.}
\label{photozspectz}
\end{figure}

Table~\ref {tab:ratios} shows the line fluxes relative to
Ly~$\alpha$. We can see C~IV , He~II and C~II] are clearly detected,
with relative values of 0.36, 0.09 and 0.09. In addition we see that
the Mg~II line is present, with a relative line flux of 0.05, while
the N~V (1240~\AA) and C~III] (1909~\AA) are not detected. The Mg~II
line must be detected at low SNR in several individual spectra to make
it detected in the composite. C~III] is only detected in one of the
objects making the composite and this could be due to the fact that at
$z \gtsimeq 2$ it is shifted to $\lambda \geq$~5700~\AA, where it
falls near the dichroic cross-over region used in the July 2004
observations. AMS04 has C~III] and is at $z = 1.78$, but it was
observed using a 5300~\AA~ dichroic. The N~V line is only present in
AMS16.  This composite spectrum is similar to that found for
high-redshift radio galaxies \citep[e.g McCarthy, 1993, and more
recently] []{2001MNRAS.322..523R,2001MNRAS.326.1563J}, except that the
relative strength of C~IV is a factor of $\sim 3$ higher here than in
\citet {1993ARA&A..31..639M}.  The C II] line is stronger by a factor
of 3 compared to \citet {1993ARA&A..31..639M} and the \citet
{2001MNRAS.326.1563J} value for radio galaxies with a projected linear
size, D, $>$ 70 kpc, but consistent with the \citet
{2001MNRAS.326.1563J} value for radio galaxies with D $<$ 70 kpc.

Comparing these ratios with
\citet{1993ARA&A..31..639M,2001MNRAS.322..523R,2001MNRAS.326.1563J} we
find that they are broadly similar to those of powerful radio galaxies
except C~IV is significantly stronger, even stronger than for the
\citet {1993ARA&A..31..639M} Seyfert-2 galaxies. The C~II] is similar
to the \citet {2001MNRAS.326.1563J} value for radio galaxies with D
$<$ 70 kpc. This might be an indication of a more complicated
line-emission mechanism than simple photoionisation from a central
engine, some of the ionisation might be happening as a result of jet
activity \citep [see for example][]{2000MNRAS.311...23B,
2001MNRAS.326.1563J}. 

Comparing our composite type-2 to four individual high-redshift
type-2s from the literature, we find that the \citet
{2002ApJ...568...71S} type-2 ($z = 3.288$) has Ly~$\alpha$, C~IV and
He~II ratios similar to our composite (100:19:9), with slightly
stronger C~IV(relative to the other two lines). The \citet
{2005MNRAS.356.1571M} object ($z = 3.660$) has a Ly-$\alpha$-to-C~IV
ratio (100:14) very similar to that of our composite and the \citet
{2005MNRAS.358L..11J} type-2 ($z = 1.65$) has a C~IV to He~II ratio
(1:1) also in very good agreement with our composite. The \citet
{2002ApJ...571..218N} type-2 at $z = 3.700$ has substantially weaker
Ly~$\alpha$ relative to C~IV and He~II (100:60:17); it is more similar
to AMS16, for example, than to the composite. We therefore find very
good agreement with the individual high-redshift type-2s from the
literature and our sample.

\begin{figure}
\hspace{-.5cm}
\centerline{\psfig{file=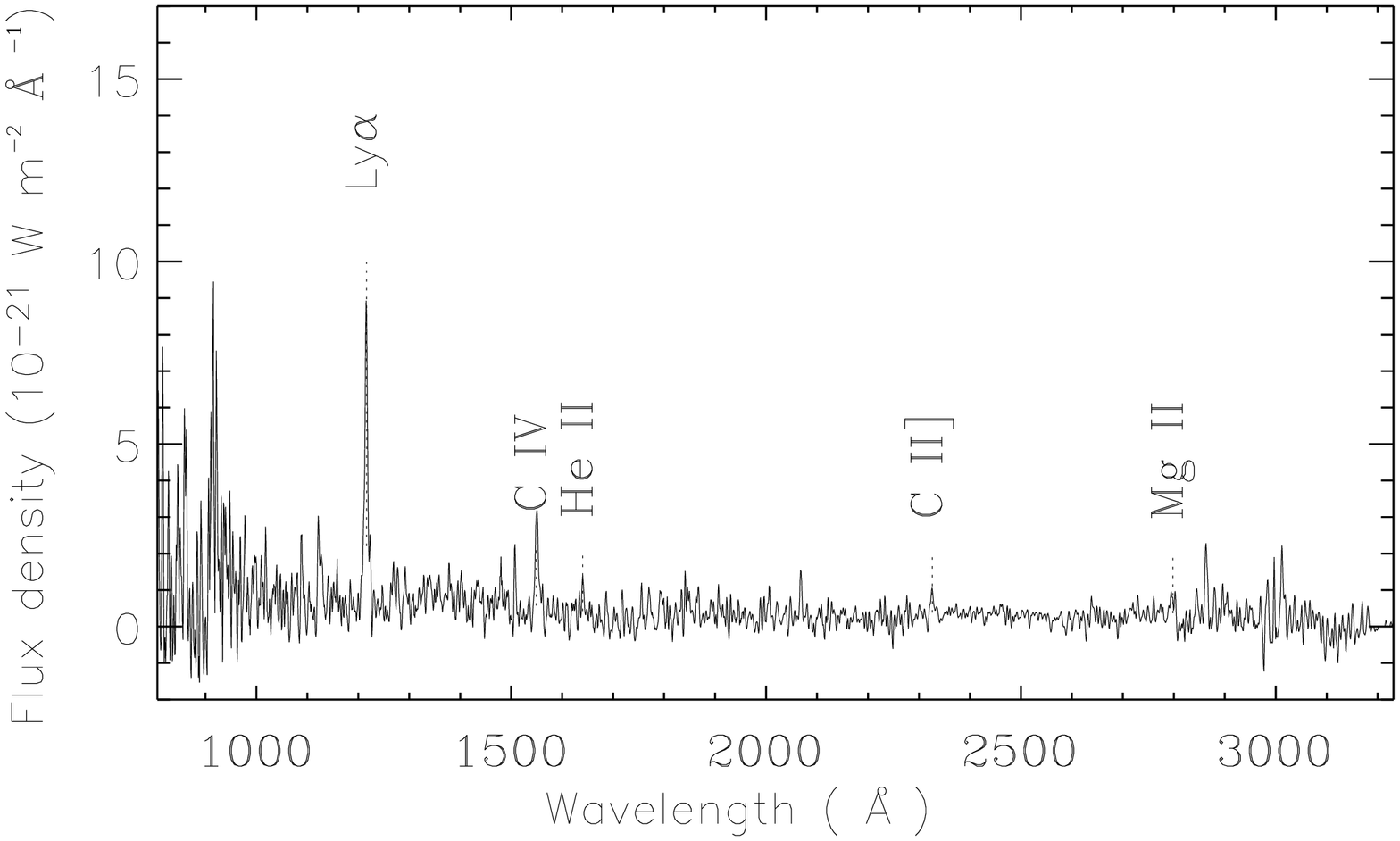,width=6cm,angle=90}}
\caption{Composite type-2 quasar spectrum. This has been made as described in
Section \ref{sec:composite}, and it shows Ly~$\alpha$, particularly
strong C~IV, He~II, C II] and Mg~II. C~III] is present in one
individual object, and might be present in more but at $z \sim 2$ it
falls around the dichroic cross-over region, so the noise might be too
high to detect it. All the lines detected in the composite and in
individual objects, except for Mg~II, are present in the observations
of IRAS FSC 10214+4724 \citep {1998MNRAS.298..321S}.  }
\label{composite}
\end{figure}

Somewhat surprisingly another characteristic in
common with radio galaxies is that all objects which show Ly~$\alpha$
have at least some extended emission (in particular AMS03, AMS04,
AMS08, AMS13 and AMS16 have very extended Ly~$\alpha$
emission). AMS12 seems to have two Ly~$\alpha$ components: a bright,
spatially-unresolved ``peak'' sitting on top of a spatially-extended
fainter ``plateau''. Also, AMS13 and AMS16 have other lines which are
spatially extended (C~IV and N~V respectively). In AMS04, the
Ly~$\alpha$ is $\sim$ 5.5 arcseconds in size (once deconvolved from
the seeing), which is larger than the VLA beam size. For the other
cases, we cannot currently tell whether the emission lines extended
over regions larger than the radio emission. These characteristics all
remind us of classic radio loud sources despite their relatively low radio flux densities, so in Section~\ref{sec:radio} we
proceed to compare our type-2 quasars with  radio loud
objects.

\section{Comparison to Radio-Loud Objects}\label{sec:radio}

The mean radio flux density  of the sample at 1.4 GHz is 780 $\mu$Jy, which at the
mean spectroscopic redshift $z = 2.33$ corresponds to a radio
luminosity of $L_{1.4 \rm ~GHz} = 2 \times 10^{24} \rm ~W ~Hz^{-1}
~sr^{-1}$ (assuming a spectral index $\alpha = 0.8$, where $L \propto \nu^{-\alpha}$). This is well
below the break in the RLF \citep{1998MNRAS.300..625W} and in the
regime where typical radio-selected objects have Fanaroff-Riley Class
I (FRI) radio structures \citep{1974MNRAS.167P..31F}. There is also little direct
evidence of associated optical quasar activity
\citep{2004MNRAS.349..503G} and weak or absent emission lines
suggesting that, whether or not obscuring tori exist in this class of
objects \citep{1999A&A...349...77C}, any associated quasar is
accreting at a very low rate \citep {1991Natur.349..138R}. However,
considering that FRIs have occasionally been found to accrete at high
rates \citep{2001ApJ...562L...5B}, we proceed to investigate the
possibility of our objects having similar properties to radio-loud
objects.

\begin{table}
\footnotesize
\begin{center}
\begin{tabular}{lcccc}
\hline\hline
\mc{1}{c}{Line} &\mc{1}{c}{This}  &\mc{1}{c}{} &\mc{1}{c}{McCarthy (1993)}    &\mc{1}{c}{} \\
\mc{1}{c}{} & \mc{1}{c}{Sample} & \mc{1}{c}{RG} &  \mc{1}{c}{SyII} & \mc{1}{c}{QSO}  \\
\hline
Ly~$\alpha$    & 100  & 100 & 100   & 100   \\
C~IV       &36 & 12& 22  &  8  \\
He~II       &9 & 10& 4 & 5   \\
C~II]        &9 & 3& - & 4   \\
Mg~II        &5 & 3& 4& 23\\
\hline \hline
\end{tabular}
\end{center}
\caption{Comparison between the line ratios of this sample and those
found by \citet {1993ARA&A..31..639M} for radio galaxies (RG)
Seyfert-2s (SyII) and QSOs. The line fluxes are quoted as a percentage
of the Ly~$\alpha$ flux. The He~II, C~II] and Mg~II are roughly
consistent with the radio galaxy values, but the C~IV is significantly
brighter, even brighter than found in Seyfert-2s.}
\label{tab:ratios}
\end{table}

One typical characteristic of radio-loud objects is the presence of
extended jets on scales larger than the host galaxy.  Figure
\ref{fig:overlays} shows that most of the type-2 quasars shown in this
paper are point-like in the VLA observations ($\sim 5$-arcsec
resolution). In the few cases where the radio flux is extended, it
generally coincides with another galaxy in the R band or 3.6 $\mu$m
images. However, if there is diffuse flux on large scales, the relatively long
baselines of the VLA B-array could miss it out, so we would not see
large jets.

The Westerbork Synthesis Radio Telescope (WSRT) has also observed at
1.4 GHz the central region of the FLS, with an angular resolution of
$\sim14$ arcsec \citep{2004A&A...424..371M}. The shortest baselines of
the WSRT are smaller than those of the VLA in B-array and this enables
it to pick up flux on more extended scales, so any extended sources
will have a significantly higher flux density in the WSRT catalogue
than the VLA one. Thirteen of our targets are within the area observed
by the WSRT: the fluxes are shown in Table~\ref{radiotable}. The rest
of our sample was outside the area covered by the survey of \citet
{2004A&A...424..371M}. Most of the objects in Table~\ref{radiotable}
have consistent fluxes and therefore are unlikely to have a
significant fraction of their radio flux in extended jets. AMS04,
however has a flux density as measured by WSRT which is 65\% higher
than that measured by the VLA, suggesting the presence of jets.  The
3.6 $\mu$m-radio overlay (Figure~\ref{fig:overlays}) shows there is
another radio source within the WSRT beam, and there is a hint of a
faint galaxy which does not match the radio position perfectly. The
companion radio source has a flux density of 331 $\mu$Jy, so the WSRT
flux density is clearly the sum of these two individual sources. If
this is indeed due to another galaxy, then the WSRT is confusing both
sources. However, we cannot be certain that the adjacent radio source
is independent and due to another galaxy, so there is a chance that
AMS04 has an extended jet with a flux density comparable to that of
the central source.

\begin{table}
\footnotesize
\begin{center}
\begin{tabular}{lcccc}
\hline\hline
\mc{1}{c}{name} & \mc{1}{c}{$S_{\rm VLA ~1.4GHz}$} &  \mc{1}{c}{$\sigma_{\rm VLA}$} & \mc{1}{c}{$S_{\rm WSRT ~1.4GHz}$} 
& $\sigma_{\rm WSRT}$ \\
\mc{1}{c}{} &\mc{1}{c}{/$\mu$Jy}  &\mc{1}{c}{/$\mu$Jy} &\mc{1}{c}{/$\mu$Jy}    &\mc{1}{c}{/$\mu$Jy} \\
\hline
AMS01 &490  & 31 & 499 & 19\\
AMS02 &1184 & 55 &1209 &17 \\
AMS03 &1986 & 87 & 2370& 16\\
AMS04 &536  & 72 & 887 & 13\\
AMS05 &1038 & 49 &1017 & 12\\
AMS06 &444  & 31 &380 & 14\\
AMS07 &354  & 27 &289 & 10\\
AMS08 &655  & 36 &679 & 12\\
AMS10 &1645 & 73 & 1911 &11 \\
AMS11 &356  & 29 &253 & 9\\
AMS13 &1888 & 83 &2076 & 14\\
AMS18 &390  & 29 &351 & 23\\
AMS19 &822  &  41 &783 & 15\\
\hline \hline
\end{tabular}
\end{center}
\caption{Comparison between the radio flux densities of the VLA in B
array (beam size $\sim$ 5 arcsec) and the WSRT fluxes (beam size
$\sim$ 14 arcsec). We can see that the two flux densities agree within
3-$\sigma$ in all but two cases. Only AMS03 and AMS04 are not
consistent, but examination of the VLA B-array images shows this is
due to confusion. In the case of AMS03 it is clearly due to another
radio source, while in the case of AMSO4, it is not yet clear whether
the confused flux density  is from another source or from a diffuse
radio component.}
\label{radiotable}
\end{table}

The type-2s with the highest flux densities (AMS03, AMS10 and AMS13)
all have more flux as observed by WSRT. Once again, however, there is
a radio source with a clear optical (and infrared) ID next to one of
the objects (AMS03), with a flux density of 203 $\mu$Jy, so the WSRT
flux is consistent with the sum of the two confused sources, within
their errors.  The flux of AMS13 is consistent in both telescopes to
within 2$\sigma$ and AMS10 is consistent within 3$\sigma$. Although
3$\sigma$ might appear poor agreement, the reader is reminded of the
highly spatially correlated nature of the noise in interferometer
images, meaning the usual interpretation that only 0.5\% of
measurements should differ by 3$\sigma$ is not appropriate here.  The
central frequencies of both observations are not exactly the same:
however, this small difference in frequency can only account for$\sim$
2\% of the flux difference.

\begin{figure}
\centerline{\psfig{file=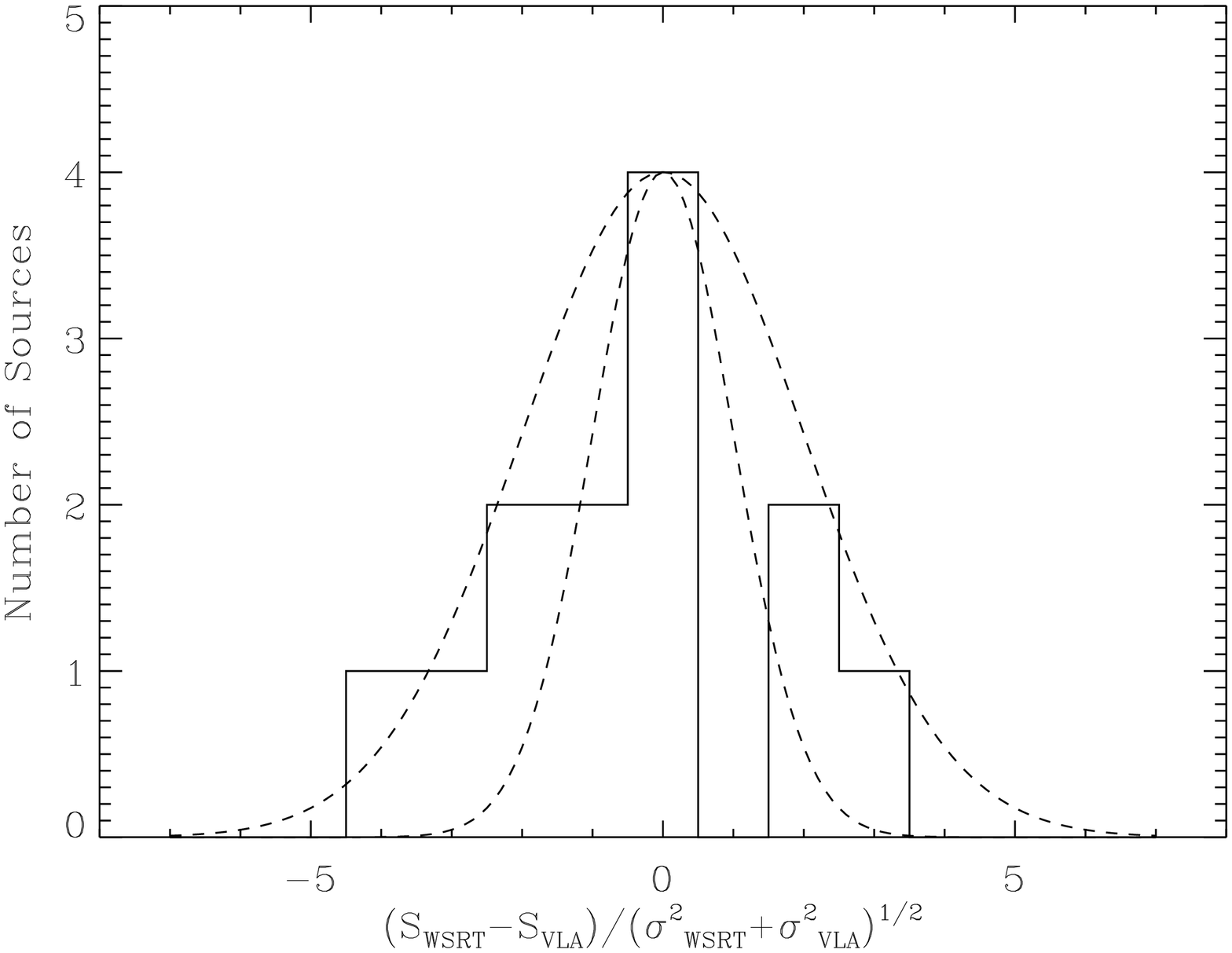,width=7cm,angle=90}}
\caption{Normalised flux density difference distribution. This is
computed as described in the test, and should match the gaussian
distribution with $\sigma = 1$ (inner gaussian). This is clearly not
the case, but it matches reasonably well somewhere between the $\sigma = 1$ and $\sigma = 2$ gaussians. }
\label{wester}
\end{figure}

To analyse the differences in flux density between both
interferometers, we computed the normalised flux density difference:
the measured flux density difference divided by the sum in quadrature
of the 1$\sigma$ errors of Table~\ref{radiotable}, $(S_{\rm WSRT} -
S_{\rm VLA})/(\sigma^{2}_{\rm WSRT} + \sigma^{2}_{\rm VLA})^{1/2}$,
and plotted it in Figure~\ref{wester}. For AMS03 and AMS04, we remove
from the WSRT flux density the VLA flux density of the adjacent radio
sources. Overplotted on the distribution are two gaussian
distributions, one with $\sigma = 1$, another with $\sigma = 2$. If
there is no systematic difference between the two measurements, the
plot should look like the gaussian distribution with $\sigma = 1$,
centred around 0. We see that this is clearly not the case, so the
flux densities from both surveys are not consistent with random noise
of variance $\sigma^{2} = \sigma^{2}_{\rm WSRT} +
\sigma^{2}_{\rm VLA}$. However, the distribution roughly agrees
somewhere between the $\sigma = 1$ and $\sigma = 2$ gaussians (with
$\sigma^{2} = 2--4\times(\sigma^{2}_{\rm WSRT} + \sigma^{2}_{\rm
VLA})$). This suggests that the measurements from the WSRT and VLA are
consistent provided the effective errors are taken to be larger (but still
less than twice) the RMS value of the noise. We therefore conclude
that there is no firm evidence of extended radio emission around any of the
13 objects observed by both interferometers (other than possibly
AMS04). The angular resolution of the VLA observations (5 arcsec) gives us
an upper limit on the size of any jets. At $z = 2$ this corresponds to
a jet size $\ltsimeq 40$ kpc.

Another known characteristic of radio-loud galaxies is the presence of
extended line emission on similar scales to those of the radio
jets. We have not only found that the composite spectrum resembles
that of radio galaxies, but several of our objects show significantly
extended Ly~$\alpha$ (AMS03, AMS04, AMS08, AMS13 and AMS16) and some
have other extended lines (AMS13 and AMS16). In some objects
(e.g. AMS16) the slit was placed along the direction of the slightly
elongated radio axes, meaning that the emission lines could be
extended along a hypothetical jet. In most cases, however, the slit PA
was chosen for other reasons, so that any alignment with hypothetical
jets would be coincidence. This suggests that any extended line
emission is more likely to be associated with the remnants of a
merger, which could have triggered the central engine, rather than an
extended radio jet \citep[something that has also been seen in radio galaxies, e.g][]{2005MNRAS.359L...5V}.

The line ratios and the spatial extent of high-excitation lines of our
objects seem to have some similarities to radio galaxies, as we have
discussed in Section \ref{sec:composite}.  The [O III] line is an
excellent indicator of the strength of the underlying quasar continuum
\citep {1998MNRAS.297L..39S}, so we now use the \citet
{1993ARA&A..31..639M} radio galaxy line ratios to convert our line
luminosities to [O III] (5007~\AA) luminosities, to compare our
results to the radio-loud samples discussed by \citet
{2004MNRAS.349..503G}. We converted the Ly~$\alpha$ line to [O~III],
and when C~IV is present we also converted that line. This yields two
different estimates for the [O~III] luminosity which are likely to be
quite different, giving us a handle on the possible systematic errors
in converting to [O~III] luminosity, given that our composite spectrum
has some differences with the \citet {1993ARA&A..31..639M} radio
galaxy composite. The \citet {1993ARA&A..31..639M} ratios are ([O
III]/C~IV)$_{\rm RG}=2.38$ and ([O III]/Ly~$\alpha$)$_{\rm RG}=0.28$.
We also convert radio flux at 1.4 GHz to luminosity at 151 MHz,  assuming a
spectral index $\alpha^{1.4}_{151} = 0.8$ (where flux density $\propto
\nu^{-\alpha}$). The values obtained are shown in Table~\ref{l151}

\begin{table}
\footnotesize
\begin{center}
\begin{tabular}{llcccc}
\hline\hline
\mc{1}{c}{name} &\mc{1}{c}{line}& \mc{1}{c}{log$_{10}L_{\rm [O III]}$}  & \mc{1}{c}{log$_{10}L_{151}$} \\
\mc{1}{c}{} &\mc{1}{c}{used}  &\mc{1}{c}{/W} &\mc{1}{c}{/W Hz$^{-1}$ sr $^{-1}$} \\
\hline
AMS03 & Ly~$\alpha$  & 35.3 & 25.6 \\
AMS03 & Ly~$\alpha$  & 34.8 & 25.6 \\
AMS04 & Ly~$\alpha$ &  35.5 & 24.7 \\
AMS05 & Ly~$\alpha$   &  35.0 & 25.1 \\
AMS05 & C~IV  &  35.9 & 25.1 \\
AMS08 & Ly~$\alpha$  &  35.6 & 24.8 \\
AMS08 & C~IV  &  35.5 & 24.8 \\
AMS12 & Ly~$\alpha$   & 35.9 & 25.3 \\
AMS12 &  C~IV   & 35.7 & 25.3 \\
AMS13 & Ly~$\alpha$   & 35.9 & 25.3 \\
AMS13 & C~IV  & 36.0 & 25.3 \\
AMS14 & Ly~$\alpha$  & 35.4 & 24.4 \\
AMS14 & C~IV  & 35.3 & 24.4 \\
AMS16 & Ly~$\alpha$  & 36.2 &25.4 \\
AMS16 & C~IV & 37.0 & 25.4 \\
AMS17 & Ly~$\alpha$ & 35.8 & 25.3 \\
\hline \hline
\end{tabular}
\end{center}
\caption{Data used to plot Figure \ref{grimes}. The second column states which lines were used to estimate the [O III] line strength. The 151 MHz luminosities were estimated from the 1.4 GHz flux densities, assuming a spectral index $\alpha^{1.4}_{151} = 0.8$ between 151 MHz and 1.4 GHz.}
\label{l151}
\end{table}

\begin{figure}
\centerline{\psfig{file=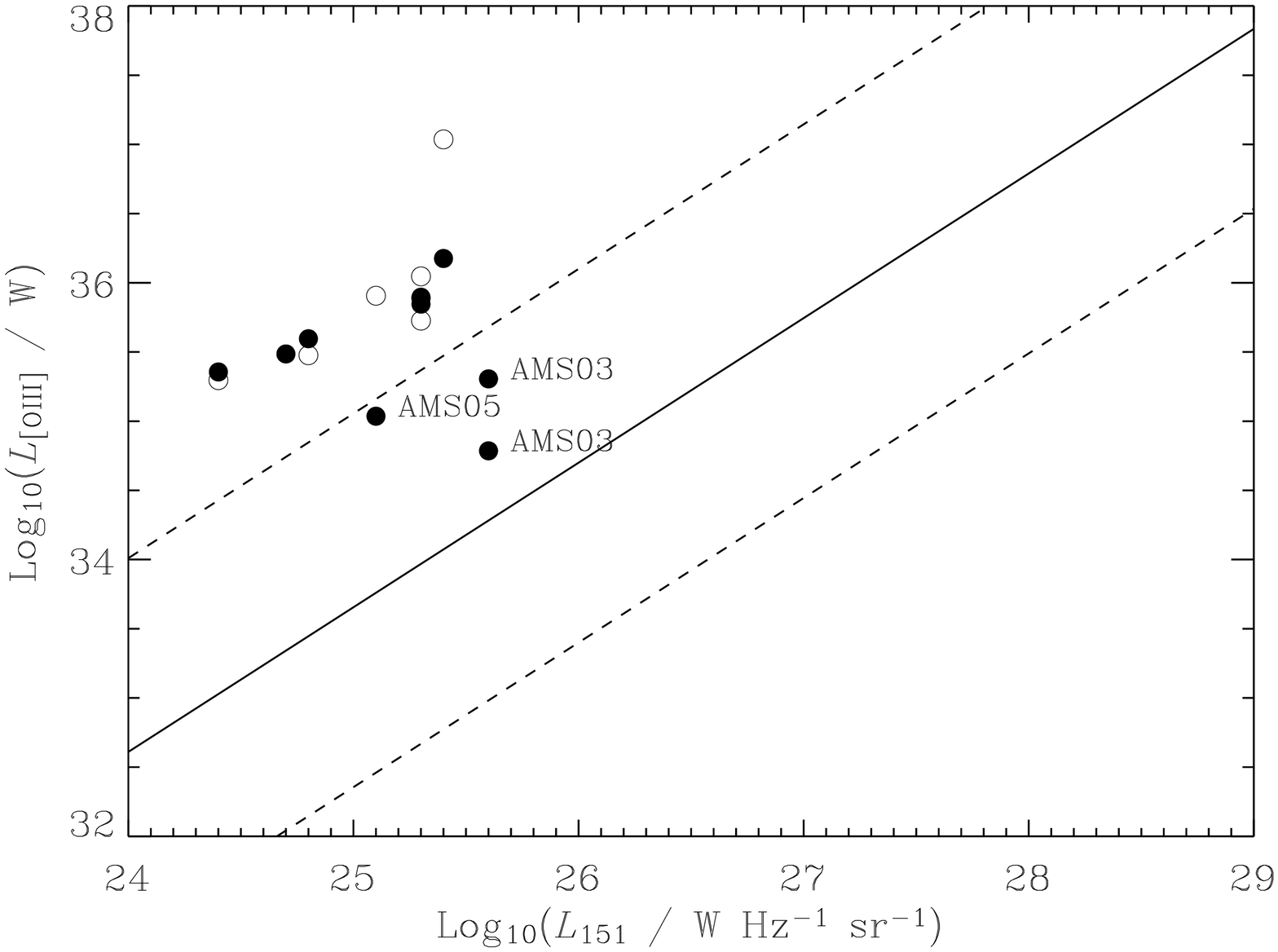,width=7cm,angle=90}}
\caption{Comparison to Figure 4 of \citet {2004MNRAS.349..503G}. The symbols represent which line has
been used to convert to [O III]: filled circles for Ly~$\alpha$, empty
circles for C~IV. The solid line represents the best fit to the
radio-loud samples (3CRR, 6CE, 7CRS) in \citet {2004MNRAS.349..503G},
with the dashed lines representing the loci which encompass all the
radio-loud sources (so similar to $\pm$3$\sigma$ contours).
}
\label{grimes}
\end{figure}

Figure \ref{grimes} shows how our objects sit on the $L_{151}-L_{[\rm
OIII]}$ plane. The solid line shows the best fit line for the
radio-loud galaxies from the 3CRR,6CE and 7CRS samples \citep[Figure 4
of][]{2004MNRAS.349..503G}. The region between the dashed lines
encompasses all of the radio loud objects, so all our objects except
AMS03 lie in a different region to them. However, some important details must
be remembered. First, we have assumed all our sources to be steep
spectrum ($\alpha^{1400}_{151} = 0.8$), something for we have no
current evidence. The true value of $\alpha^{1.4}_{151}$ is unlikely
to be much larger (i.e. steeper) but objects with flatter true spectra
will actually have lower $L_{151}$ than shown here. In addition, the
$L_{[\rm OIII]}$ for all of the objects has been estimated using the
radio-galaxy ratios from \citet {1993ARA&A..31..639M} which have
Ly~$\alpha$ relative to C~IV or He~II systematically brighter than in
our sample. The objects for which we have estimated $L_{[\rm OIII]}$
from Ly~$\alpha$ (such as AMS03) are likely to have true $L_{[\rm
OIII]}$ values brighter than those shown here. Finally, the dashed
lines in Figure \ref{grimes} are not $\pm1\sigma$ loci, they are the
lines that encompass the entire radio-loud population (i.e. they could
be considered $\sim3\sigma$ contours). Once these details are taken
into consideration, one sees that all the possible systematic errors
would only lead to brighter [O III] lines and fainter luminosity at
151 MHz. Only two objects are anywhere near the radio-loud samples:
AMS03 and AMS05. The first object has two different Ly~$\alpha$ lines,
and although we have used both for comparative reasons, the weaker of
the two lines (the narrower line of the two) is unlikely to originate
from an AGN, the stronger line should be looked at. The Ly~$\alpha$ line 
of AMS05 has a large uncertainty in it's flux measurement and is likely to be 
missing some flux since it was observed at a high airmass. 

The sample presented here has typical [O III] luminosity (in logs) of
35-36. This is comparable to the radio-loud quasars in Figure 4 of
\citet {2004MNRAS.349..503G}, however, for a similar value of
$L_{[\rm OIII]}$, the sample presented here has a $L_{151}$ fainter by
a factor of 100. It therefore differs from typical radio-selected
objects in several respects: once the obscuration of the quasar
nucleus is accounted for, they have the high accretion rates of
typical quasars, but relatively low radio luminosities. The radio luminosities 
of these objects show that these are the radio-bright end of the radio-quiet 
population, or `so called' radio-intermediate quasars.

\section{Discussion}\label{sec:discussion}

We have found a set of selection criteria that yields a population of
radio-quiet type-2 quasars at $z\sim2$. About half of the objects in
this sample show emission lines in their optical spectra. The ratios
of these emission lines are similar to those of radio galaxies with
the main difference that the C~IV line of type-2 quasars is much
stronger and consistent with individual high-redshift type-2 quasars from
the literature. The Ly~$\alpha$ is often spatially-extended, as are
sometimes the high ionisation lines. This suggests that some of the
line emission originates outside the classic narrow-line region, but whether
this is due to jet activity or remnants of a merger cannot be deduced
here.  To further compare the type-2s to radio galaxies and radio-loud
quasars, we compare their location on the $L_{151}-L_{[\rm OIII]}$
plane, and find significant differences. The type-2s have $L_{[\rm
OIII]}$ comparable to radio-loud type-1 quasars, while being fainter
at radio wavelengths by a factor of $\sim 100$. In addition, the type-2s
presented here show no signs of large extended jets (except possibly
AMS04).  Finally, comparison between our rough photometric redshift
estimation, which assumes $2.6L^{*}$ host galaxies and our
spectroscopic redshifts suggests that the objects in our sample which
yield optical spectra are consistent with having host galaxies which
are massive ellipticals. In particular changing our model elliptical
from $2.6L^{*}$ to $2L^{*}$ would bring agreement to the median
$z_{\rm phot}$ and $z_{\rm spec}$, and close to agreement for the mean
$z_{\rm phot}$ and $z_{\rm spec}$.  We have therefore found a
population of classic (radio-intermediate) type-2 quasars, showing
only narrow-lines with, presumably the torus blocking direct view of the
quasar along our line-of-sight. We refer to these objects as
`torus-obscured' type-2s.

The lack of objects with spectroscopic redshifts around $z \sim 1.5$
could plausibly be due to the Ly~$\alpha$ line being below the
atmospheric cut-off, but C~IV is detected in most  of the higher-$z$
objects (and He~II and C~II] are sometimes detectable) so the lack of
objects at this redshift may instead be associated with the silicate
absorption feature falling on the 24-$\mu$m band, as discussed in
Section~\ref{sec:Xray}.

Of the remaining 11 objects, 10 show completely blank spectra and only
one shows faint red continuum (AMS20). Their faint 3.6 $\mu$m flux
densities and large R-band magnitudes suggest these are high redshift
objects just like the `torus-obscured' objects. In our 30-minute 
integrations, type-1 quasars would have shown extremely bright
continuum and lower-redshift ($z \ltsimeq 1$) starbursts would have
shown [O~II] line emission or at least some strong continuum. Thus our
sample does not suffer from any contamination: these blank objects are
probably also high-redshift type-2 quasars.

Some of these 11 objects could be at $1.4 \leq z \leq 1.7$ with the
C~IV or He~II lines too faint to be detected, although we have seen
that 24-$\mu$m selection might disfavour this particular range of
redshifts. The remaining objects are presumably obscured by dust on a
large scale ($\sim$10pc-1kpc), which hides the narrow-line region as
well as the broad-line region. Such concentration of dust is
characteristic of star-forming regions, so we are presumably seeing
either a nuclear or a galaxy-scale starburst. From their blank optical
spectra we know that these objects are indeed at high redshift and
therefore their mid-infrared flux density is characteristic of objects
accreting at a high rate, while their radio flux density is too high
to originate only in star-forming regions and must be due to an AGN.
In two cases (AMS06 and AMS19), published mid-infrared spectra confirm
these objects to lie in the correct redshift range. In the other nine
cases, despite the lack of lines to pinpoint their exact redshift, we
are confident that these blank objects are type-2 quasars at high
redshift. The obscuration, however, is not entirely due to an
orientation effect, like for `torus-obscured' objects, but due to the
host galaxy, and so we refer to these objects as `host-obscured'
type-2s. At this stage, however, we cannot differenciate between dust
at scales of $\sim$10 pc or $\sim$1 kpc. Such objects, lacking AGN
emission lines in the optical spectra, have been found at lower
redshift in the samples of \citet
{2004ApJS..154..166L,2005MmSAI..76..154L} and \citet
{2005A&A...440L...5L}.

\begin{figure}
\centerline{\psfig{file=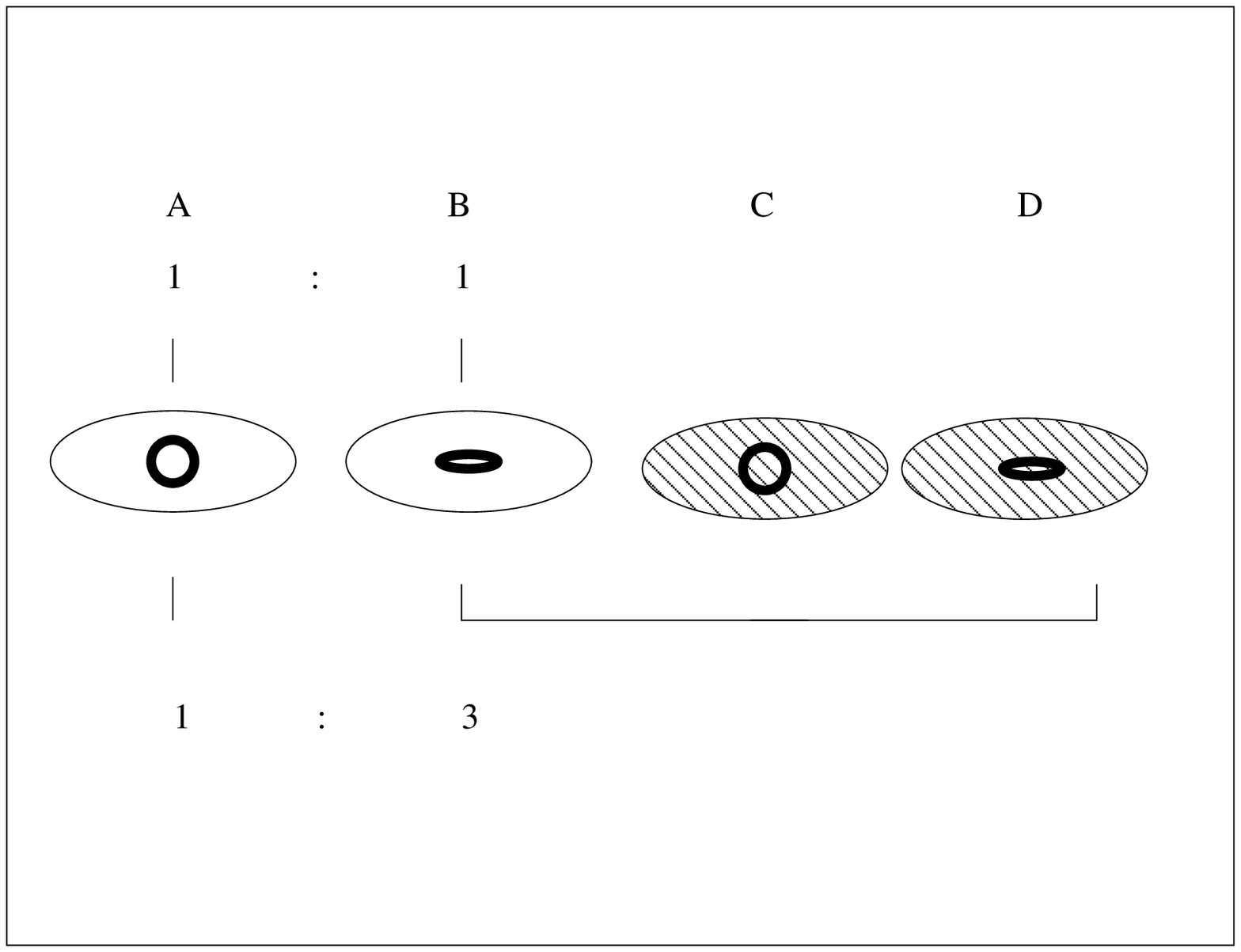,width=7cm,angle=0}}
\caption{Schematic Representation of the three types of quasar. The
empty ellipses represent a transparent host galaxy, while the filled
ellipses representing dusty ones. The solid circle represents the
torus around the quasar. A is hosted by a transparent galaxy, and
since we are viewing the torus face on, we can see the broad line
region. It is a type-1 quasar. B is also hosted by a transparent
galaxy, but since the torus is edge on, we only see the narrow line
region. This is a torus-obscured type-2. C has a favourable geometry,
while in D the torus is in the line of sight, but since both are
obscured by large scale dust in the galaxy, we cannot see the central
region, so they are both host-obscured type-2s. From \citep
{2005Natur.436..666M} the ratio of B to A is $\sim$1:1, while the
ratio of B,C and D (together) to A is $\sim$2-3:1. We note that
although we are calling C and D `host'-obscured, the dust could
plausibly be due to a nuclear starburst, with a characteristic scale
$\sim 100$ pc, as well as a galaxy-scale starburst.}
\label{sketch}
\end{figure}

It is therefore not surprising that the Ly~$\alpha$ (whether from the
narrow-line or star-forming regions) in the host-obscured type-2s is
too obscured for our exposure times (30 minutes) with a 4-m telescope:
Successful spectroscopy of sub-mm galaxies in the blue requires very
long integrations (1.5-6 hours) with 8-m or 10-m class telescopes
\citep[e.g.][] {2005ApJ...622..772C}. We do note that AMS03 shows
spatially-extended Ly~$\alpha$, no other lines and no continuum, so
there is a chance this object is host-obscured and the second
Ly~$\alpha$ line traces the starburst: this requires further
study.

In \citet {2005Natur.436..666M}, the type-2 to type-1 ratio at $z \geq
2.00$ was modelled by predicting the number of type-1 quasars which
would follow the 24-$\mu$m and radio criteria. The 3.6-$\mu$m
criterion was simply assumed as a way of rejecting type-1s as well as
providing a rough photometric redshift. The number of $z \geq 2.00$
type-1 quasars meeting the 24-$\mu$m was predicted from the B-band
luminosity function of \citet {2003A&A...408..499W}, converting from
B-band to 24-$\mu$m using the \citet {1995MNRAS.272..737R} SED. The
optical-to-radio correlation of \citet {2003MNRAS.346..447C} was then
used to predict what fraction of the 24-$\mu$m detected type-1s would
also meet the radio criterion. This modelling predicted
4.3$^{+2.2}_{-1.1}$ such type-1 quasars, at $z \geq 2.00$ and in our
3.8 deg$^{2}$ field. There are 5 narrow-line objects in our sample
with spectroscopically confirmed $z \geq 2.00$, and the quasar type-2
to type-1 ratio for narrow-line (torus-obscured) type-2s and type-1
quasars is therefore $\sim$1:1. For the blank objects, the crude
photometric redshifts can be used to estimate, on average, how many
blank (host-obscured) objects are at $z \geq 2$. There are 6 such
objects, and so in \citep {2005Natur.436..666M} the ratio of total
number of type-2s (host- and torus-obscured) to type-1s was estimated
to be 11/4 or $\sim$3:1. Since then AMS06 has been shown
spectroscopically to lie at $z \leq 2.00$ the current number of
objects at $z \geq 2.00$ is 10. From the errors in the modelled number
of type-1 quasars, and the Poisson errors in our sample, the ratio of
type-2 to type-1 quasars is found to be 2.3$^{+1.8}_{-1.2}$. In a
future paper we will calculate better photometric redshifts, to try
and obtain a more reliable estimate of the quasar fraction at $z \geq
2$.

The `two types of type-2' quasar, and a type-1 quasar are schematically
represented in Figure \ref {sketch}.  For a quasar to be of type-1 it
has to lie in a relatively dust free galaxy (a `transparent' galaxy)
as well as having a favourable orientation to the torus (case A in
Figure \ref {sketch}). A quasar in a transparent galaxy, with an
unfavourable angle to the torus will not outshine its host galaxy, but
the narrow-line region will still be visible (case B: a torus-obscured
type-2). When the host galaxy is dusty enough to obscure the quasar
(cases C and D) it is no longer possible to see the central region,
whether we have a favourable angle to the torus (C) or not
(D). Objects of case C and D are likely to yield blank spectra, and
are indistinguishable from each other with the current
dataset. Observations at radio frequencies, however, could shed some
light on the orientation if radio jets are present.

This simple picture is likely to be applicable only to radio-quiet
objects, since the powerful radio jets are likely to be capable of
removing dust on large scales
\citep{2002ApJ...568..592B,2002MNRAS.331..435W}. Therefore, although
the ratio of type-2 to type-1 is likely to be $\sim$2-3:1 for
radio-quiet samples (and for all non-radio-selected samples, since
radio-quiet objects will dominate), it should be closer to 1:1 for
radio-loud samples. Thus radio-loud samples will have few
`host-obscured' quasars \citep [cases C and D, although some are
probably present in the sample of ] [] {2001MNRAS.324....1W} and the
quasar fraction measured in these radio-loud samples will be dominated
by the geometry of the torus. This picture is therefore consistent
with the ratio of narrow-line to broad-line quasars predicted by
unified models and found for radio-loud objects \citep [$\sim$1:1
e.g][] {2000MNRAS.316..449W}.

Inferring a type-2 to type-1 ratio $\sim$2-3:1 for the sample presented
in this article does not disagree with the receding-torus models
\citep[e.g.][] {1991MNRAS.252..586L,2005MNRAS.360..565S}, since the obscuring mechanism
for `host-obscured' type-2s is unrelated to the torus orientation or
geometry. The ratio of torus-obscured (B) to type-1 (A) quasars
predicted for our sample is $\sim$1:1 \citep [and indeed consistent
with the receding-torus: Figure 2 of ][]{2005Natur.436..666M}. This
phenomenon of host-obscuration might also be present in the Seyfert-2s
at $z \sim 1$ that dominate the hard X-ray background, in which case
care must be taken to separate the obscuring mechanism necessary to
explain the  AGN dominating the X-ray background from
modifications to the unified schemes such as the receding torus.

The ratio of $\sim$2-3:1 found for this sample is similar to the ratio
required for Seyfert-2s to Seyfert-1s at $z \sim 1$ to fit the hard
X-ray background, although a ratio $\sim$2-3:1 for the quasars is not
required by the hard X-ray background \citep {2003ApJ...598..886U,
2005ApJ...630..115T}.  We do note, that due to the range in
gas-to-dust ratios, some of the X-ray obscured quasars that contribute
to the hard X-ray background are not necessarily type-2s in the
optical sense (they might only be reddened quasars with $A_{\rm V}
\sim 1-3$ but a large gas-to-dust ratio). However, at least in our
sample, we find the ratio of all type-2s (B, C and D) to type-1 (A) is
$\sim$2-3:1: a higher type-2 to type-1 ratio than is found in deep
X-ray surveys.  This suggests that indeed, 24-$\mu$m selection can
find type-2 quasars that X-ray selection in the 2-10 keV band has not
been able to detect, and therefore a number of the type-2 quasars
presented here should be Compton-thick. This, of course, can be tested
via X-ray follow-up of this or a similar sample.

\section*{Acknowledgments}

We would like to warmly thank Richard Wilman for access to his absorbed X-ray
spectra and for useful comments, the anonymous referee for valuable
suggestions, and Caroline van Breukelen for help with the
overlays. AMS would like to thank the Council of the European Union
for support, SR and CS would like to thank the UK PPARC for a Senior
Research Fellowship and an Advanced Fellowship respectively. This work
is based on observations made with the Spitzer Space Telescope, which
is operated by the Jet Propulsion Laboratory, California Institute of
Technology. The William Herschel Telescope (WHT) is operated on the
island of La Palma by the Isaac Newton Group in the Spanish
Observatory del Roque de los Muchachos of the Instituto de Astrof\'\i
sica de Canarias.

\label{lastpage}

\end{document}